\documentclass[12pt,a4paper,english]{article}
\usepackage[T1]{fontenc}
\usepackage[latin1]{inputenc}
\usepackage{float}
\usepackage{amsmath}
\usepackage{amsfonts}
\usepackage{latexsym}
\usepackage{graphicx}

\newcommand{\ZZ}{\mathbb{Z}}
\newcommand{\RR}{\mathbb{R}}

\newcommand{\ket}[1]{|#1 \rangle}
\newcommand{\braket}[2]{\langle #1 | #2 \rangle}

\newcommand{\brakettt}[3]{\langle #1 | #2 |#3 \rangle}

\newcommand{\rme}{\mathrm{e}}
\newcommand{\rmi}{\mathrm{i}}
\newcommand{\rmd}{\mathrm{d}}

\usepackage{babel}

\begin{document}

\thispagestyle{empty}
\vspace*{0.5cm}
{
\LARGE
\bf
\noindent
A study of truncation effects in boundary flows\\[2mm] 
of the
Ising model on a strip
}\\[1cm]

\noindent
\hspace{2cm}{\bf \large G.\ Zs.\ T\'oth}\\[-3mm]

\noindent
\hspace{2cm}{\small Theoretical Physics Research Group of the Hungarian Academy of 
Sciences}\\
\hspace*{2cm}{\small at E\"otv\"os University,
 1117 Budapest, P\'{a}zm\'{a}ny P\'{e}ter S\'{e}t\'{a}ny
1/A, Hungary}\\[-2mm]

\noindent
\hspace{2cm}{\small E-mail: tgzs@ludens.elte.hu}\\

\noindent
\hspace{2cm}{\small 23 April 2007}\\
\hspace*{2cm}{\small Journal reference: J.\ Stat.\ Mech.\ (2007) P04005}\\[1.5cm]

\noindent
{\small 
{\bf Abstract.}
We investigate the idea that the effect of the truncation applied in the TCSA
method on the spectrum coincides with the effect of a suitable changing
of the coefficients of the terms in the Hamiltonian operator. The
investigation is done in the case of the critical Ising model on a strip with an
external magnetic field on one of the boundaries.  A detailed quantum field
theoretical description of this model is also given, and we propose a
description as a perturbation of the infinite coupling limit. The investigation is also
carried out for a truncation method which preserves the
solvability of the model. The results of perturbative and numerical
calculations presented support the above idea and show that the qualitative behaviour of the truncated 
spectrum as a function of the coupling constant depends on the truncation 
method.} \\[3mm]

\noindent
{\small
{\it Keywords\/}: integrable quantum field theory, conformal field
theory, finite-size scaling, renormalization group }

\newpage
\tableofcontents

\section{Introduction}

The main subject of this paper is an  investigation of the  TCSA
(truncated conformal space approach) method, which is a numerical method 
introduced in \cite{YZ1}
for the
calculation of the spectra and eigenvectors of Hamiltonian operators of 
perturbed conformal field 
theories in finite volume in two spacetime dimensions.
The Hamiltonian operators of these theories have the 
form $H=H_0+H_I$, where $H_0$ has a known discrete spectrum. The method
consists in (numerically)
diagonalizing the finite matrix $H^{TCSA}(n_c)=P(n_c)(H_0+H_I)P(n_c)$, where $P(n_c)$ is the
projection on the subspace spanned by the eigenstates of $H_0$ belonging to
the lowest $n_c+1$ energy levels, and $n_c$ is the truncation parameter. This gives an approximation to the
eigenvalues and eigenvectors of $H$.

The uses of the data obtained by TCSA include the  verification of results obtained by
other methods, for example in 
\cite{FRT1},
the extraction of resonance widths \cite{PT},  the mapping of the phase structure of
certain quantum field theories as in \cite{BPTW,sajat2}, and the finding of renormalization group
flow fixed points as in \cite{Kormos}.  Our
investigation is related to the  use of TCSA
for the study of boundary renormalization group flows (i.e.\ of flows generated by boundary perturbations) between minimal boundary 
conformal
field theories  defined on the strip $[0,L]\times
\RR$.
A brief
review of the areas where boundary
conformal field theory and flows between them play important role can be found in \cite{RRS}.

By boundary flow we mean a one-parameter family of models, the parameter being the
width (or 'volume') $L$ of the strip.  In the simplest case the parameter
can  be taken to be  a coupling constant $h$ instead of $L$ and the models have the Hamiltonian operators
$H=H_0+hH_I$. $H_0$ is  the Hamiltonian operator of a boundary conformal field theory, $h$ is allowed to vary from $0$ to $\infty$ or from $0$ to
$-\infty$. $H_I$ is a relevant boundary field taken at a certain initial time. 

A particular problem of interest is that of finding values of $h$ other than $0$,
called fixed points,  where the model
corresponding to  $H_0+hH_I$ is a conformal field theory, and identifying these
conformal field theories.  The  TCSA can be used for this purpose if the Hilbert space of the  conformal field theory
corresponding to $h=0$ consists of finitely many (or countably many)
irreducible representations of the
Virasoro algebra. Boundary conformal minimal models 
satisfy
this condition. We refer the reader to \cite{Watts3,Watts2,Watts4,Watts1,Kormos} for boundary flows and the application of
the TCSA to study them.

Regarding the spectra at nonzero values of $h$,
conformal symmetry can be recognized by  the equal distances between neighbouring energy levels; 
the representation content can be identified by the degeneracy of
the energy levels.  It should be
noted that perturbation
theory and other methods can also be used (see e.g.\ \cite{RRS, GW, Fr, LSS,
  CAZ, Kl-M}) to explore flows. An important 
problem regarding the
TCSA is that it gives  approximate data, and our knowledge of the precise relation
between this data and the exact spectrum is still limited (see \cite{CLM} for already existing results).  
A good understanding of the
effect of the truncation could be used to improve the TCSA data and to explain
the qualitative features of TCSA pictures of flows, and generally it would make the
results obtained using TCSA data better founded. 

An idea proposed by G.M.T.\ Watts and K.\ Graham \cite{prc,talk}
recently is that the
effect of the truncation on the spectrum is equivalent to a
suitable change of the coefficients of the terms in $H$, i.e.\ the spectrum of
$H^{TCSA}(h,n_c)$  is equal, in an approximation at least, to the spectrum of
\begin{equation}
\label{eq.i}
H^r(h,n_c)=s_0(h,n_c)H_0+s_1(h,n_c)H_I+s_2(h,n_c)H_{I,2}+\dots\ ,
\end{equation}
which we shall call
renormalized Hamiltonian operator. 
$s_0,s_1,\dots$ are suitable functions and $H_{I,2},\dots$ are suitable operators. $H_{I,2},\dots$
should be primary or descendant bulk or boundary operators. Our main purpose
in this paper is to investigate
the validity of this picture.            

We do this investigation in the case of a relatively simple model:
the  perturbed boundary conformal field theory on the strip
$[0,L]\times \RR$
with Hamiltonian operator  
\begin{equation}
\label{eq.yyy}
H=\frac{\pi}{L}L_0+hL^{-1/2}\phi_{1/2}(x=L,t=0)\ .
\end{equation}
The unperturbed ($h=0$) model is the $c=1/2$ unitary conformal minimal model, i.e.\ the
continuum limit of the critical Ising model, 
with the Cardy boundary condition
$0$ on the left and $1/16$ on the right. $\frac{\pi}{L}L_0$ is the Hamiltonian
operator of this model. $L_0$ is the 'zero index' Virasoro generator. $L$ will be kept fixed at the value $L=1$.  
The Hilbert space of the unperturbed model is the single $c=1/2$, $h=1/16$
irreducible highest weight representation of the  Virasoro
algebra. (Here and later on the coupling constant and the highest
weight are both denoted by
$h$, but it should be clear from the context which one is meant.) 
The field $\phi_{1/2}(L,t)$ is the
weight $1/2$  boundary primary field on the right boundary, which is
also known in the literature as the  boundary spin operator
 \cite{Cardy1,CL,GZ}.
 The coupling constant $h$ can also be regarded as a constant external
boundary magnetic field, which is coupled to the boundary spin operator. The
model (\ref{eq.yyy}) is also referred to as the critical Ising model on the
strip with boundary magnetic field. 
The perturbation $h\phi_{1/2}(L,t=0)$ violates the conformal symmetry, which is nevertheless
restored in the $h\to \pm\infty$
limit. It is known that in the $h \to \infty$ limit the $c=1/2, h=1/2$
representation is realized; in the $h \to -\infty$ limit the $c=1/2, h=0$
representation is realized (see e.g.\ \cite{GW,Fr,GZ,AL2}). 
This can be written in a shorthand form as 
\begin{eqnarray}
\label{eq.flow1}
(1/2,1/16)+\phi_{1/2} &  \to &  (1/2,1/2)\\
\label{eq.flow2}
(1/2,1/16)-\phi_{1/2} &  \to &  (1/2,0)\ .
\end{eqnarray}
We have chosen the model (\ref{eq.yyy})  because it is integrable, furthermore
it is relatively easy to handle;
in particular the spectrum can be calculated analytically in terms of simple
functions and the application of
Rayleigh-Schr\"odinger perturbation theory  is also relatively
easy. Investigations in other perturbed conformal minimal models, especially
in the case of the tricritical Ising model and generally in the case of a
perturbation by the field $\phi_{(13)}$, are carried
out in \cite{FGPW} (see also \cite{talk}). 

It should be noted that the Ising model with boundary magnetic field has been
studied much, especially on the lattice and on the half-line.
We refer the reader to \cite{C2,GZ,C1,CZ} and the references in them for further
information. 

To recognize conformal symmetry and to identify the representation content it is sufficient to look at the ratios of the energy gaps; therefore
one often considers normalized spectra  which are obtained by subtracting 
the ground state energy and dividing by the lowest energy gap. 
The normalized exact and  TCSA spectra for the flows (\ref{eq.flow1}), (\ref{eq.flow2}) as a function of the logarithm of $h$ can
be seen in Figure \ref{fig.tcsa2}. 
An interesting feature of these TCSA spectra is that
they appear to correspond to the  flows $(1/2,1/16) \to (1/2,1/2) \to (1/2,
1/16)$ and   $(1/2,1/16) \to (1/2,0) \to (1/2,
1/16)$, i.e.\ second flows appear to be present after the normal flows. 
Flows in other minimal models  also show this behaviour \cite{talk,FGPW}.   
One application of the picture (\ref{eq.i}) could be the explanation of this phenomenon.

Following Watts' proposal  based on the look of the TCSA spectra shown
in
Figure \ref{fig.tcsa2}  we
assume that only the first two terms are nonzero in (\ref{eq.i}): 
\begin{equation}
\label{eq.i2}
H^r(h,n_c)=s_0(h,n_c)H_0+s_1(h,n_c)H_I\ .
\end{equation}

In summary,  we look for answers for the following
questions: 1.\ Does the spectrum of (\ref{eq.i2}) agree with the TCSA spectrum in some
approximation with a suitable choice of
the functions  $s_0$ and $s_1$? 2.\ How can we explain the 'second' flows in
the TCSA spectra? 
   
We determine the  exact spectrum of  (\ref{eq.yyy}) using an essentially  known (see e.g.\
\cite{GZ,C1,CZ,LMSS,Kon,KLeM})  quantum field theoretic representation of
this operator. In this representation the operator
(\ref{eq.yyy}) is a quadratic expression of fermionic fields. We extract
the spectrum from the (quantum) field equations, which are linear.  

Concerning the TCSA Hamiltonian operator, the main difficulty we encounter is that it is hard to handle the TCSA spectra 
analytically
even if the non-truncated model is exactly solvable. Therefore,
hoping that we can gain some insight by looking at a similar but exactly solvable
truncation method, we tried another
truncation method which we call mode truncation. The mode truncated model can
be solved exactly, but it turns out, rather unexpectedly, that the behaviour
of the spectrum for large values of $h$ is different from the behaviour of the
TCSA spectrum, namely the  qualitative behaviour of the mode truncated
spectrum is very similar to that of the exact spectrum; the second flows are
not present. This, besides leaving the second question open, raises the
problem of finding the possible   behaviours  for large values of $h$ and their dependence on the truncation
method, and whether the mode truncation method can be generalized to other models.

We also  apply the Rayleigh-Schr\"odinger perturbation theory  to verify
the validity of (\ref{eq.i2}) for both the TCSA and mode truncation methods.

A further calculation that we do is
a  numerical comparison of the exact and TCSA spectra. We perform the same
calculation for the mode truncation scheme as well.

Our secondary aim in this paper, which is partially independent of the problem
of the TCSA approximation,  is to provide a 
detailed description of the 
quantum field theoretic
representation of (\ref{eq.yyy}).
 Although this  was studied in the literature (e.g.\ in \cite{LMSS} and especially  in
\cite{C1} (at finite temperature), and on the half-line in several papers),
we give a description in which the perturbed Hamiltonian operator structure, on
which the TCSA is based, is emphasized and observed consistently. In a
calculation using the 
Bethe-Yang equations (which we also present in the appendix) or in the approach of \cite{C1} where  boundary
conditions play central role,  the link with the TCSA formulation would not be
entirely obvious. Furthermore, the formulation we present  is
also suitable for Rayleigh-Schr\"odinger perturbation theory and for the treatment of the mode
truncated version. 
We do not consider the classical level of the field theoretic model, mainly
because it is irrelevant to our problem. We remark that  the same model with massive
unperturbed part (see e.g.\ \cite{C2}) could be studied  along the same lines.
The  field theoretic approach also raises the problem of
defining distributions (or similar objects) on a closed interval. We do not know of a systematic
exposition of this subject (neither for a closed interval nor for the half-line), although it would be needed for  boundary
field theory.  
In this paper we use distributions on  closed intervals, 
restricting to the most necessary formulae only. These formulae are collected
in Appendix \ref{sec.distr}.

We remark that our TCSA program relies on the conformal transformation
properties only and does   
not make use of the 
representation mentioned above.

The contents of the paper are the following:

Section \ref{sec.eel} contains the field theoretical description of the model
(\ref{eq.yyy}), in particular the calculation of the exact spectrum.
Further details concerning the boundary conditions, the normalization of interacting
creation and annihilation operators, 
the relation between free and
interacting creation and annihilation operators, matrix elements of the
interacting fields,
identities for the Dirac delta on the strip and the expression of eigenstates in
terms of the unperturbed eigenstates are deferred to Appendix \ref{app.folyt}.

In Section \ref{sec.reverse}
we propose a description of (\ref{eq.yyy}) as a perturbation of the $h\to
\pm \infty$ limiting case, which cannot be found in the literature. An
interesting feature of this description, which we call reverse description, is that
the  perturbing operator is non-relevant. We
calculate the exact spectrum in a similar way as in the case of the standard
description mentioned above.  This reverse description is motivated by the presence of the second (reverse) flows
in the TCSA spectra. 

In Section \ref{sec.modetr} we describe  the mode truncated model and 
calculate  its spectrum. This is done along the same lines as in the
non-truncated case.

Section \ref{sec.pertres} contains the perturbative results for the $s_0$ and $s_1$
functions.

Section \ref{sec.numres} contains the results of the numerical test of the approximation by
(\ref{eq.i2}) for the TCSA and the mode truncation schemes. 

The paper contains further appendices beyond those mentioned above.

In Appendix \ref{app.by} we  describe  the spectrum using
the Bethe-Yang equations. We find that they  give the exact result in
this case. 

Appendix \ref{sec.RS} contains the power series for the exact, TCSA and mode
truncated energy levels up to third
order in $h$ which we obtained by means of the Rayleigh-Schr\"odinger perturbation theory.    

Conclusions and some directions for future work are given in Section \ref{sec.disc2}.

\section{The exact spectrum}
\label{sec.eel}

In this section we present a quantum field theoretic description of the
model (\ref{eq.yyy}) and calculate its energy spectrum.

\subsection{The free model}
\label{sec.freemod}

The defining constituents of the unperturbed model are the following:
two  fermion fields $\Psi_1(x,t)$ and $\Psi_2(x,t)$ and a fermionic  operator $A_2(t)$ with the anticommutators
\begin{align}
\label{ac1}
\{ \Psi_1(x,t),\Psi_1(y,t)\} &=  \{ \Psi_2(x,t),\Psi_2(y,t)\} = 4L\delta(x-y)\\
\{ \Psi_1(x,t),\Psi_2(y,t)\} &=  -4L[\delta(x+y)+\delta(x+y-2L)]\\
\{ A_2(t), \Psi_1(x,t) \} &= \{ A_2(t), \Psi_2(x,t) \} =  0\\
\{ A_2(t),A_2(t) \} &=  2\  \label{ac2}
\end{align}
and the relations
\begin{equation}
\Psi_1(x,t)^\dagger =\Psi_1(x,t)\qquad \Psi_2(x,t)^\dagger =\Psi_2(x,t)\qquad
A_2(t)^\dagger =A_2(t)\ ,
\end{equation}
the Hamiltonian operator
\begin{equation}
H_0=-\frac{\rmi}{8L}\int_0^L\ \rmd x\ \Psi_1(x,0)\partial_x \Psi_1(x,0)+\frac{\rmi}{8L}\int_0^L\
\rmd x\ \Psi_2(x,0)\partial_x \Psi_2(x,0)\ ,
\end{equation}
the equations of motion
\begin{align}
\label{mozg1}
\frac{\rmd}{\rmd t}A_2(t) &=  [\rmi H_0,A_2(t)]=0\\ 
\label{mozg2}
\partial_t\Psi_1(x,t) & =   [\rmi H_0,\Psi_1(x,t)]\nonumber\\ 
& =   -\partial_x
\Psi_1(x,t)+\frac{1}{2}[\Theta(-x)+\Theta(x-L)][-\partial_x\Psi_2(x,t)+\partial_x\Psi_1(x,t)]\nonumber\\
&\ \ \ \ +\frac{1}{2}[-\Psi_1(0,t)-\Psi_2(0,t)]\delta(x)+\frac{1}{2}[\Psi_1(L,t)+\Psi_2(L,t)]\delta(x-L)\ ,
\end{align}
\begin{align}
\label{mozg3}
\partial_t\Psi_2(x,t) &=  [\rmi H_0,\Psi_2(x,t)]\nonumber\\
&=  \partial_x
\Psi_2(x,t)+\frac{1}{2}[\Theta(-x)+\Theta(x-L)][-\partial_x\Psi_2(x,t)+\partial_x\Psi_1(x,t)]\nonumber\\
&\ \ \ \  +\frac{1}{2}[\Psi_1(0,t)+\Psi_2(0,t)]\delta(x)+\frac{1}{2}[-\Psi_1(L,t)-\Psi_2(L,t)]\delta(x-L)\ .
\end{align}
The fermion fields, which are one-component real fermion fields with
zero mass,  have the following expansion:
\begin{align}
\Psi_1 (x,t) & =  \sum_{k\in\frac{\pi}{L}\ZZ,\ 
  k>0}\sqrt{2}[a(k)\rme^{\rmi k(t-x)}+a^\dagger (k)\rme^{-\rmi k(t-x)}]+A_1\\
\Psi_2 (x,t) & =  \sum_{k\in\frac{\pi}{L}\ZZ,\ 
  k>0}\sqrt{2}[-a(k)\rme^{\rmi k(t+x)}-a^\dagger (k)\rme^{-\rmi k(t+x)}]-A_1
\end{align}
\begin{align}
\label{a1}
a(k)^\dagger &= a(-k) &  A_1^\dagger &=A_1\\ 
\{ a(k_1),a(k_2)\} &= \delta_{k_1,-k_2} & 
\{ a(k),A_1 \} &= 0\\
\{A_2,A_1 \} &=  0 &
\{ A_1,A_1 \} &=  2\\
\{ A_2, a(k) \} &= 0\ .
\label{a2}
\end{align}
The fermion fields and $A_2$ are dimensionless.

A unitary representation for the above operator algebra is defined by the
following formulae:
an orthonormal basis of the Hilbert space is
\begin{equation}
\label{eq.freebasis}
a(k_1)a(k_2)a(k_3)\dots a(k_n)\ket{u}\qquad 
a(k_1)a(k_2)a(k_3)\dots a(k_n)\ket{v}\ ,
\end{equation}
where $k_1>k_2>k_3>\dots >k_n>0$, $k_i \in \pi \ZZ/L$ and $n\ge 0$,
\begin{align}
A_1a(k_1)a(k_2)a(k_3)\dots a(k_n)\ket{u} &=  (-1)^n a(k_1)a(k_2)a(k_3)\dots a(k_n)\ket{v}\\
A_1a(k_1)a(k_2)a(k_3)\dots a(k_n)\ket{v} &=  (-1)^n a(k_1)a(k_2)a(k_3)\dots a(k_n)
 \ket{u}\\
A_2a(k_1)a(k_2)a(k_3)\dots a(k_n)\ket{u} &=  (-1)^n (-\rmi ) a(k_1)a(k_2)a(k_3)\dots a(k_n) \ket{v}\\
A_2a(k_1)a(k_2)a(k_3)\dots a(k_n) \ket{v} &=  (-1)^n (+\rmi )
a(k_1)a(k_2)a(k_3)\dots a(k_n)\ket{u}\ .
\end{align}
The condition  $k_1>k_2>k_3>\dots >k_n>0$, $k_i \in \pi \ZZ/L$ and $n\ge
0$ also applies below  if not stated otherwise. $a(k)\ket{u}=0$,
$a(k)\ket{v}=0$ if $k<0$. 
The energy eigenvalue of a basis vector $a(k_1)a(k_2)a(k_3)\dots a(k_n)
  \ket{u}$ is
$\sum_{i=1}^n k_i$, 
and the same applies if $u$ is replaced by $v$.

The following boundary conditions are satisfied by the fermion fields:
\begin{equation}
\brakettt{E_1}{\Psi_1(0,t)+\Psi_2(0,t)}{E_2}=0\qquad
\brakettt{E_1}{\Psi_1(L,t)+\Psi_2(L,t)}{E_2}=0\ ,
\end{equation}
where $\ket{E_1}$ and $\ket{E_2}$ are arbitrary energy eigenstates.

A representation of the Virasoro algebra can  be defined on the Hilbert
space in the following way (which is well known essentially in conformal field
theory; see \cite{FMS,Ginsparg}):
\begin{align}
\label{eq.ln1}
L_{N} & =  \frac{L}{2\pi}\sum_{k\in\frac{\pi}{L}\ZZ}
-ka(-k)a(k-\frac{N\pi}{L})\\
\label{eq.ln1b}
L_0 & =  \frac{L}{2\pi}\sum_{k\in\frac{\pi}{L}\ZZ} [-k:a(-k)a(k):]+\frac{1}{16}\ ,
\end{align}
where $N=\pm 1, \pm 2, \dots $, $a(0)=(1/\sqrt{2})A_1$, 
$::$ denotes the normal ordering for fermionic creation and annihilation
operators, $:a(-k)a(k):=a(-k)a(k)$ if $k<0$,  $:a(-k)a(k):=-a(k)a(-k)$ if 
$k>0$.
It can be verified that in the above representation of the $a(k)$ and $A_2$ operators $L_N$, $N\in \ZZ$ satisfy the usual relations of the
generators of the  Virasoro algebra with central charge  $c=1/2$.
$A_2$ commutes with the $L_N$. $\ket{v}$ and $\ket{u}$ are highest weight states with weight
$1/16$ and the Hilbert space decomposes into two copies of the $M(c=1/2, h=1/16)$
unitary highest weight representation of the Virasoro algebra. The 
invariant subspace belonging to $u$  is spanned by the vectors 
$a(k_1)a(k_2)\dots a(k_n)\ket{u}$ with $n$ even and   
$a(k_1)a(k_2)\dots a(k_n)\ket{v}$ with $n$ odd. The 
invariant subspace belonging to $v$  is spanned by the vectors 
$a(k_1)a(k_2)\dots a(k_n)\ket{u}$ with $n$ odd and   
$a(k_1)a(k_2)\dots a(k_n)\ket{v}$ with $n$ even. These subspaces will be called
$u$ and $v$ sectors. 
The relation between $H_0$ and $L_0$ is 
\begin{equation}
H_0=\frac{\pi}{L}L_0-\frac{1}{16}\frac{\pi}{L}\ .
\end{equation}
The fields $\Psi_1$ and $\Psi_2$ can be written as
$\Psi_1(z)  =  \sqrt{2} \sum_{n \in \ZZ} \tilde{a} (n) z^n$,
$\Psi_2(\bar{z})  =  \sqrt{2}\sum_{n\in\ZZ} -\tilde{a}(n) \bar{z}^n$,
where $\tilde{a}(n)=a(k)$, $k=n\pi/L$, $z=\exp[\rmi\frac{\pi}{L}(t-x)]$, $\bar{z}=\exp[\rmi\frac{\pi}{L}(t+x)]$.
For 
$\epsilon(z)=\Psi_1(z)/\sqrt{z}$ and $\bar{\epsilon}(\bar{z})=\Psi_2(\bar{z})/\sqrt{\bar{z}}$ we have 
$[L_N,\epsilon(z)]=z^{N+1}\frac{\rmd\epsilon}{\rmd z}(z)+\frac{1}{2}(N+1)z^N\epsilon(z)$
and
$[L_N,\bar{\epsilon}(\bar{z})]=\bar{z}^{N+1}\frac{\rmd\bar{\epsilon}}{\rmd\bar{z}}(\bar{z})+\frac{1}{2}(N+1)\bar{z}^N\bar{\epsilon}(\bar{z})$,
i.e.\ $\epsilon(z)$ and  $\bar{\epsilon}(\bar{z})$ are chiral Virasoro primary
fields of weight $1/2$. 
The same relations apply to the fields 
$A_2\epsilon$ and $A_2\bar{\epsilon}$, in particular $A_2\epsilon$ and
$A_2\bar{\epsilon}$ are also chiral primary fields of weight $1/2$.
$A_2\epsilon$ and
$A_2\bar{\epsilon}$ have zero matrix elements between the $u$ and $v$ sector,
whereas  $\epsilon(z)$ and  $\bar{\epsilon}(\bar{z})$ have zero matrix
elements within the $u$ and $v$ sectors. 
In the above equations the domains of $z$ and $\bar{z}$ are extended to the
whole complex plane.

The  operator $A_2$ is an auxiliary operator and if it is  omitted, then it is possible to represent
the fields so that the Hilbert space is a single Virasoro module $(1/2,1/16)$.
It is the next section where the presence of $A_2$ will be really useful.

\subsection{The perturbed model}
\label{sec.pertmod}

\noindent
The Hamiltonian operator is
\begin{multline}
\label{eq.pertham}
H  =  -\frac{\rmi}{8L}\int_0^L\  \Psi_1(x,0)\partial_x \Psi_1(x,0)\ \rmd x\
+\frac{\rmi}{8L}\int_0^L\ \Psi_2(x,0)\partial_x \Psi_2(x,0)\ \rmd x\\
 +h\rmi A_2(0)[\Psi_2(L,0)-\Psi_1(L,0)]\ ,
\end{multline}
where $h$ is a coupling constant of dimension $mass$.
The perturbing term
\begin{equation}
hH_I=h\rmi A_2(0)[\Psi_2(L,0)-\Psi_1(L,0)]
\end{equation}
has zero matrix elements
between vectors belonging to different sectors,
which means that $H$ can be
restricted to the $u$ and $v$ sectors separately.
These restrictions are denoted by $H|_u$ and $H|_v$.
$H_I$ is also  a primary boundary
field of weight $1/2$ with respect to the Virasoro algebra representation defined in the
previous section  taken at $t=0$. The matrix elements of
$H_I$, i.e.\ of $H_I|_u$ and $H_I|_v$, are uniquely determined by this property and
by the values of the matrix elements  $\brakettt{u}{H_I}{u}$ and
$\brakettt{v}{H_I}{v}$. It is easy to verify that
$2=\brakettt{u}{H_I}{u}=-\brakettt{v}{H_I}{v}$, and there exists an
intertwiner $Y$
of the Virasoro algebra representations on the $u$ and $v$ sectors so that $Yu=v$, 
$YH_0|_uY^{-1}=H_0|_v$. This also implies that $YH_I|_u Y^{-1}=-H_I|_v$. This
means finally that we can restrict to $0\le h \le \infty$, and the $u$ sector and
$H|_u$ will
correspond to the $h\ge 0$ case of (\ref{eq.yyy}) and to (\ref{eq.flow1}), the
$v$ sector and $H|_v$ will
correspond to the $h\le 0$ case of (\ref{eq.yyy}) and to (\ref{eq.flow2}). For $0\le h \le
\infty$ the operator 
(\ref{eq.pertham}) describes in the two sectors the two flows mentioned in the
Introduction. The precise relation between the $h$ in (\ref{eq.yyy}) in the Introduction and the
$h$ in (\ref{eq.pertham}) is the following: $h_{Introd}=2L^{1/2}h$ in the $u$ sector,
$h_{Introd}=-2L^{1/2}h$ in the $v$ sector.
Further on we shall assume that  $0<h<\infty$.

The eigenvalues of the Hamiltonian operator (\ref{eq.pertham}) are ultraviolet
divergent in perturbation theory; the divergence can be removed by adding
a term $ch^2I$ with appropriate value of the logarithmically divergent (as a
function of the cutoff energy) coefficient $c$. This means that
the differences of the eigenvalues of $H$ are not ultraviolet divergent. We shall
assume that the ground state energy is set to zero, and
the $ch^2I$ term will not be written explicitly.

The equations of motion are obtained by adding the following terms to the
right-hand side of the unperturbed equations of motion (\ref{mozg1}), (\ref{mozg2}), (\ref{mozg3}):
\begin{align}
[\rmi hH_I,\Psi_1(x,t)]& =  
8LhA_2(t)\delta(x-L)\\ {}
[\rmi hH_I,\Psi_2(x,t)] & = 
-8LhA_2(t)\delta(x-L) \\ {}
[\rmi hH_I,A_2(t)] & =  2h(\Psi_2(L,t)-\Psi_1(L,t))\ .
\end{align}
The initial condition for the fields $\Psi_1(x,t)$, $\Psi_1(x,t)$ and $A_2(t)$
is that at $t=0$ they are equal to the unperturbed fields (of the previous section).  
The above terms are linear in the fermion fields and $A_2$, so the equations of
motion for the perturbed theory are linear and by sandwiching these equations
between energy eigenstates we get a system of three first-order differential
equations for the expectation values of the fields and $A_2$.  
These expectation values can be assumed to take the form
\begin{align}
\label{eq.ans1}
\brakettt{E_1}{\Psi_1(x,t)}{E_2} & =   -\rme^{\rmi
  k(t-x)}-\Theta(x-L)C_1(k)\rme^{\rmi kt}\\
\label{eq.ans2}
\brakettt{E_1}{\Psi_2(x,t)}{E_2} & =  \rme^{\rmi
  k(t+x)}-\Theta(x-L)C_2(k)\rme^{\rmi kt}\\
\label{eq.ans3}
\brakettt{E_1}{A_2(t)}{E_2} & =  C_3(k)\rme^{\rmi kt}\ ,
\end{align}
where $C_1(k)$, $C_2(k)$, $C_3(k)$ are finite constants, $\ket{E_1}$ and
$\ket{E_2}$ are eigenstates of $H$ and $k=E_1-E_2$. $\ket{E_1}$ and
$\ket{E_2}$ are not necessarily normalized to 1 here.
Substituting (\ref{eq.ans1})-(\ref{eq.ans3}) into the equations of motion we get algebraic equations for
$C_1(k)$, $C_2(k)$, $C_3(k)$, which have the following solution:

\begin{equation}
\label{spektr}
kL\tan(kL)=16L^2h^2
\end{equation}
\begin{alignat}{2}
\psi_1(k)(x,t) &= \brakettt{E_1}{\Psi_1(x,t)}{E_2} & &= -\rme^{\rmi
  k(t-x)}-\Theta(x-L)\rmi \sin(kL)\rme^{\rmi kt}\\
\psi_2(k)(x,t) &= \brakettt{E_1}{\Psi_2(x,t)}{E_2} & &= \rme^{\rmi
  k(t+x)}-\Theta(x-L)\rmi \sin(kL)\rme^{\rmi kt}\\
a_2(k)(t) &= \brakettt{E_1}{A_2(t)}{E_2} & &= -\frac{\rmi
  \sin(kL)}{4Lh}\rme^{\rmi kt}\ .
\end{alignat}
Equation (\ref{spektr}) is the formula that determines the possible values of $k$ at a
given value of $h$ and $L$ and so the spectrum of $H$ up to an undetermined
additive overall constant. 

The assumption (\ref{eq.ans1}), (\ref{eq.ans2}) is motivated by the fact that
the equations of motion for $x \in (0,L)$ are $(\partial_t+\partial_x)\Psi_1(x,t)=0$,
$(\partial_t-\partial_x)\Psi_2(x,t)=0$. We remark that the equations of motion could also be
solved directly without making any assumptions on the form of the expectation values.

Introducing the notation
\begin{equation}
n(k)=(\psi_1(k),\psi_2(k),a_2(k))
\end{equation}
the mode expansion of $(\Psi_1,\Psi_2,A_2)$ is
\begin{equation}
\label{eq.mode}
(\Psi_1,\Psi_2,A_2)=\sum_{k\in S} b(k)n(k)\ ,
\end{equation}
where the $b(k)$ are creation/annihilation operators and the summation is done
over the set $S$ of all real solutions of (\ref{spektr}). The  $b(k)$
satisfy the relations (see Appendix \ref{app.folyt} for details)
\begin{equation}
\{ b(k_1),b(k_2)\} = \delta_{k_1+k_2,0}\frac{4Lk_1}{2Lk_1+\sin(2Lk_1)}
\end{equation}
and $b(k)^\dagger=b(-k)$.

The Hilbert space is
spanned by the orthogonal eigenstates
$b(k_1)b(k_2)b(k_3)\dots b(k_n)\ket{0_h}$ where $k_1>k_2>\dots >k_n>0$, $\ket{0_h}$
is the ground state, which is unique, and  
$b(k)\ket{0_h}=0$ if $k<0$.
The eigenvalues of these states are 
$\sum_{i=1}^n k_i$.
The eigenvector $b(k_1)b(k_2)\dots b(k_n)\ket{0_h}$ belongs to the $v$ sector if
$n$ is even and to the $u$ sector if $n$ is odd. 
The first few energy gaps (i.e.\ energies relative to the lowest energy) within the two
sectors are shown in Figure \ref{fig.tcsa1} as functions of $\ln(h)$ with $L=1$.

In the 
$h: 0 \to \infty$ limit
\begin{equation}
k(n,h)\to
k(n,0)+\frac{1}{2}\frac{\pi}{L}=\frac{n\pi}{L}+\frac{1}{2}\frac{\pi}{L}\ ,
\end{equation}
where $k(n,h)$ is the $n$-th nonnegative root of (\ref{spektr}) as a function of $h$,
$k(n,0)=n\pi/L$, $n=0,1,2,\dots $;   
and
$\{ b(k_i),b(k_j) \} \to 2 \delta_{k_i+k_j,0}$.
It can be verified that in the $h\to \infty$ limit the $(c=1/2,h=0)$
representation of the Virasoro algebra can be introduced in the $v$ sector and
the $(c=1/2,h=1/2)$ representation can be introduced in the $u$ sector. One
can write   expressions (which are well known essentially, see \cite{FMS,Ginsparg}) for the
generators in terms of the $b(k)$ similar to
(\ref{eq.ln1}), (\ref{eq.ln1b}). Therefore $h:0\to \infty$ corresponds to the $(1/2,1/16)\to (1/2,0)$
flow in the $v$ sector and to the $(1/2,1/16)\to (1/2,1/2)$ flow in the $u$ sector. 
We remark that the easiest way to determine which representations are realized
in the $h\to \infty$ limit is by counting the degeneracies of the first few
energy levels (separately in the two sectors).

\section{Reverse description}
\label{sec.reverse}

In this section we propose the description of  the model (\ref{eq.pertham}) as a perturbation of its $h \to
\infty$ limit.  This is motivated by the presence of the second (reverse) flows
in the TCSA spectra (Figure \ref{fig.tcsa2}). It should be noted that the meaning of the notation $H_0$ or $H_I$ etc 
differs from that in  Section \ref{sec.eel}. The precise correspondence between
the quantities in this section and in the previous section
will be given
explicitly for  the coupling constant and for the spectrum. 

\subsection{The free model}

The fundamental objects of the  model at $h=\infty$  are two one-component
real massless
fermion fields
$\Phi_1(x,t)$, $\Phi_2(x,t)$ with the anticommutation relations
\begin{align}
\{ \Phi_1(x,t),\Phi_1(y,t)\} & =  \{ \Phi_2(x,t),\Phi_2(y,t)\} =  4L\delta(x-y)\\
\{ \Phi_1(x,t),\Phi_2(y,t)\} & =   -4L[\delta(x+y)-\delta(x+y-2L)]
\end{align}
and reality property
$\Phi_1(x,t)^\dagger =\Phi_1(x,t)$, $\Phi_2(x,t)^\dagger =\Phi_2(x,t)$.
$\Phi_1$ and $\Phi_2$ are dimensionless.
The Hamiltonian operator is 
\begin{equation}
H_0=-\frac{\rmi}{8L}\int_0^L\ \rmd x\ \Phi_1(x,0)\partial_x \Phi_1(x,0)+\frac{\rmi}{8L}\int_0^L\
\rmd x\ \Phi_2(x,0)\partial_x \Phi_2(x,0)\ .
\end{equation}
The equations of motion are
\begin{eqnarray}
\label{eq.rmo1}
\partial_t\Phi_1(x,t)& =& [\rmi H_0,\Phi_1(x,t)]\ =\  -\partial_x\Phi_1(x,t) \nonumber\\
& & +\frac{1}{2}\delta(x-L)
[\Phi_1(L,t)-\Phi_2(L,t)]+\frac{1}{2}\delta(x) [-\Phi_1(0,t)-\Phi_2(0,t)]\\
& & +\frac{1}{2} \Theta(-x) [\partial_x \Phi_1(x,t)-\partial_x \Phi_2(x,t)]
+\frac{1}{2} \Theta(x-L) [\partial_x \Phi_1(x,t)+\partial_x \Phi_2(x,t)]\nonumber
\end{eqnarray}
\begin{eqnarray}
\label{eq.rmo2}
\partial_t\Phi_2(x,t) & =& [\rmi H_0,\Phi_2(x,t)]\ =\  \partial_x\Phi_2(x,t) \nonumber\\
& & +\frac{1}{2}\delta(x-L)
[\Phi_1(L,t)-\Phi_2(L,t)]+\frac{1}{2}\delta(x) [\Phi_1(0,t)+\Phi_2(0,t)]\\
& & +\frac{1}{2} \Theta(-x) [\partial_x \Phi_1(x,t)-\partial_x \Phi_2(x,t)]
+\frac{1}{2} \Theta(x-L) [-\partial_x \Phi_1(x,t)-\partial_x \Phi_2(x,t)]\ . \nonumber
\end{eqnarray}
The fermion fields have the following mode expansion:
\begin{align}
\Phi_1(x,t) &  =  \sum_{k \in \frac{\pi}{L} \ZZ+\frac{\pi}{2L}} \sqrt{2} a(k)
  \rme^{\rmi k(t-x)}\\
\Phi_2(x,t) & =  \sum_{k \in \frac{\pi}{L} \ZZ+\frac{\pi}{2L}} -\sqrt{2} a(k)
  \rme^{\rmi k(t+x)}\\
\end{align}
\begin{equation}
\{ a(k_1),a^\dagger (k_2)\} =\delta_{k_1,k_2}\qquad a(k)^\dagger =a(-k)\ .
\end{equation}
An orthonormal basis for the Hilbert space is formed by the vectors
$a(k_1)a(k_2)\dots a(k_n)\ket{0}$, 
where $k_i>0$, $k_i\in  \pi\ZZ/L+\pi/(2L)$, $n\ge 0$.
$a(k)\ket{0}=0$ if $k<0$.
The eigenvalue of the eigenvector $a(k_1)a(k_2)\dots a(k_n)\ket{0}$ is
$\sum_{i=1}^n k_i$. 
The fermion fields satisfy the following boundary conditions:
\begin{equation}
\brakettt{E_1}{\Phi_1(0,t)+\Phi_2(0,t)}{E_2}=0\qquad
\brakettt{E_1}{\Phi_1(L,t)-\Phi_2(L,t)}{E_2}=0\ ,
\end{equation}
where $\ket{E_1}$, $\ket{E_2}$ are eigenstates of $H_0$. 

One can define a representation of the Virasoro algebra on the Hilbert space
in the same way as in Section \ref{sec.freemod} (see also \cite{FMS,Ginsparg}). This representation
is $(1/2,0)\oplus (1/2,1/2)$. The fields $\Phi_1(x,t)$ and $\Phi_2(x,t)$ can
be converted to 
weight $1/2$ primary fields by multiplying them by a suitable simple exponential factor.

\subsection{The perturbed model}

The perturbed Hamilton operator is
\begin{multline}
H  = -\frac{\rmi}{8L}\int_0^L\ \rmd x\ \Phi_1(x,0)\partial_x \Phi_1(x,0)+\frac{\rmi}{8L}\int_0^L\
\rmd x\ \Phi_2(x,0)\partial_x \Phi_2(x,0)\\
+g[\rmi(\Phi_1+\Phi_2)(L,0) \lim_{x\to L} \partial_x
(\Phi_2-\Phi_1)(x,0) ]\ ,  
\end{multline}
where $g$ is a dimensionless coupling constant. In the same way as in Section \ref{sec.pertmod}, the
perturbing term has zero matrix elements between vectors belonging to  different irreducible
representations. 

The limit prescription in the perturbing term is important, and the limit
$\lim_{x\to L}$ should be taken at the end of any calculation. 
It should  be assumed that $\lim_{x \to L} \delta (x-L)=0$, for example,
and similarly for the derivatives of $\delta (x-L)$.   

It should be noted that the only
primary field in the $0$ or in the $1/2$ representation is the identity
operator, all other fields are descendant and non-relevant fields.

The equations of motion are obtained by adding the following terms to the
right-hand side of the unperturbed equations of motion (\ref{eq.rmo1}), (\ref{eq.rmo2}): 
\begin{align}
\label{eq.rint1}
[\rmi gH_I,\Phi_1(x,t)] & =  g8L\delta(x-L)\lim_{x\to L}
\partial_x(\Phi_2-\Phi_1)(x,t)\\ {}
\label{eq.rint2}
[\rmi gH_I,\Phi_2(x,t)] & =  g8L\delta(x-L)\lim_{x\to L}
\partial_x(\Phi_2-\Phi_1)(x,t)
\end{align}
where
\begin{equation}
H_I= [\rmi(\Phi_1+\Phi_2)(L,0) \lim_{x\to L} \partial_x
(\Phi_2-\Phi_1)(x,0) ]\ .
\end{equation}
Similar steps to those in Section \ref{sec.pertmod} can now be taken to obtain
$\brakettt{E_1}{\Phi_1(x,t)}{E_2}$ and\\ 
$\brakettt{E_1}{\Phi_2(x,t)}{E_2}$. 
In the same way as in Section \ref{sec.pertmod}, the forms 
\begin{align}
\brakettt{E_1}{\Phi_1(x,t)}{E_2} & =  \rme^{\rmi
  k(t-x)}+\Theta(x-L)D_1(k)\rme^{\rmi kt}\\
\brakettt{E_1}{\Phi_2(x,t)}{E_2} & =  -\rme^{\rmi
  k(t+x)}+\Theta(x-L)D_2(k)\rme^{\rmi kt}
\end{align}
can be assumed, where 
$D_1(k)$ and $D_2(k)$ are finite constants, $\ket{E_1}$ and
$\ket{E_2}$ are eigenstates of $H$ and $k=E_1-E_2$. 

Solving the equations of motion for
$D_1(k)$, $D_2(k)$ we get
$D_2(k) = -D_1(k) =  \cos(kL)$
and
\begin{equation}
\label{spektrrev}
(kL)\tan(kL)=\frac{-1}{16g}\ .
\end{equation}
Equation (\ref{spektrrev}) is the formula that determines the spectrum of $H$ up to an
overall additive constant. 
The eigenvalues of $H$ are 
$k_1+k_2+\dots +k_n$, 
where $n \ge 0$, $k_i\ge 0$, $k_i\ne k_j$ if $i \ne j$, the $k_i$ are
real roots of  (\ref{spektrrev}) and the lowest eigenvalue is assumed to be
set to zero by adding a constant which is not written explicitly.
The substitution
\begin{equation}
g=\frac{-1}{256 L^2h^2}
\end{equation}
converts (\ref{spektrrev}) into (\ref{spektr}). Thus  $g:0\to -\infty$ corresponds to the flow
$ (0 \to 1/16) \oplus (1/2 \to 1/16)$. 

In perturbation theory there are divergences if we take $x=L$ in $H_I$ at the
beginning. However, if one allows $x$ to take general values, then in
Rayleigh-Schr\"odinger perturbation theory  for the differences of the energy
levels one can expect to get sums
at any fixed order  which are possible to evaluate. We expect
that the evaluation yields, besides non-singular parts, step functions, $\delta (x-L)$ and its
derivatives, and the result remains finite after the $x \to L$
limit.

\section{Exact spectrum in the Mode Truncated version}
\label{sec.modetr}

\subsection{The free model}
\label{sec.modetrfree}

Let $n_c$, called the truncation level, be a positive integer. The mode truncated version of the free model
described in Section \ref{sec.freemod} is the following:
\begin{align}
\{ \Psi_1(x,t),\Psi_1(y,t)\} & =   2[1+2\sum_{k\in\frac{\pi}{L} \{ 1\dots n_c \} } \cos(k(x-y)) ]\\
\{ \Psi_2(x,t),\Psi_2(y,t)\} & =   2[1+2\sum_{k\in\frac{\pi}{L} \{ 1\dots n_c \} } \cos(k(x-y)) ]\\
\{ \Psi_1(x,t),\Psi_2(y,t)\} & =   -2[1+2\sum_{k\in\frac{\pi}{L} \{ 1\dots n_c
  \} } \cos(k(x+y)) ]\ ,
\end{align}
\begin{align}
\Psi_1(x,t) & =   \sum_{k\in\frac{\pi}{L} \{ 1\dots n_c \}
}\sqrt{2}[a(k)\rme^{\rmi k(t-x)}+a^+(k)\rme^{-\rmi k(t-x)}]+A_1\\
\Psi_2(x,t) & =  \sum_{k\in\frac{\pi}{L}\{ 1\dots n_c   \}
}\sqrt{2}[-a(k)\rme^{\rmi k(t+x)}-a^+(k)\rme^{-\rmi k(t+x)}]-A_1\ .
\end{align}
Equations (\ref{a1})-(\ref{a2}) apply unchanged. 
The Hamiltonian operator is 
\begin{equation}
H_0=\sum_{k\in\frac{\pi}{L} \{ 1\dots n_c \} }\ k [a(k)a^+(k)]\ .
\end{equation}
The fields satisfy the following equations of motion:
\begin{alignat}{2}
\partial_t \Psi_1(x,t) &= [\rmi H_0,\Psi_1(x,t)]   &&=  -\partial_x \Psi_1(x,t)\\
\partial_t \Psi_2(x,t) &= [\rmi H_0,\Psi_2(x,t)]  &&= \partial_x \Psi_2(x,t)\\
\frac{\rmd}{\rmd t} A_2 (t)     &= [\rmi H_0, A_2(t)] && = 0\ .
\end{alignat}
The Hilbert space and the energy eigenstates are similar to those in Section
\ref{sec.freemod}, but 
\begin{equation}
k\in \frac{\pi}{L} \{ -n_c,-n_c+1,\dots ,n_c-1,n_c\}
\end{equation}
applies instead of $k \in \pi\ZZ /L$. The Hilbert space is $2\times 2^{n_c}$ dimensional.

\subsection{The perturbed model}

The perturbed Hamiltonian operator is $H_0+hH_I$, where
\begin{equation}
H_I=\rmi A_2(0)[\Psi_2(L,0)-\Psi_1(L,0)]\ .
\end{equation}

The equations of motion are
\begin{alignat}{2}
\partial_t \Psi_1(x,t) &=[\rmi (H_0+hH_I),\Psi_1(x,t)] && =-\partial_x \Psi_1(x,t)-hA_2(t)C_1(x)\\
\partial_t \Psi_2(x,t) &=[\rmi (H_0+hH_I),\Psi_2(x,t)] && =\partial_x \Psi_2(x,t)-hA_2(t)C_2(x)\\
\frac{\rmd}{\rmd t}A_2(t) &=[\rmi (H_0+hH_I),A_2(t)] && =2h(\Psi_2(L,t)-\Psi_1(L,t))\ ,
\end{alignat}
where
\begin{align}
C_1(x) & =  \{ \Psi_1(x,t),\Psi_2(L,t)-\Psi_1(L,t) \} \nonumber\\ 
& =  -4[1+2 \sum_{k\in\frac{\pi}{L} \{ 1\dots n_c \} } \cos (k(x+L)) ]\\
C_2(x) & =  \{ \Psi_2(x,t),\Psi_2(L,t)-\Psi_1(L,t) \} =  -C_1(x)\ .
\end{align}
These equations are linear as in the non-truncated case, so sandwiching them
between energy eigenstates gives a system of three first-order linear partial
differential equations for the expectation values. The analogue of
(\ref{spektr}) can be obtained from these equations in the following way: 
we can eliminate $A_2$:
\begin{align}
\label{wr1}
\partial_t^2 \Psi_1(x,t) & =  -\partial_{xt}\Psi_1(x,t)-2h^2(\Psi_2(L,t)-\Psi_1(L,t))C_1(x)\\
\partial_t^2 \Psi_2(x,t) & =  \partial_{xt}\Psi_1(x,t)-2h^2(\Psi_2(L,t)-\Psi_1(L,t))C_2(x)
\label{wr2}
\end{align}
and introduce the functions $f_1(x)$, $f_2(x)$:
\begin{equation}
\brakettt{E_1}{\Psi_1(x,t)}{E_2}  =  f_1(x)\rme^{\rmi kt}\qquad
\brakettt{E_1}{\Psi_2(x,t)}{E_2}  =  f_2(x)\rme^{\rmi kt}\ ,
\end{equation}
where $k=E_1-E_2$ and the dependence of $f_1$ and $f_2$ on $k$ is not denoted
explicitly. Sandwiching  (\ref{wr1}) and (\ref{wr2}) between
$\ket{E_1}$ and $\ket{E_2}$  
gives two  first-order inhomogeneous linear ordinary differential
equations (with the deviation that the inhomogeneity depends on the unknown
functions). 
We also have the boundary conditions
$f_1(0) =  -f_2(0)$,
$f_1(L) =  -f_2(L)$. 
Elementary calculation yields that a solution of the differential equations
and the boundary conditions  exists if and only if $k$
satisfies the equation
\begin{equation}
\label{spektrmt}
16h^2\left[\frac{1}{k^2}+\sum_{k_0\in\frac{\pi}{L} \{ 1\dots n_c \} }
  \frac{2}{k^2-k_0^2}   \right]=1\ ,
\end{equation}
which is the analogue of (\ref{spektr}) and determines the energy of the modes
as functions of $h$.

(\ref{spektrmt}) as an algebraic equation for $k$  has finitely many real
roots, and if $k$ is a root, then $-k$ is a root as well. All real roots converge
to finite values as $h \to \infty$ except for the pair with the largest
absolute value. This pair of roots diverges linearly as $h \to \infty$. This
pair has the largest absolute value already at $h=0$.  A 
 consequence of this behaviour is that the lower half of the spectrum, namely
 those states which do not contain the mode with the highest energy, remains
 finite as $h\to\infty$, whereas the higher half of the spectrum, i.e.\ the
 states which contain the mode with the highest energy, diverges linearly as $h
 \to \infty$ with a common slope. Here it is assumed that the ground state
 energy is set to zero. In the subsequent sections we shall consider the lower half of the spectrum. The
 look of this half as a function of $\ln(h)$ is very similar to that of the exact
 spectrum  shown in Figure \ref{fig.tcsa1}. 

Applying the formula 
\begin{equation}
1+\sum_{n=1}^\infty \frac{2k^2}{k^2-4n^2\pi^2}=\frac{k/2}{\tan (k/2)}
\end{equation}
we can easily verify that the limit of (\ref{spektrmt}) as $n_c\to \infty$ is (\ref{spektr}).

\section{Perturbative results}
\label{sec.pertres}

The functions $s_0(h,n_c)$ and $s_1(h,n_c)$ in 
$H^{r}=s_0(h,n_c)H_0+s_1(h,n_c)H_I$\  are determined by the
condition that the differences  of those  eigenvalues of
$H^{r}$ that are low compared to the truncation
level  should be equal to those of the truncated Hamiltonian operator
$H^t(n_c)$.  $H^t(n_c)$ is, in particular, the TCSA Hamiltonian operator
$H^{TCSA}(n_c)$, or the Hamiltonian operator $H^{MT}(n_c)$ of the mode
truncated model. This condition, which we shall call renormalization condition, applies separately and independently within the $u$ and $v$
sectors, and we have in fact a   pair $s_0^u$, $s_1^u$ for the $u$ sector and another
pair 
 $s_0^v$, $s_1^v$ for the $v$ sector.

The renormalization condition is a very strong condition on $s_0$ and $s_1$
and generally we cannot expect that it can be satisfied. It is possible,
however, that it can be  satisfied in certain approximations.

\subsection{Mode Truncation scheme}

Using (\ref{eq.rspert}) in Appendix \ref{sec.RS} we can obtain the following
results:

The renormalization conditions have a solution if the eigenfunctions are
expanded into a power series in $h$ and terms that are higher order than 3 are omitted: 
\begin{equation}
\label{eq.crv1}
s_0(h,n_c) =  1+O(h^4)\qquad
%\label{eq.crv2}
s_1(h,n_c)  =  h+   \frac{1}{2}(S-S(n_c))L^2 h^3+O(h^4)\ ,
\end{equation}
where
\begin{equation}
S=\sum_{n=1}^\infty \frac{32}{n^2\pi^2} \qquad
S(n_c)=\sum_{n=1}^{n_c} \frac{32}{n^2\pi^2}\ .
\end{equation}
This solution applies to both the $u$ and $v$ sectors and it is exact in $n_c$.
We remark that 
\begin{equation}
S-S(n_c) = \frac{32}{\pi^2 n_c}+O(1/n_c^2) \ .
\end{equation}

In the MT scheme we can obtain another result that is non-perturbative in $h$:
doing power series expansion in $1/n_c$ we obtain the formula 
\begin{equation}
\frac{1}{k^2}+\sum_{n=1}^{n_c} \frac{2}{k^2-\frac{\pi^2}{L^2} n^2} =
\frac{L}{k\tan(kL)}+\frac{2L^2}{\pi^2}+\frac{1}{n_c} +O(1/n_c^2)\ .
\end{equation}
Omitting the terms which are second or higher order in $1/n_c$, equation (\ref{spektrmt})  takes the form
of (\ref{spektr}) if 
\begin{equation}
h_{\mathrm{eff}}=\frac{h}{\sqrt{1-\frac{32L^2h^2}{\pi^2
      n_c}}}=h+16h^3\frac{L^2}{\pi^2}\frac{1}{n_c} +O(1/n_c^2)
\end{equation}
is introduced:
\begin{equation}
kL\tan(kL)=16L^2h_{\mathrm{eff}}^2\ .
\end{equation}
This means that rescaling by 
\begin{equation}
s_0(h,n_c)=1+O(1/n_c^2)\qquad
s_1(h,n_c)=h+16h^3\frac{L^2}{\pi^2}\frac{1}{n_c}+O(1/n_c^2)
\end{equation}
improves the convergence of the mode truncated energy gaps to the exact
energy gaps from order $1/n_c$
to order $1/n_c^2$ (at least), i.e.\ the difference between the energy gaps of
$s_0(h,n_c)H_0+s_1(h,n_c)H_I$ and the energy gaps of $(H_0+hH_I)^{MT}$ tends to zero
as $1/n_c^2$ (at least) for any fixed finite value of $h$, whereas the difference between the
energy gaps of $H_0+hH_I$ and  $(H_0+hH_I)^{MT}$ tends to zero as $1/n_c$.

\subsection{TCS scheme}

Using (\ref{eq.rspert}) in Appendix \ref{sec.RS} we  obtain the following
results:

The renormalization conditions have a solution if the eigenfunctions are
expanded into a power series in $h$ and in $1/n_c$ and terms that are of order
higher than 3 in $h$ and 1 in $1/n_c$  are omitted: 
\begin{eqnarray}
\label{eq.crv3}
s_0(h,n_c)  =  1+x_1h^2+y_1h^3 + O(h^4)\\
\label{eq.crv4}
s_1(h,n_c)  =  h+y_2h^2+x_2h^3+O(h^4)
\end{eqnarray}
\begin{align}
x_1 &=  \frac{8L^2}{\pi^2 n_{c}}+O(1/n_{c}^2) &
y_1 & =  0 + O(1/n_c^2) \\
x_2 & =  \frac{1}{2}(S-S(n_c))L^2+O(1/n_{c}^2)  &
y_2  &=  0 + O(1/n_c^2)\ . 
\end{align}
This solution applies to both sectors. 
If terms of order $1/n_c^2$ are also taken into consideration, then the
renormalization conditions do not have a solution. 
A significant difference between the TCS and mode truncation schemes is that
 the value of the coefficient $x_1$ is zero in the mode truncation scheme but
 non-zero in the TCS scheme.

\section{Numerical results}
\label{sec.numres}

In the numerical calculations described in this section the value of $L$ is
set equal to 1. This does not affect the generality of the results. 
In the calculations we used the same normalizations as in Section \ref{sec.eel}.

$s_0(h,n_c)$ and $s_1(h,n_c)$ are calculated from the lowest three energy
levels and
are given by the following formulae:
\begin{align}
\label{eq.s1}
\frac{s_1}{s_0}(h) & =  \left( \frac{E_2-E_0}{E_1-E_0} \right)^{-1}\left(  \left(
  \frac{E_2-E_0}{E_1-E_0}\right)^{t}(h) \right)  \\
\label{eq.s2}
s_0(h) & =  \frac{(E_1-E_0)^{t}(h)}{(E_1-E_0)(\frac{s_1}{s_0}(h))}\ ,
\end{align}
where the superscriptless quantities are the non-truncated ones and those with
the ${}^t$ superscript are obtained by a truncation. $E_0,E_1,E_2$
denote the three lowest energy eigenvalues.

\subsection{Mode Truncation Scheme}
\markright{\thesubsection.\ \ NUMERICAL RESULTS - MODE TRUNCATION SCHEME}

\begin{figure}
\begin{tabular}{lr}
\includegraphics[clip=true,height=8.5cm,width=7.5cm,
]{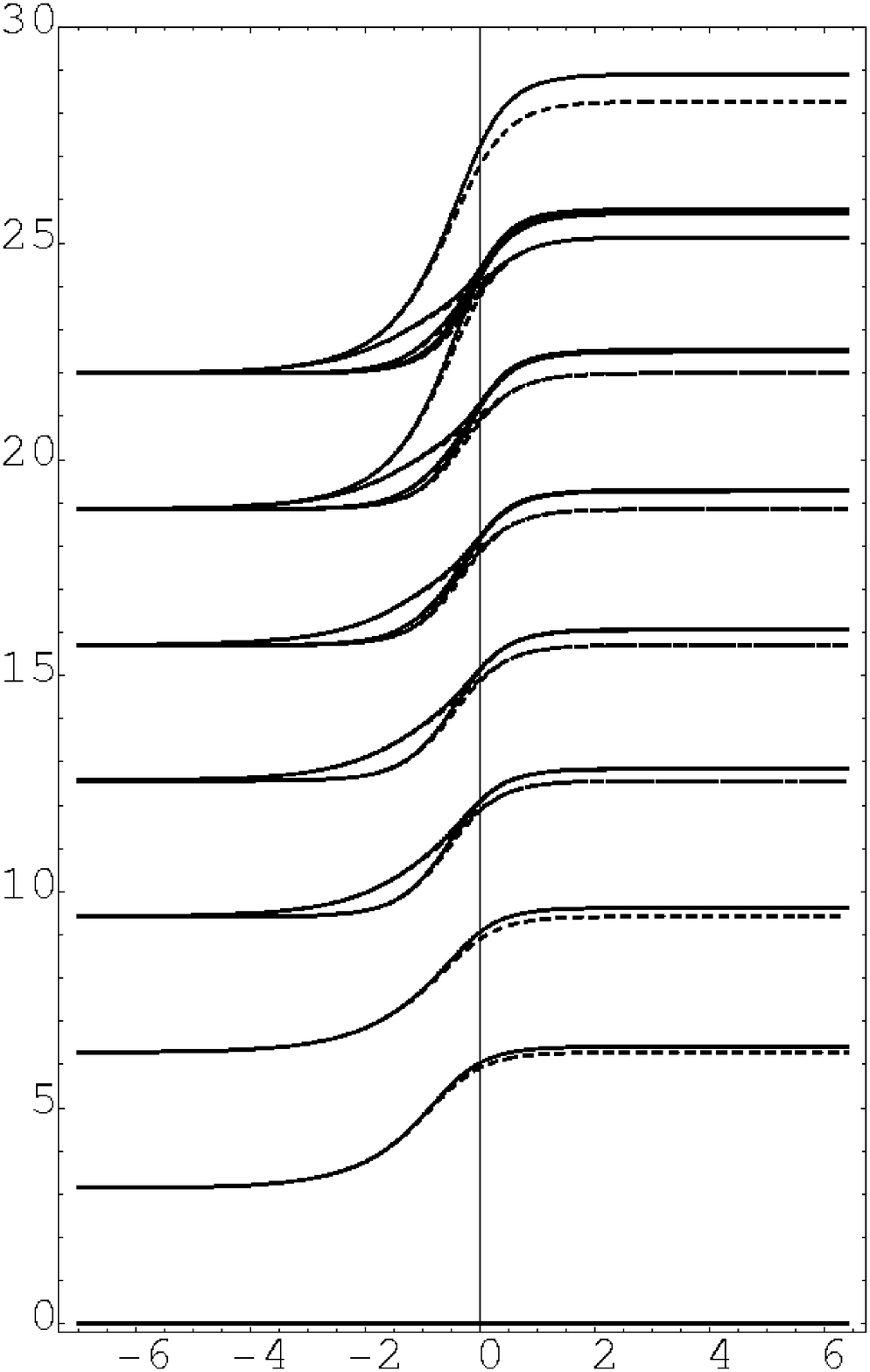}
&
\includegraphics[clip=true,height=8.5cm,width=7.5cm,
]{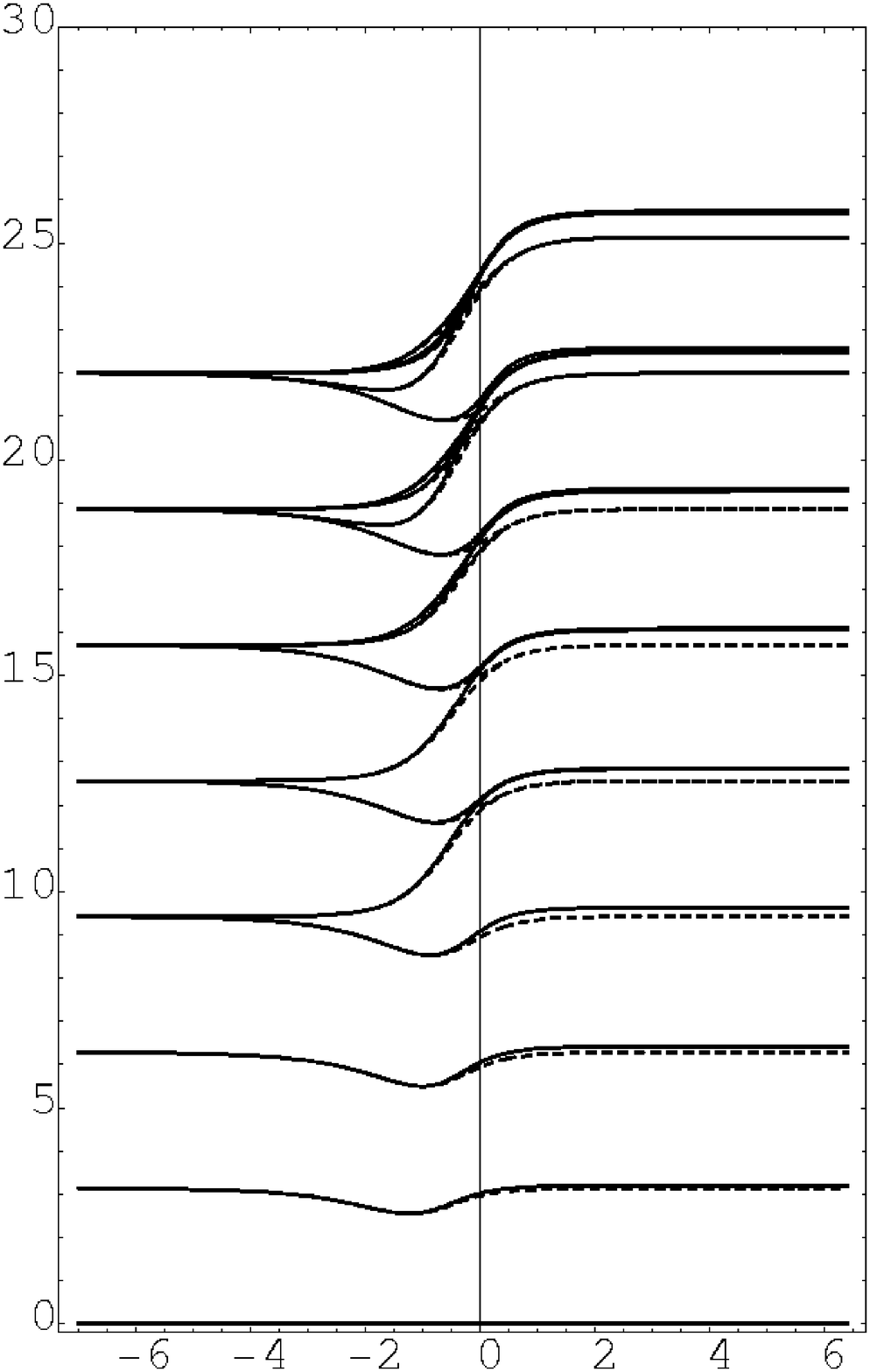}
\end{tabular}
\caption{\label{fig.mt1}Exact (dashed lines) and mode truncated (solid lines) energy
  gaps ($E_i-E_0$)
  in the $v$ and $u$ sectors respectively as a function of $\ln(h)$ at truncation level $n_c=9$}
\end{figure}

Figure \ref{fig.mt1} shows the exact and mode truncated spectra as a function of the
logarithm of the coupling constant. The truncation level is $n_c=9$, and the
dimension of the Hilbert space is 512 in each sector. It is remarkable that
there is a good qualitative agreement between the mode truncated and exact
spectra for all values of $h$.

\begin{figure}
\begin{tabular}{lr}
\includegraphics[clip=true,height=8.5cm,width=7.5cm,
]{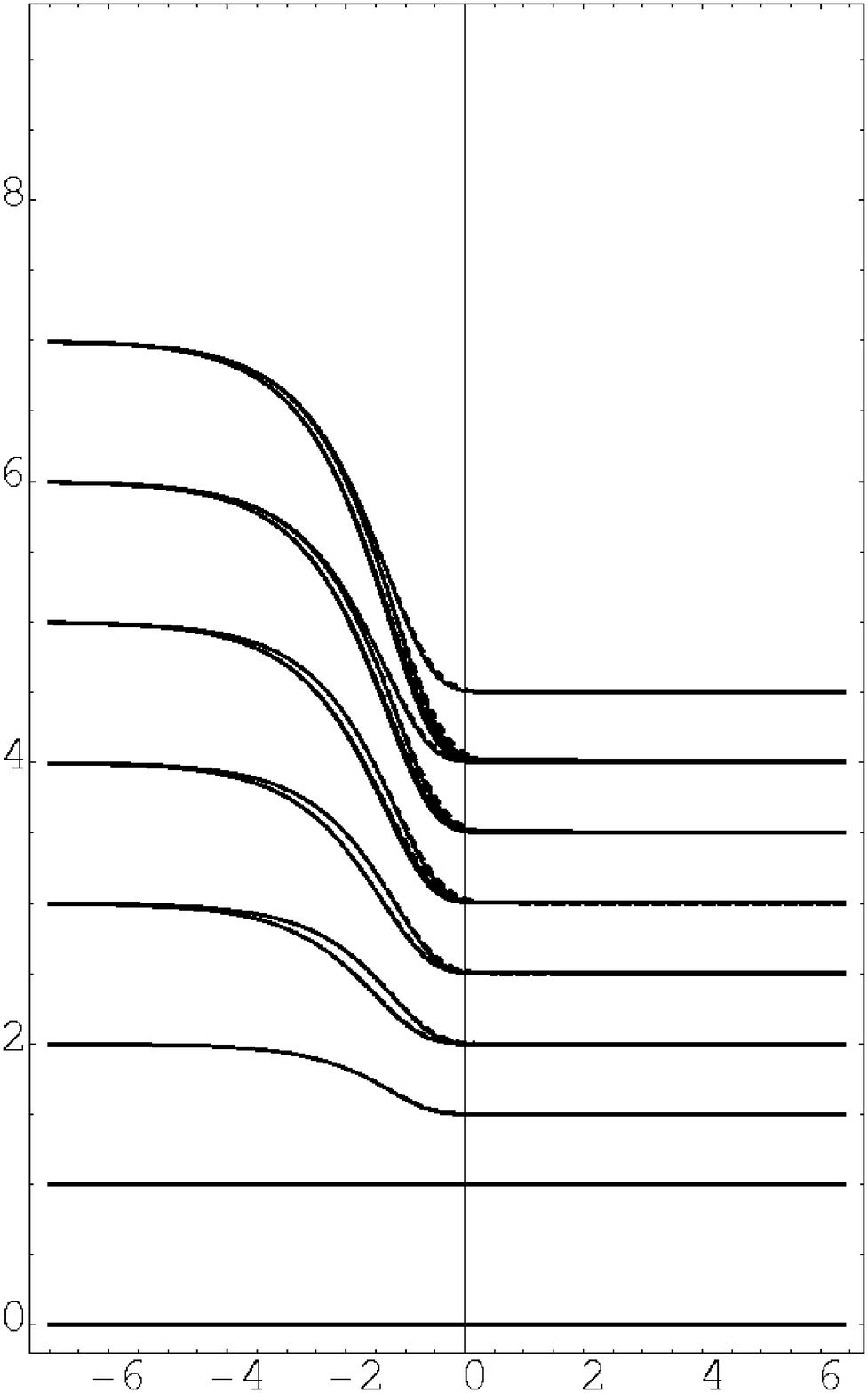}
&
\includegraphics[clip=true,height=8.5cm,width=7.5cm,
]{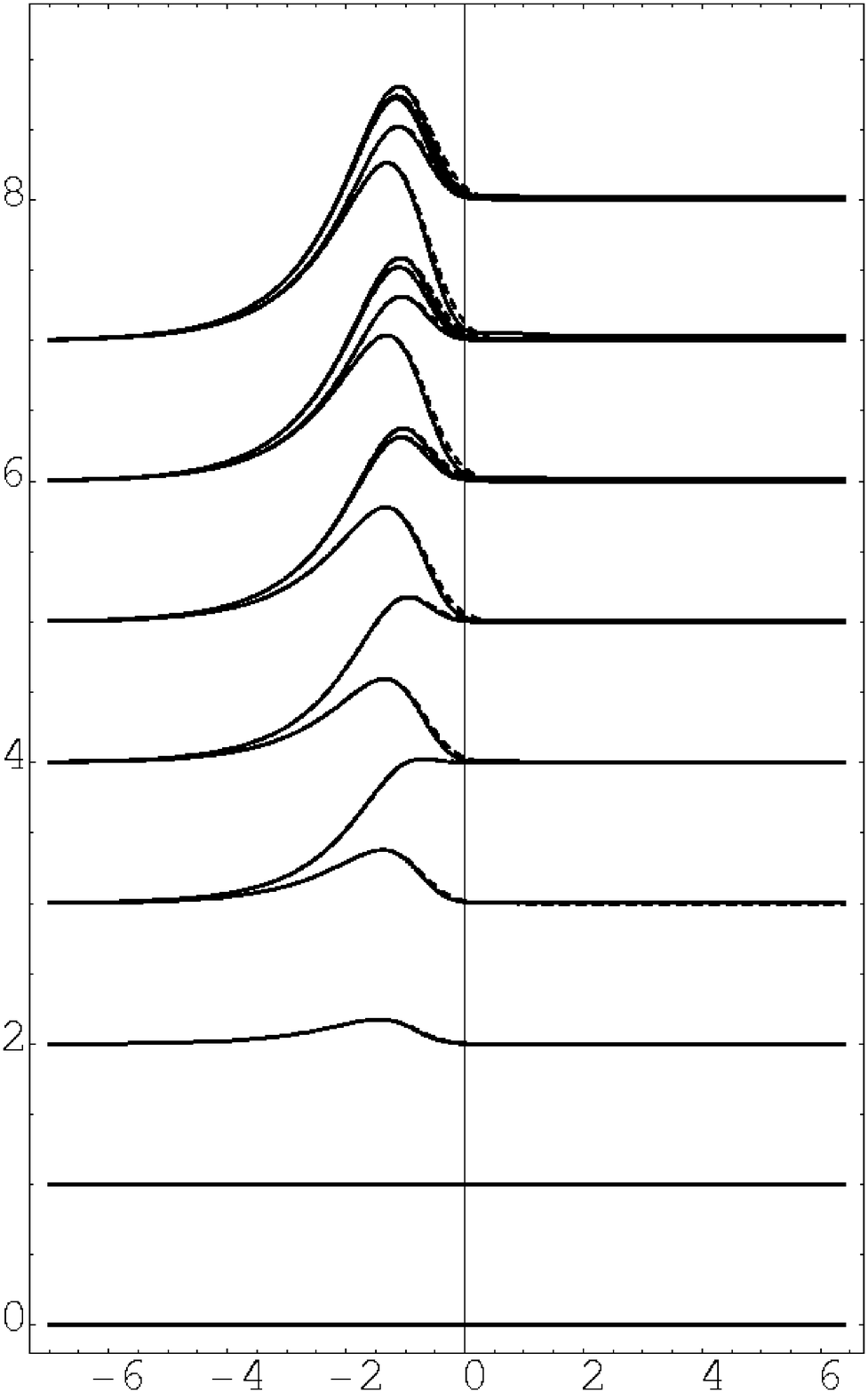}
\end{tabular}
\caption{\label{fig.mt2}Exact (dashed lines) and mode truncated (solid lines) normalized
  spectra
  in the $v$ and $u$ sectors respectively as a function of $\ln(h)$ at truncation level $n_c=9$}
\end{figure}

Figure \ref{fig.mt2}  shows the  same spectra,  but the  lowest gap is normalized to
one, i.e.\ the functions $\frac{E_i(h)-E_0(h)}{E_1(h)-E_0(h)}$ are shown. It is
remarkable that the agreement between the exact and mode truncated spectra looks
considerably better than in the case of not normalized spectra.

\begin{figure}
\begin{tabular}{lll}
\includegraphics[clip=true,height=4.5cm,
]{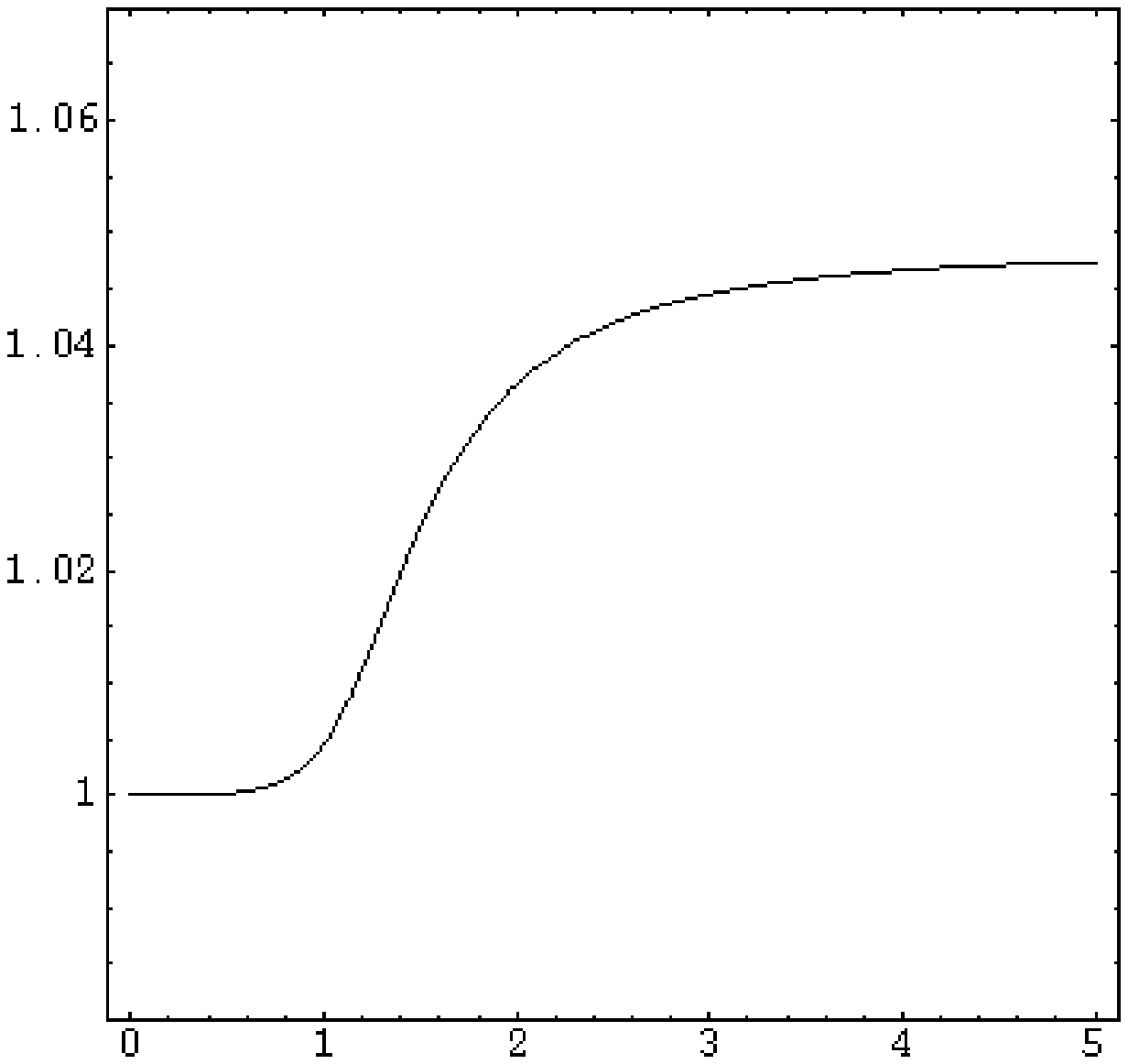}
&
\includegraphics[clip=true,height=4.5cm,
]{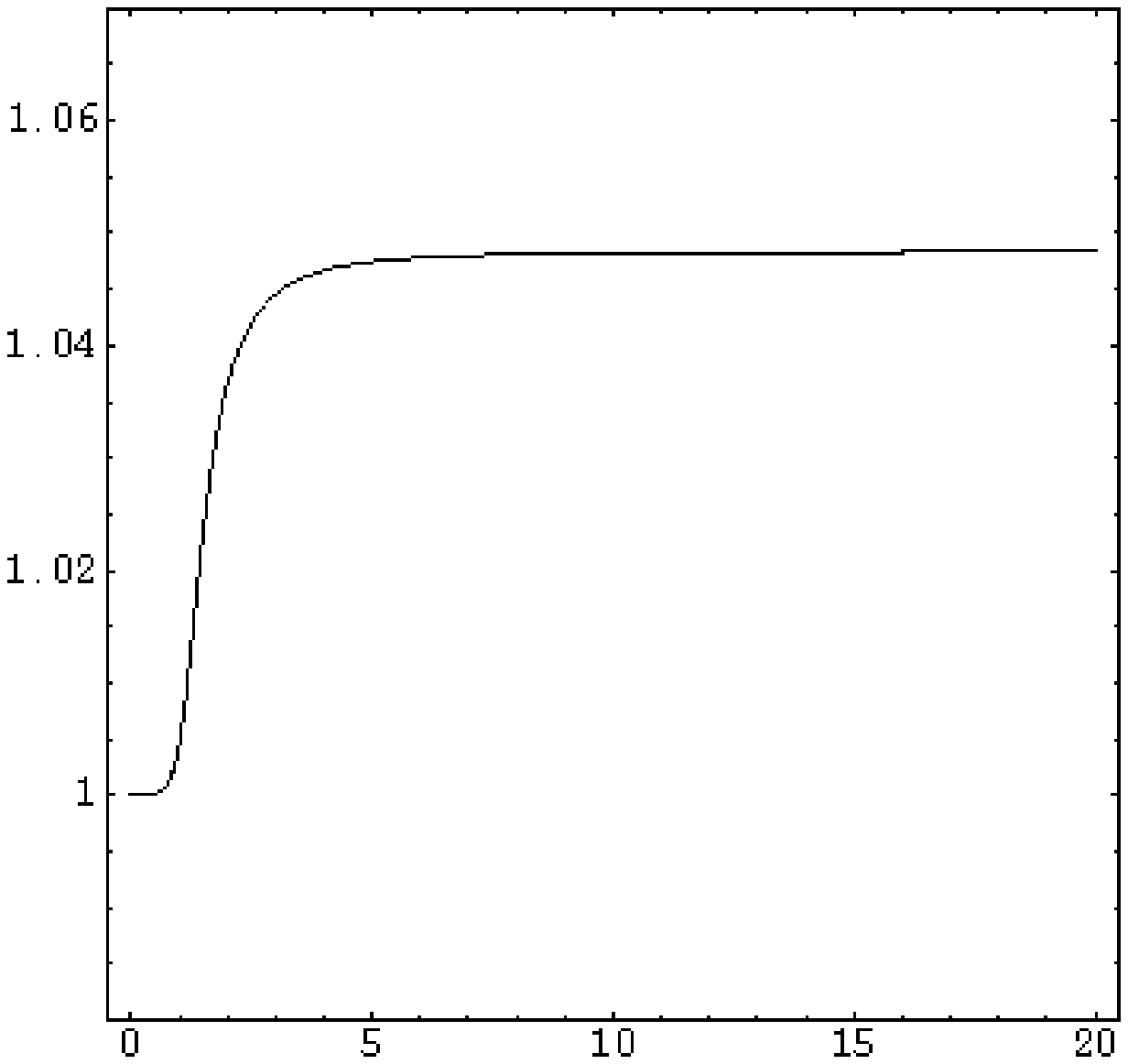}
&
\includegraphics[clip=true,height=4.5cm,
]{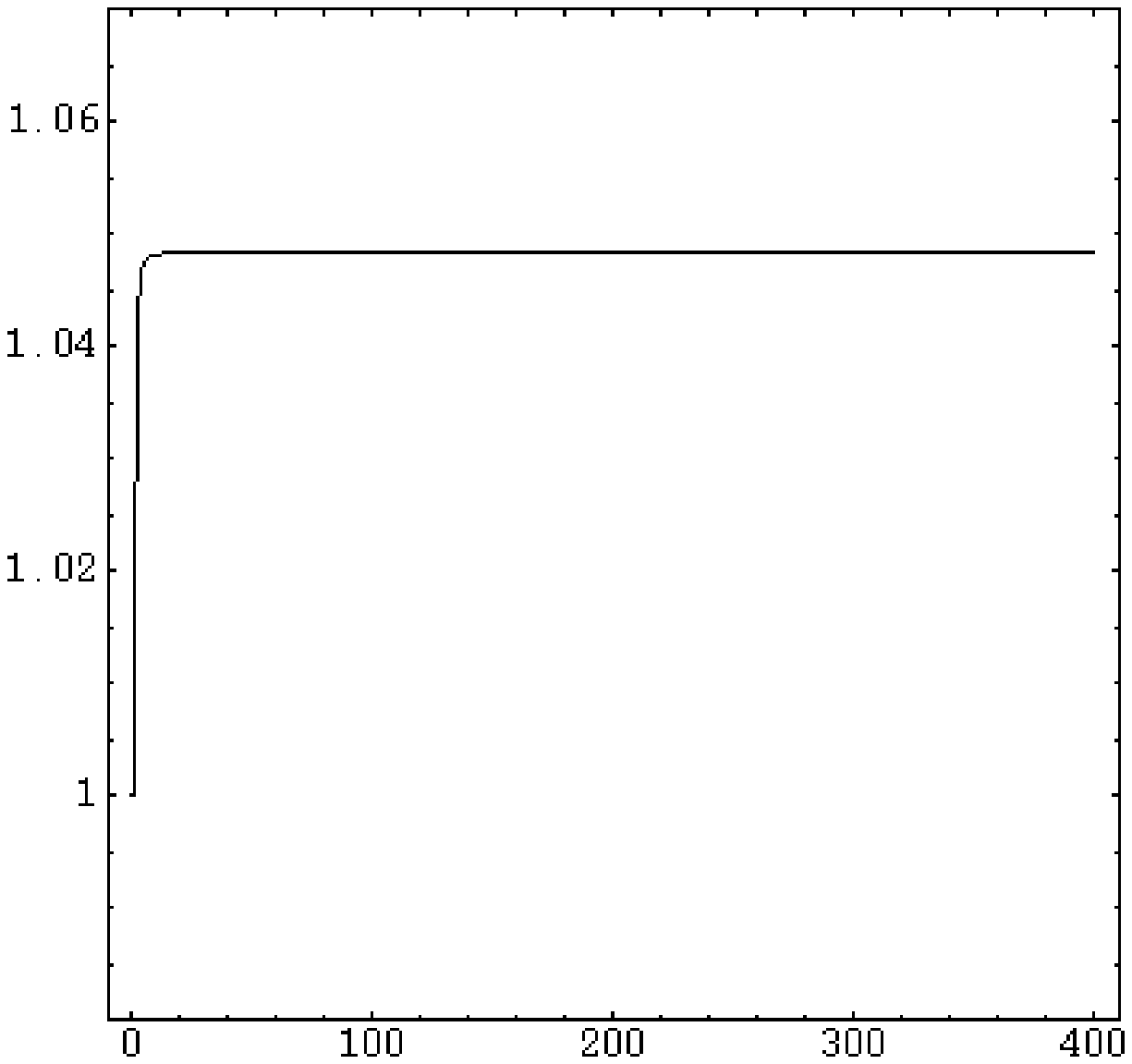}
\end{tabular}
\caption{\label{fig.mt3}The function $s_0(h)$ for the $v$ sector in the ranges $h\in [0,3]$, $h\in [0,20]$, $h\in [0,400]$ and
  $s_0\in [0.95,1.05]$ at truncation level $n_c=9$}
\end{figure}

\begin{figure}
\begin{tabular}{lll}
\includegraphics[clip=true,height=4.5cm,
]{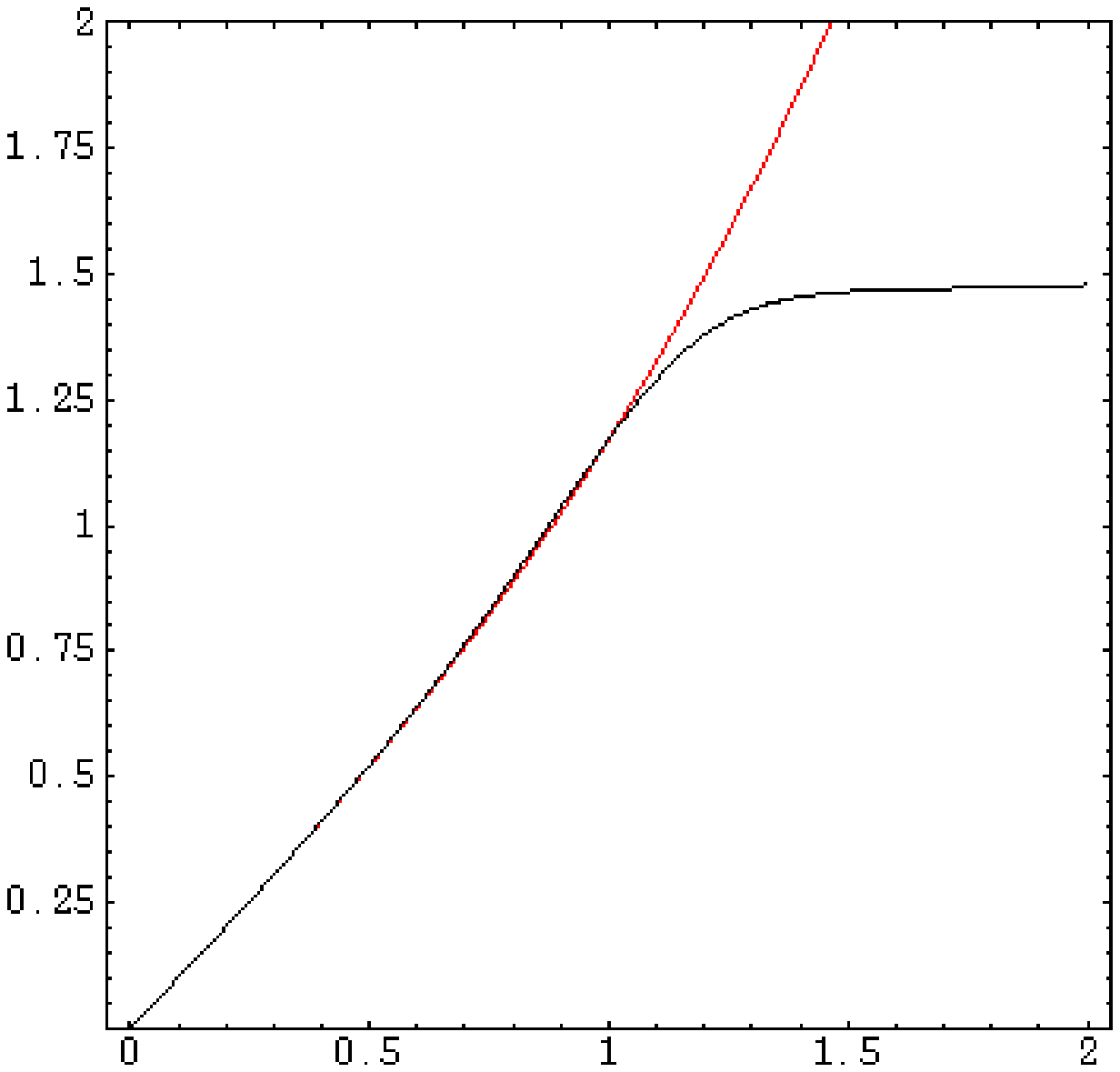}
&
\includegraphics[clip=true,height=4.5cm,
]{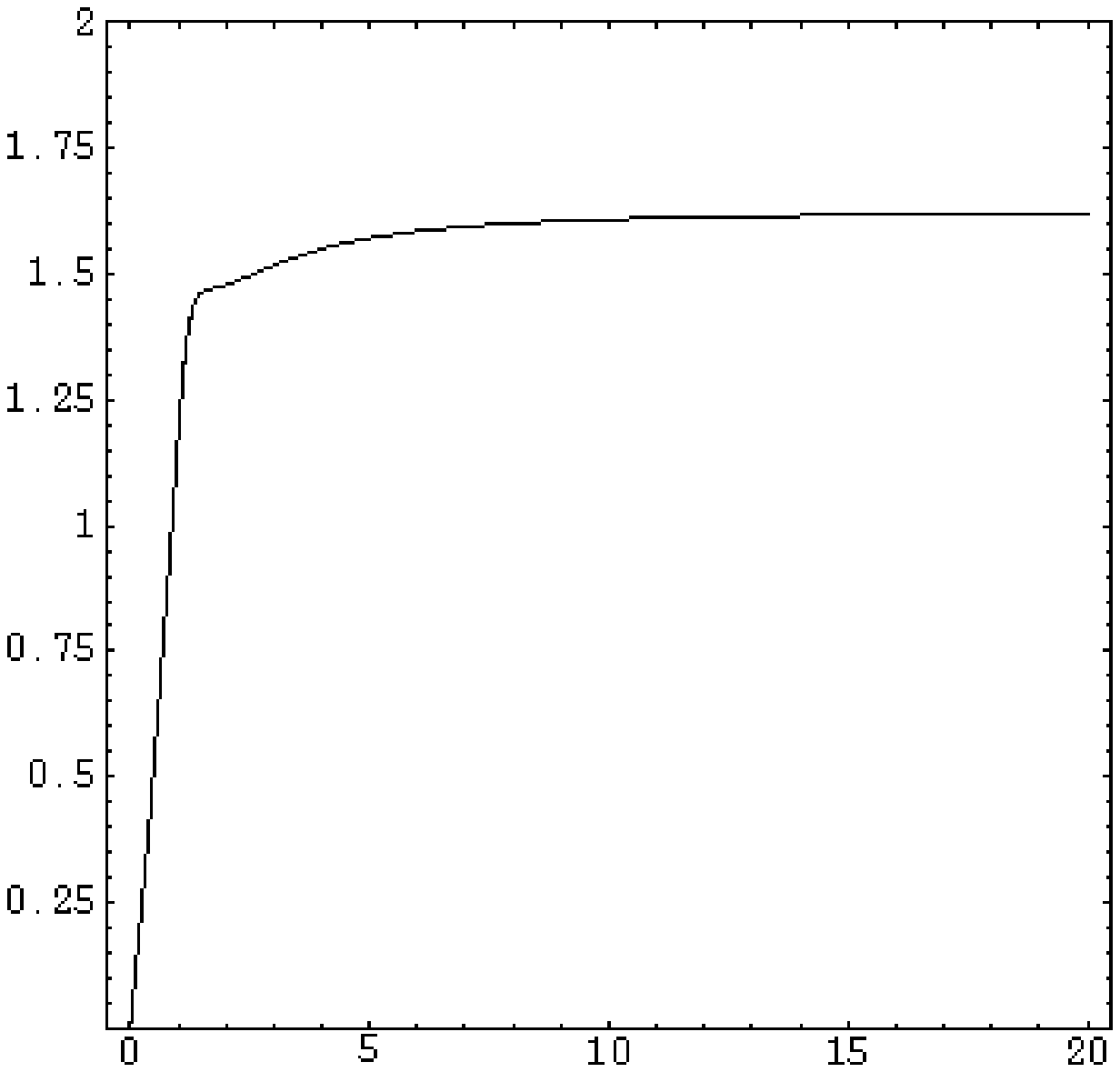}
&
\includegraphics[clip=true,height=4.5cm,
]{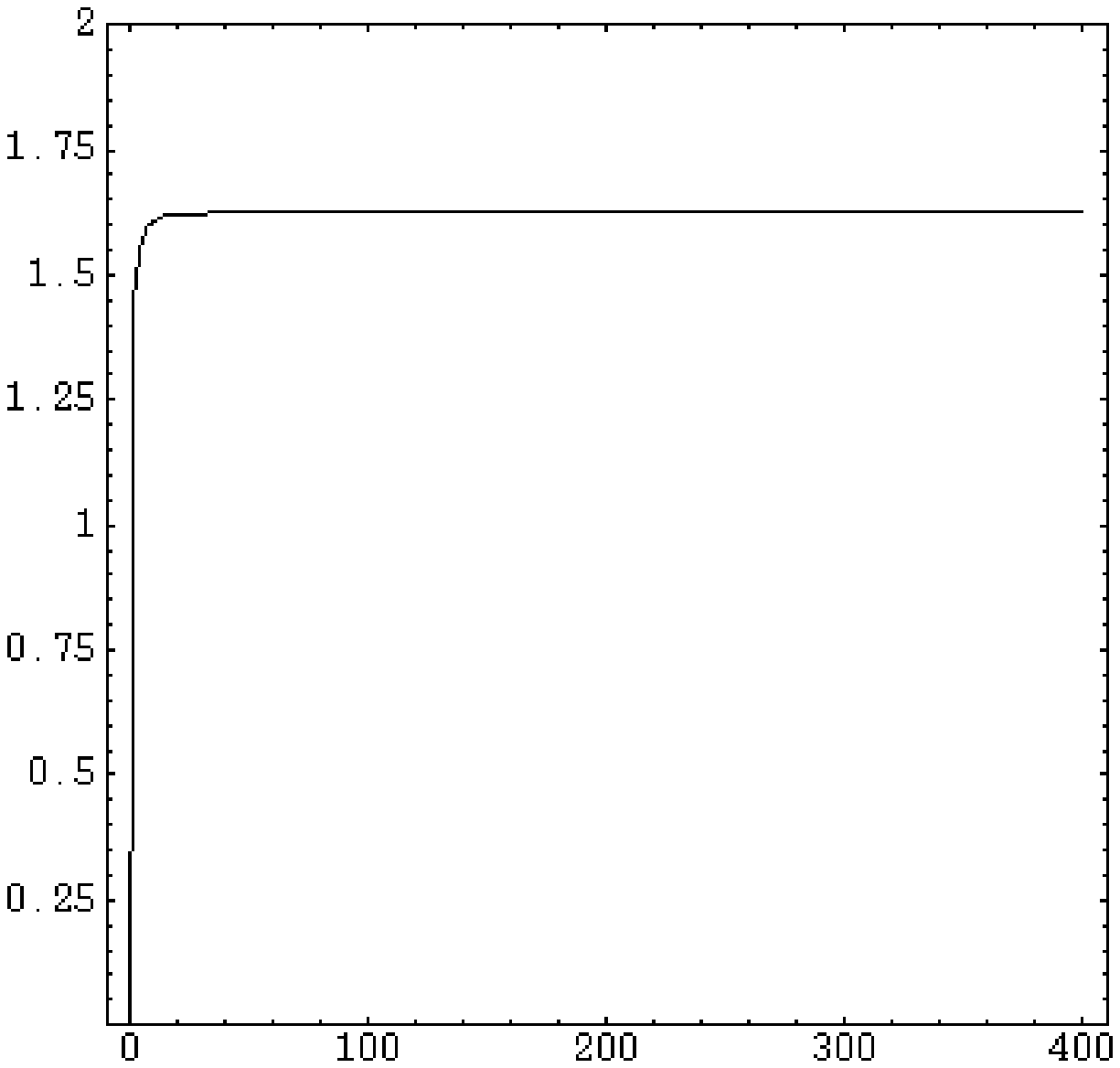}
\end{tabular}
\caption{\label{fig.mt4}The function $s_1(h)$ for the $v$ sector in the ranges $h\in [0,2]$, $h\in [0,20]$, $h\in [0,400]$ and
  $s_1\in [0,2]$ at truncation level $n_c=9$}
\end{figure}

\begin{figure}
\begin{tabular}{lll}
\includegraphics[clip=true,height=4.5cm,
]{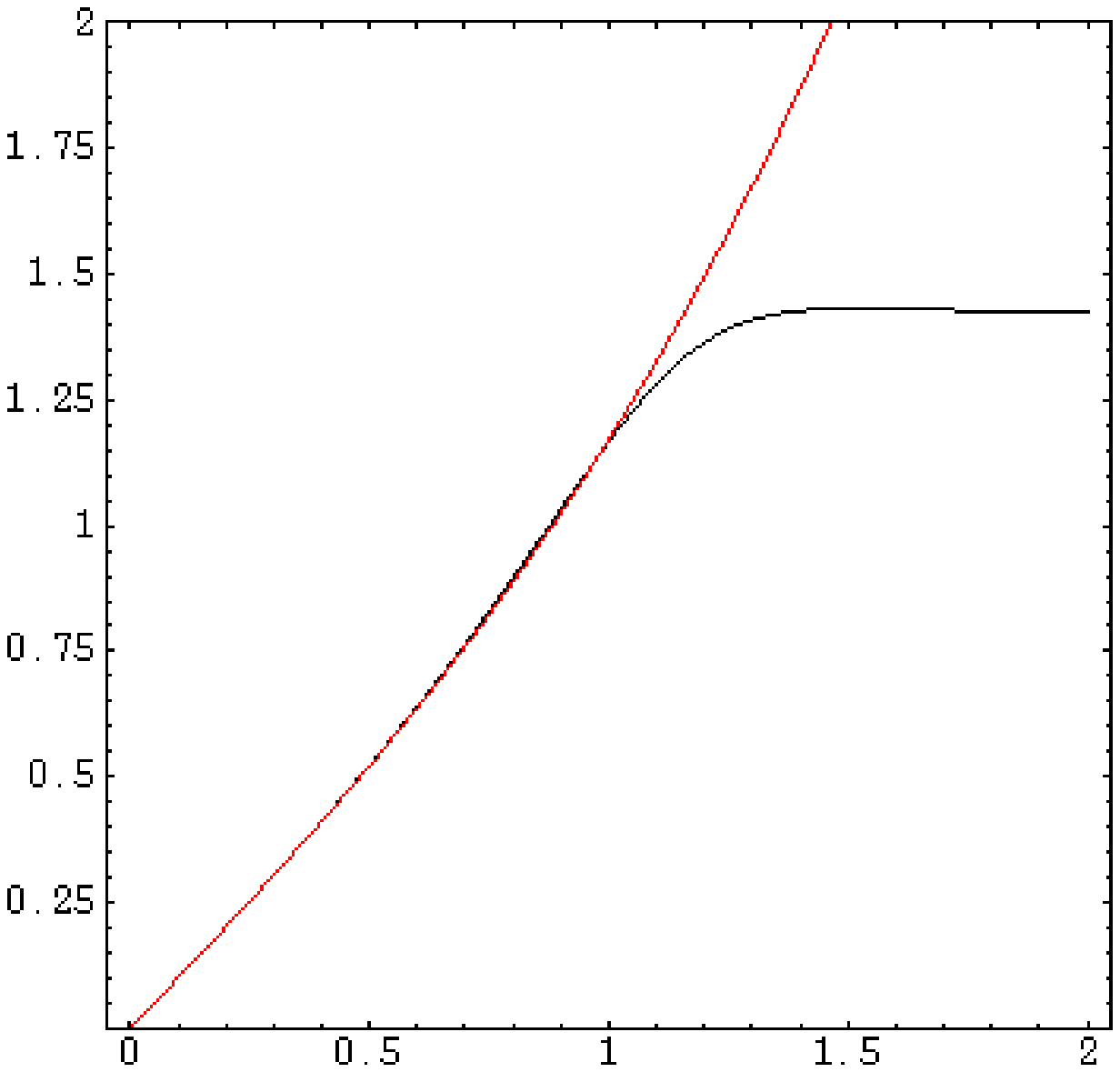}
&
\includegraphics[clip=true,height=4.5cm,
]{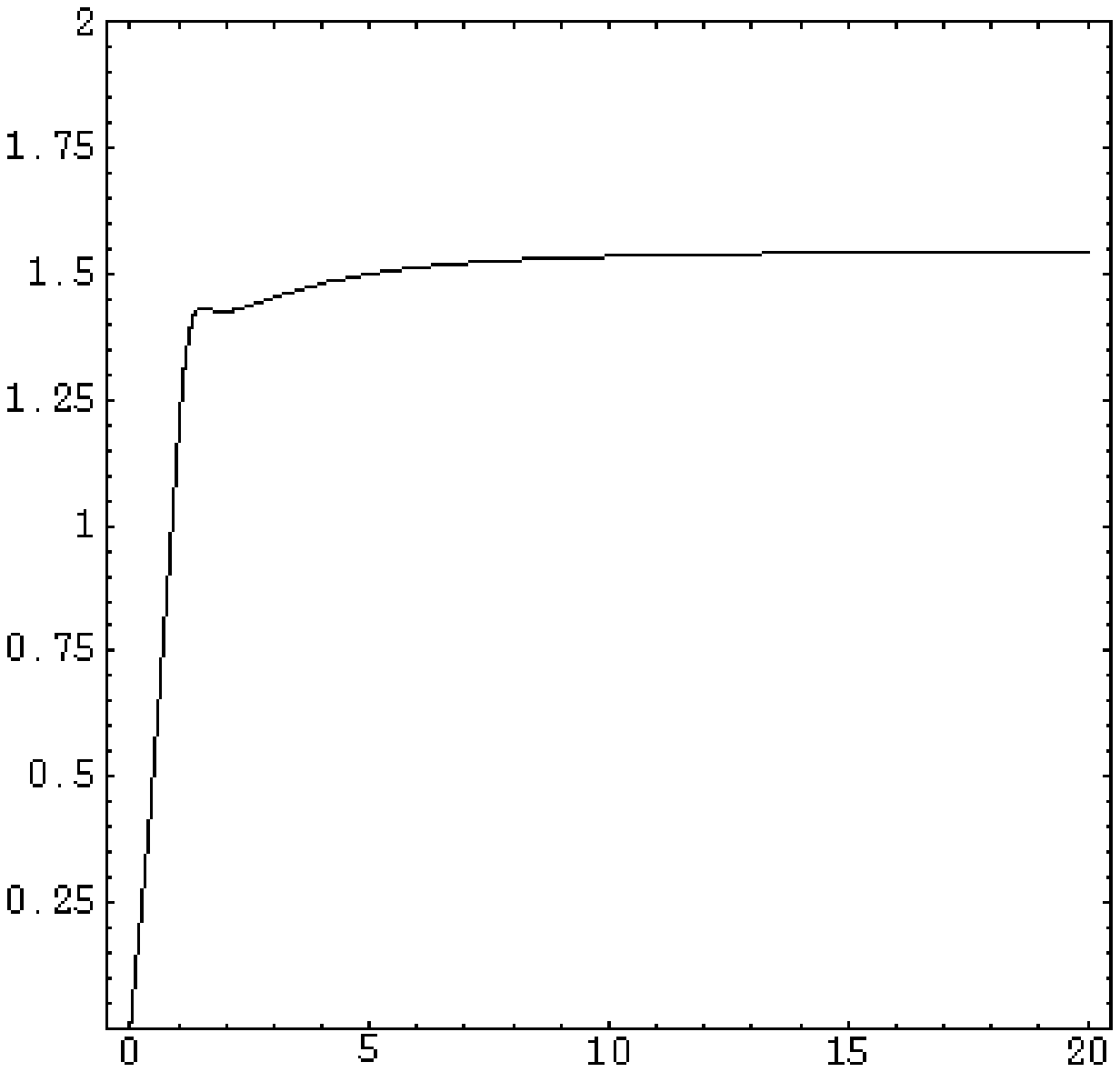}
&
\includegraphics[clip=true,height=4.5cm,
]{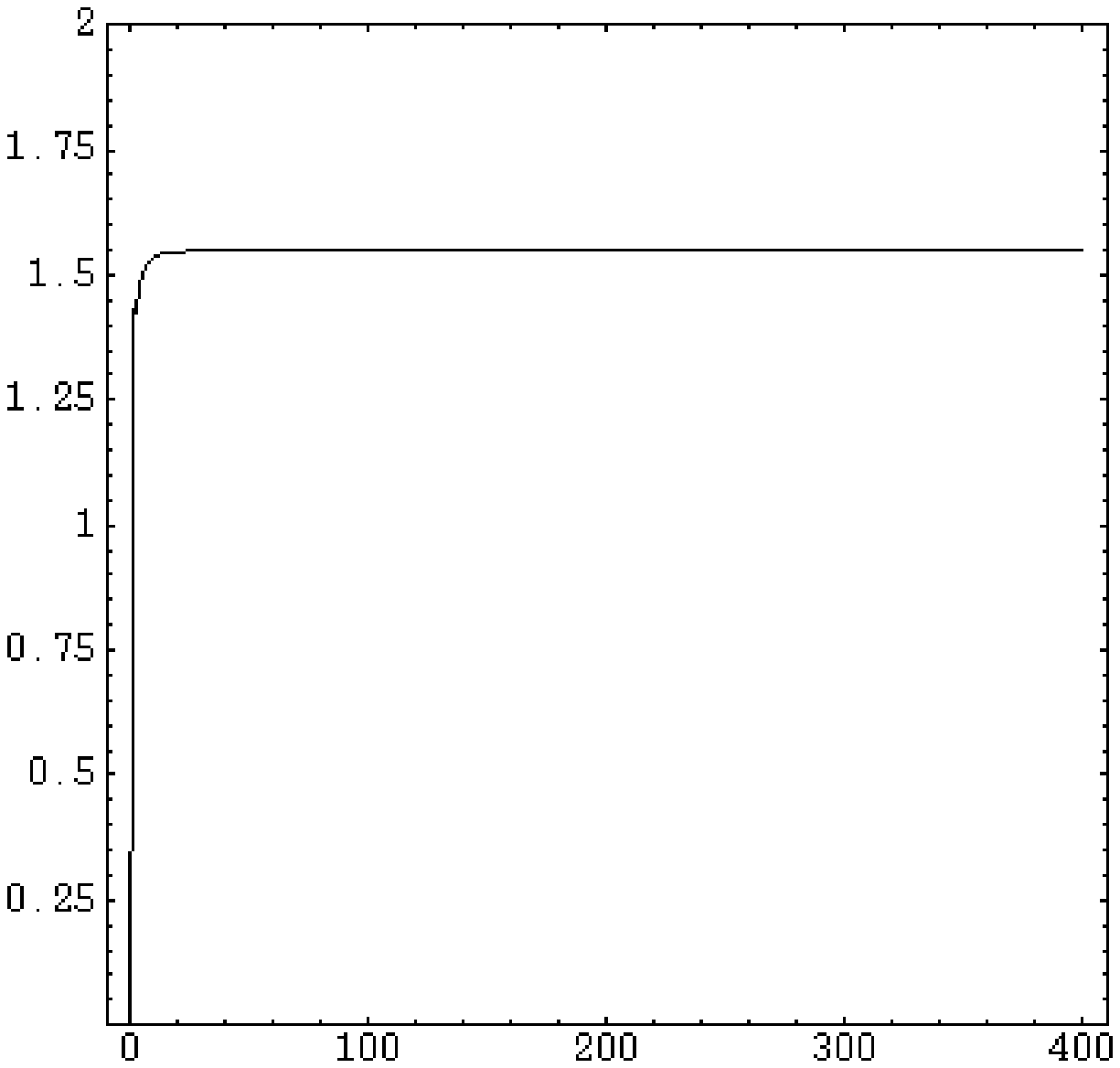}
\end{tabular}
\caption{\label{fig.mt5}The function $s_1(h)/s_0(h)$ for the $v$ sector in the ranges $h\in [0,2]$, $h\in [0,20]$, $h\in [0,400]$ and
  $s_1/s_0\in [0,2]$ at truncation level $n_c=9$}
\end{figure}

Figures \ref{fig.mt3}-\ref{fig.mt5}   show the functions $s_0(h)$, $s_1(h)$,
$s_1(h)/s_0(h)$ determined by the lowest three energy levels in the $v$ sector
via the formulae (\ref{eq.s1}), (\ref{eq.s2}) in
various ranges.
Figures \ref{fig.mt4} and \ref{fig.mt5} also show the curves given by 
(\ref{eq.crv1})  on the left-hand side  (red/grey line).
 $s_0(h)$ remains close to 1 and for large values of $h$ it
tends to a constant which can be expected to converge to 1 as $n_c \to \infty$. 
$s_1(h)$ also tends to a constant for large values of $h$ which can be
expected  to  increase to infinity as $n_c \to \infty$. 
The behaviours of $s_1(h)/s_0(h)$ and $s_1(h)$ are similar.

\begin{figure}
\begin{tabular}{cc}
\includegraphics[clip=true,height=12cm,
]{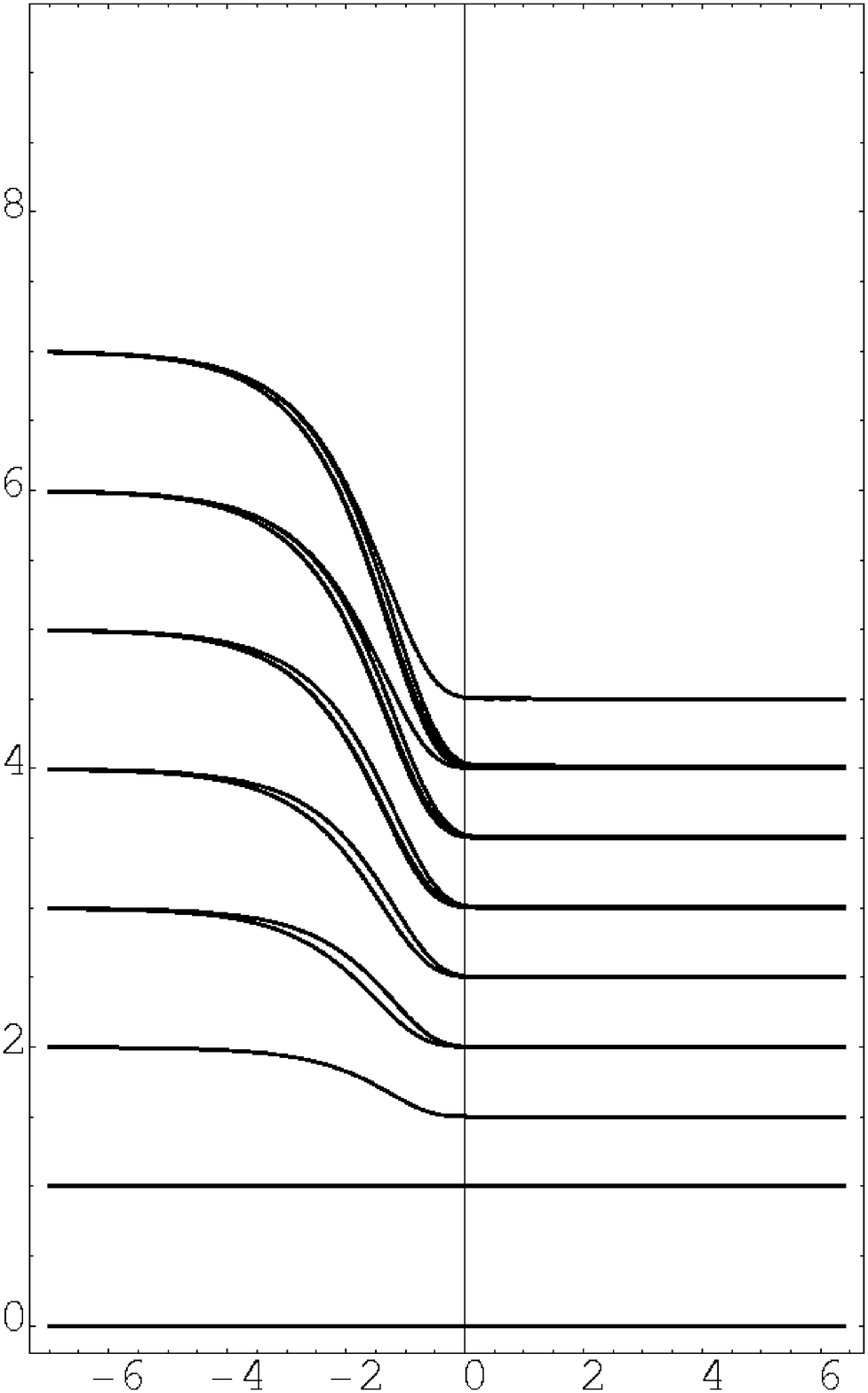}
&
\includegraphics[clip=true,height=12cm,
]{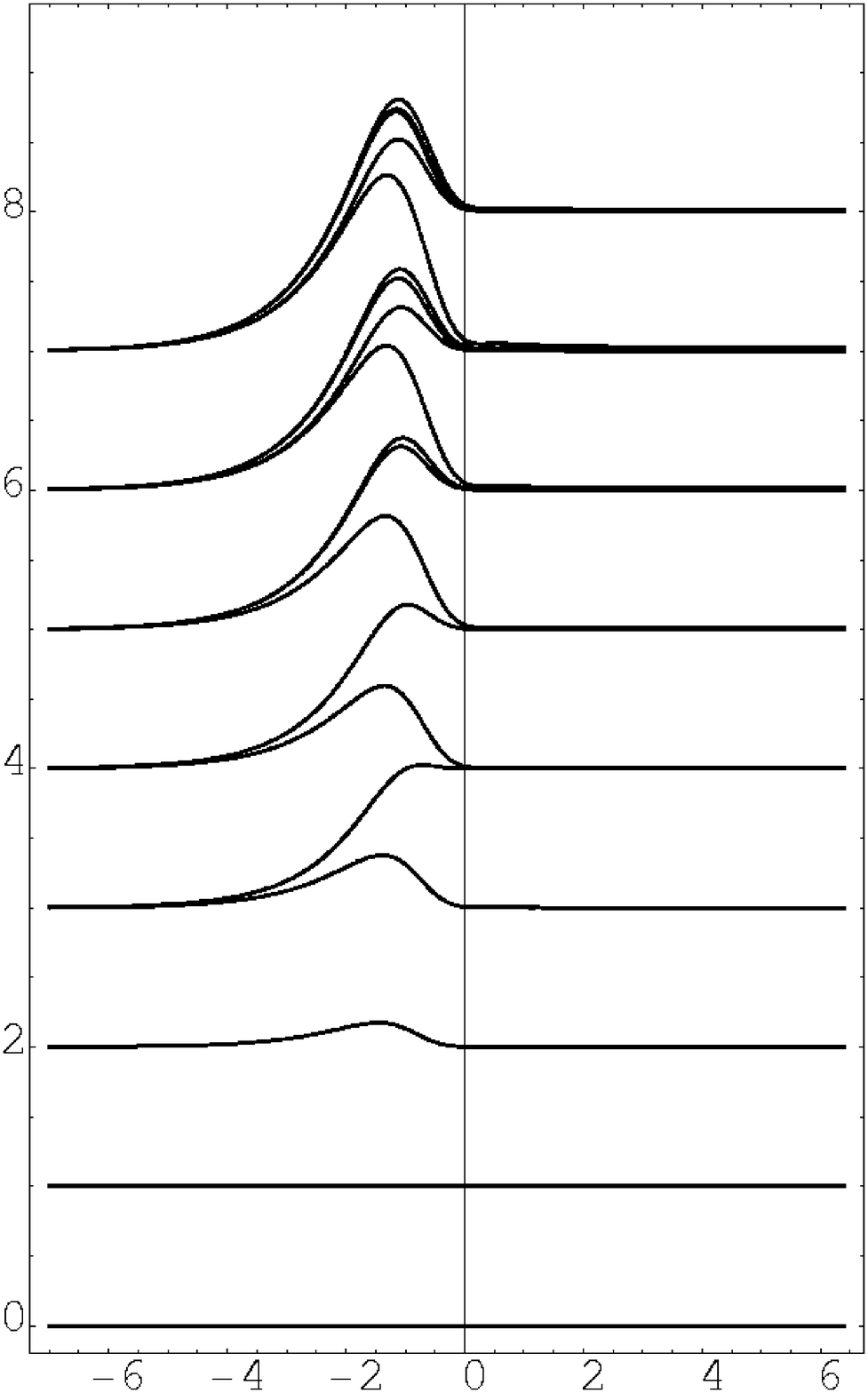}\\
(a)&(b)
\end{tabular}
\caption{\label{fig.mt6}The mode truncated (solid lines) and rescaled exact (dashed
  lines) normalized spectra in the $v$ and $u$ sectors respectively as a function of $\ln (h)$ at truncation level $n_c=9$}
\end{figure}

Figure  \ref{fig.mt6}.a  shows the normalized mode truncated spectrum and the normalized exact spectrum
rescaled by $s_0(h)$ and $s_1(h)$ (i.e.\ the normalized spectrum of $H^r$) in the $v$ sector. No
difference between the two is visible. In Table \ref{tab.mt1} values of the
fifth normalized energy gap $\frac{k(3,h)+k(0,h)}{k(1,h)+k(0,h)}$ of the $v$ sector are listed:
the values in the non-truncated case are listed in the first column, the
values in the mode truncated case are listed in the second column and the
rescaled non-truncated values are listed in the third column (which would
 be the same as the values in the second column if the renormalization
 condition could
 be satisfied exactly).

Figure  \ref{fig.mt6}.b  shows the normalized mode truncated spectrum and the
normalized exact spectrum
rescaled by $s_0(h)$ and $s_1(h)$ in the $u$ sector. The $s_0(h)$, $s_1(h)$
obtained in the $v$ sector were used for the rescaling, which corresponds to the assumption that
$s_0(h)$  and $s_1(h)$ are the same for both sectors. The difference  between the
mode truncated and rescaled exact spectra is not visible in the figure.

We also see from  Table \ref{tab.mt3} in which values of the
fourth normalized energy gap $\frac{k(3,h)-k(0,h)}{k(1,h)-k(0,h)}$ of the $u$
sector are listed  that the  rescaling together with the above
assumption works well.

We have not tried to calculate $s_0$ and $s_1$ for the $u$ sector because
$\frac{E_2-E_0}{E_1-E_0}(h)$ is not invertible in this case. One way to
circumvent this difficulty would be to use other energy levels $E_i,E_j,E_k$
for which $\frac{E_i-E_k}{E_j-E_k}(h)$ is invertible.

\clearpage

\clearpage

\clearpage

\clearpage

\begin{table}
\caption{\label{tab.mt1}The normalized energy gap
  $\frac{k(3,h)+k(0,h)}{k(1,h)+k(0,h)}$  in the $v$ sector:  exact, MT $(n_c=9)$ 
   and rescaled  exact values.}\vspace{2mm}
\begin{tabular}{@{}clll}
\hline
$\ln(h)$ & Exact & MT & Rescaled exact \\
\hline
$-7$ & 2.997677 & 2.997677 & 2.997677  \\
$-6$ & 2.993681  & 2.993681 & 2.993681 \\
$-5$ & 2.982797  & 2.982797 & 2.982797\\
$-4$ & 2.953071  & 2.953068  & 2.953068 \\
$-3$ & 2.8719584  & 2.8719043  & 2.8719043\\
$-2$ & 2.66042177 & 2.6594328 & 2.6594328 \\
$-1$ & 2.25064420  & 2.24043886  & 2.24043637 \\
  0 & 2.00954942  & 2.00440950  & 2.00435155 \\
  1 & 2.00003474  & 2.00145821  & 2.00136637 \\
  2 & 2.00000008  & 2.00106850  & 2.00100783  \\
  3 & 2.0000000  & 2.0009872  & 2.0009201 \\
  4 & 2.0000000  & 2.0009756  & 2.0009202  \\
  5 & 2.0000000  & 2.0009740  & 2.0009187  \\
  6 & 2.0000000  & 2.0009738  & 2.0009185  \\
\hline
\end{tabular}
\vspace{5mm}

\caption{\label{tab.mt3}The normalized energy gap
  $\frac{k(3,h)-k(0,h)}{k(1,h)-k(0,h)}$  in the $u$ sector:  exact, MT  $(n_c=9)$ 
   and rescaled  exact values.}\vspace{2mm}
\begin{tabular}{@{}clll}
\hline
$\ln(h)$ & Exact & MT & Rescaled exact \\
\hline
$-7$ & 3.002321 & 3.002321 & 3.002321\\
$-6$ & 3.006305 & 3.006305 & 3.006305\\
$-5$ & 3.017105 & 3.017105 & 3.017105\\
$-4$ & 3.046201 & 3.046203 & 3.046203\\
$-3$ & 3.1225067 & 3.1225559 & 3.1225559\\
$-2$ & 3.29172833 & 3.29237422 & 3.29237405\\
$-1$ & 3.32029283 & 3.31272102 & 3.31271418\\
 0  & 3.01716419 & 3.00801374 & 3.00788791\\
 1  & 3.00006366 & 3.00269022 & 3.00249222\\
 2  & 3.0000000 & 3.0019707 & 3.0018400\\
 3  & 3.0000000 & 3.0018211 & 3.0017004\\
 4  & 3.0000000 & 3.0017997 & 3.0016804\\
 5  & 3.0000000 & 3.0017968 & 3.0016776\\
 6  & 3.0000000 & 3.0017964 & 3.0016773\\
\hline
\end{tabular}
% \end{indented}
\end{table}

\clearpage

\subsection{TCS scheme}

\begin{figure}
\begin{tabular}{lr}
\includegraphics[clip=true,height=8.5cm,width=7.5cm,
]{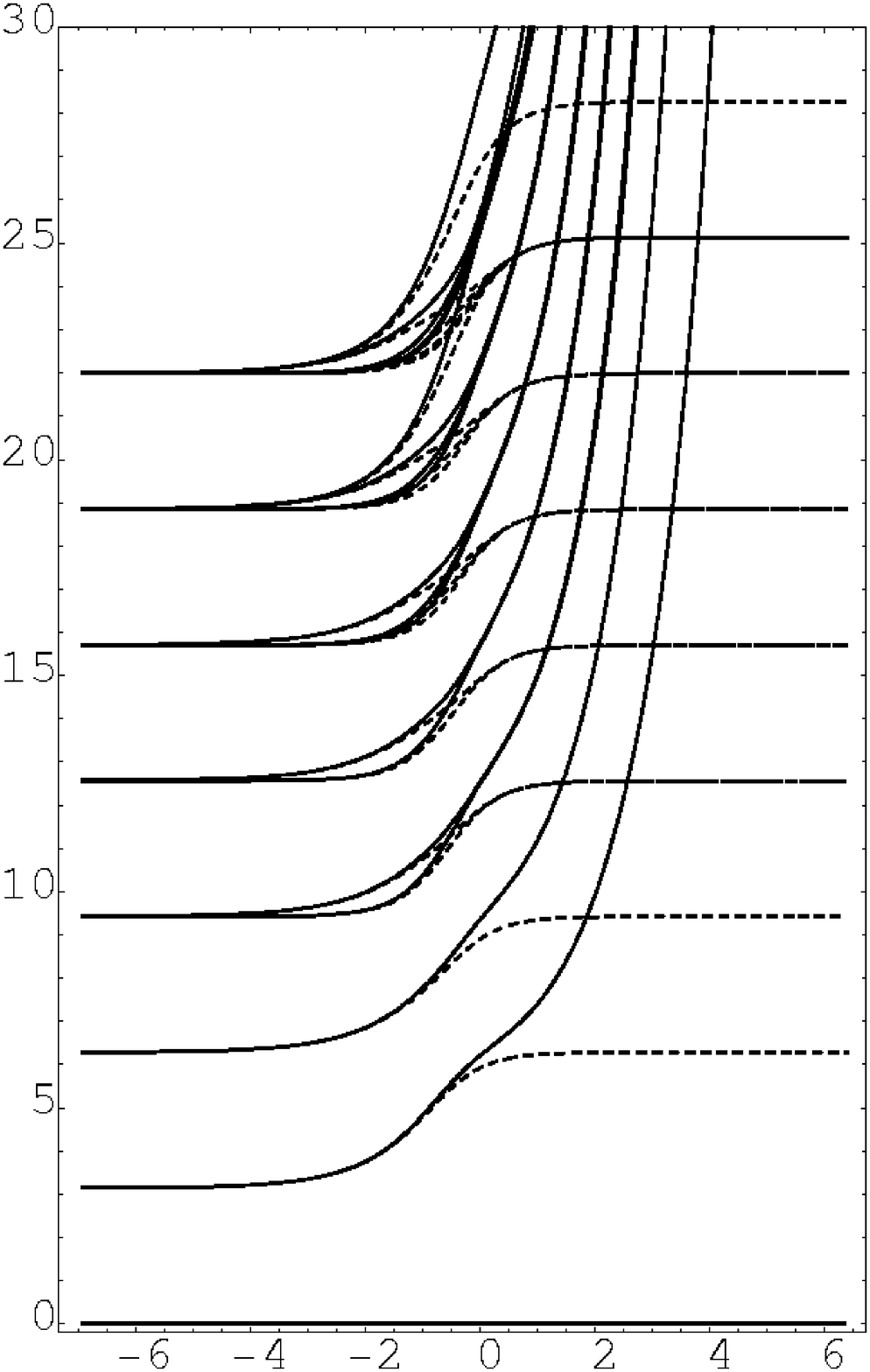}
&
\includegraphics[clip=true,height=8.5cm,width=7.5cm,
]{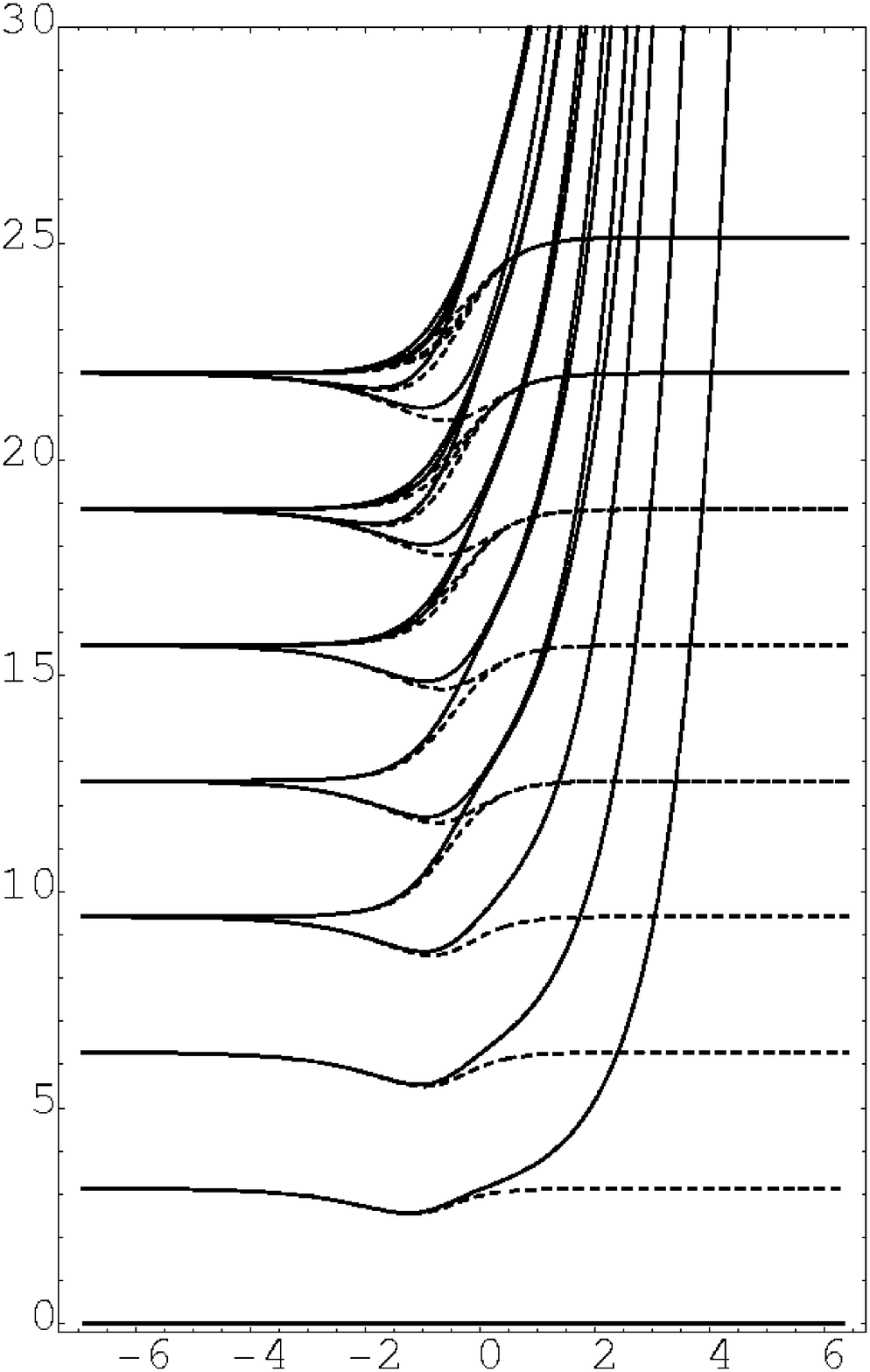}
\end{tabular}
\caption{\label{fig.tcsa1}Exact (dashed lines) and TCSA (solid lines) energy gaps ($E_i-E_0$)
  in the $v$ and $u$ sectors respectively as a function of $\ln(h)$ at
  truncation level $n_c=14$}
\end{figure}

Figure  \ref{fig.tcsa1} shows the exact and TCSA spectra as a function of the
logarithm of the coupling constant. The truncation level is $n_c=14$, and the
dimension of the Hilbert space is 110 in each sector. It is remarkable that
there is strong deviation between the TCSA and exact
spectra for large values of $h$. The behaviour of the TCSA energy gaps is
$E_i(h)-E_0(h) \propto h$ for large values of $h$.

\begin{figure}
\begin{tabular}{lr}
\includegraphics[clip=true,height=8.5cm,width=7.5cm,
]{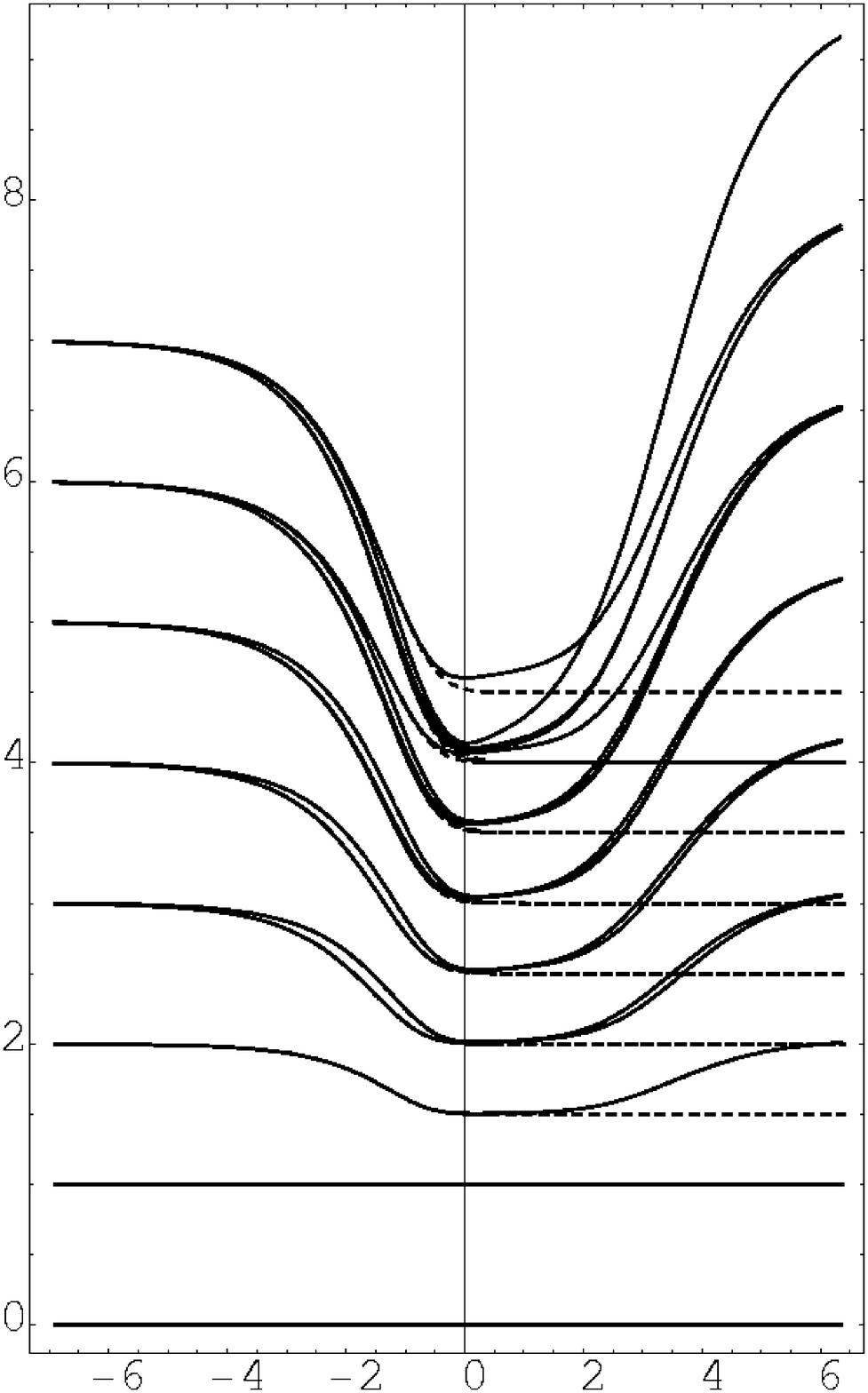}
&
\includegraphics[clip=true,height=8.5cm,width=7.5cm,
]{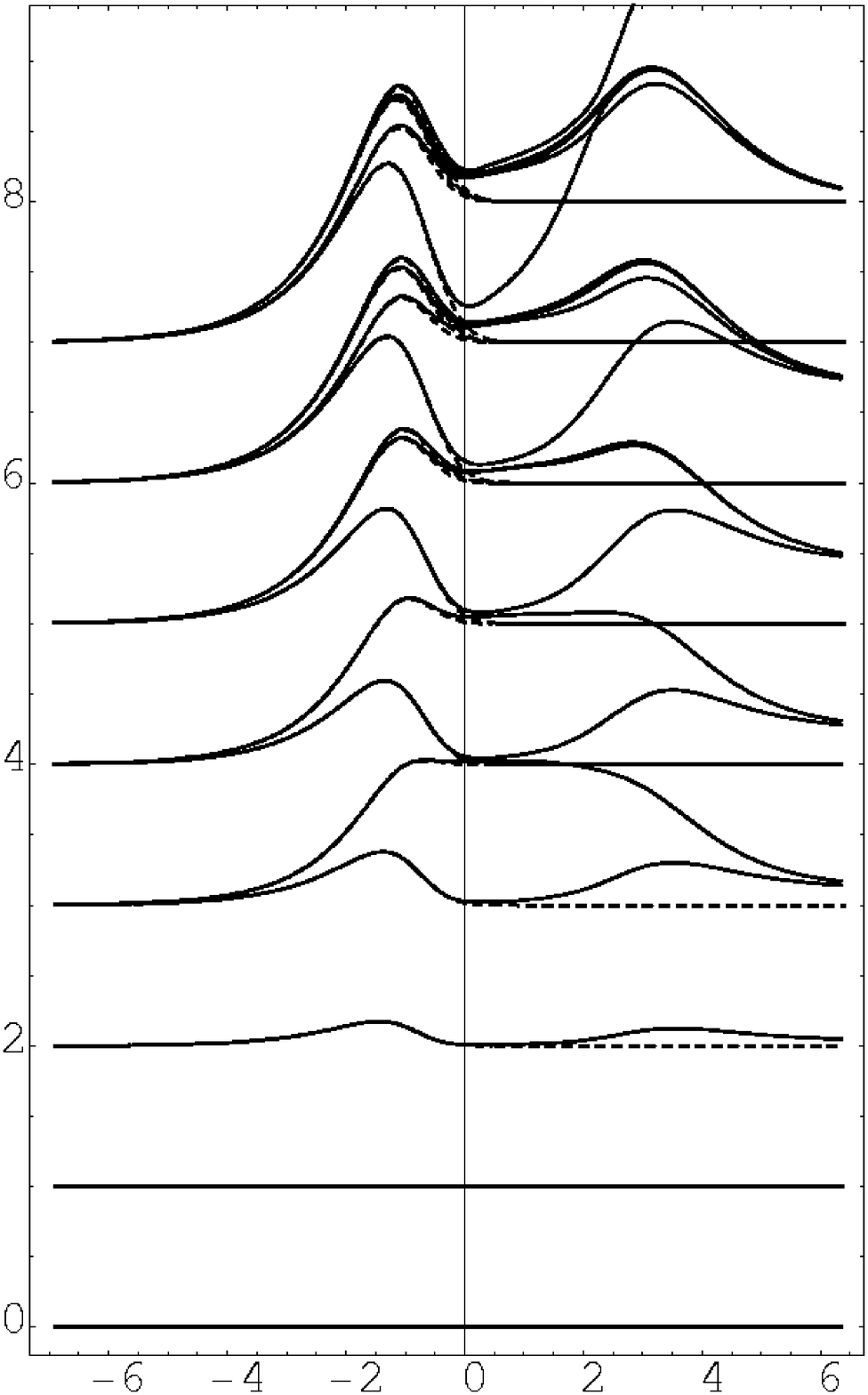}
\end{tabular}
\caption{\label{fig.tcsa2}Exact (dashed lines) and TCSA (solid lines) normalized spectra
  in the $v$ and $u$ sectors respectively as a function of $\ln(h)$ at
  truncation level $n_c=14$}
\end{figure}

Figure   \ref{fig.tcsa2} shows  the same spectra,  but the  lowest gap is normalized to
$1$, i.e.\ the functions $\frac{E_i(h)-E_0(h)}{E_1(h)-E_0(h)}$ are shown. It is
remarkable that the agreement between the exact and TCSA spectra looks better 
than in the case of not normalized spectra.  The functions
$\frac{E_i(h)-E_0(h)}{E_1(h)-E_0(h)}$ have finite limit as  $h \to \infty$ and the
degeneracy pattern in this limit appears to  correspond to the $c=1/2,h=1/16$ representation of the
Virasoro algebra.
The correspondence improves as  $n_c$ is increased (this
improvement is
not illustrated). 
At any fixed finite value of $h$,
however, the TCSA data are expected to converge to the exact values as   $n_c \to
\infty$.  

\begin{figure}
\begin{tabular}{lll}
\includegraphics[clip=true,height=4.5cm,
]{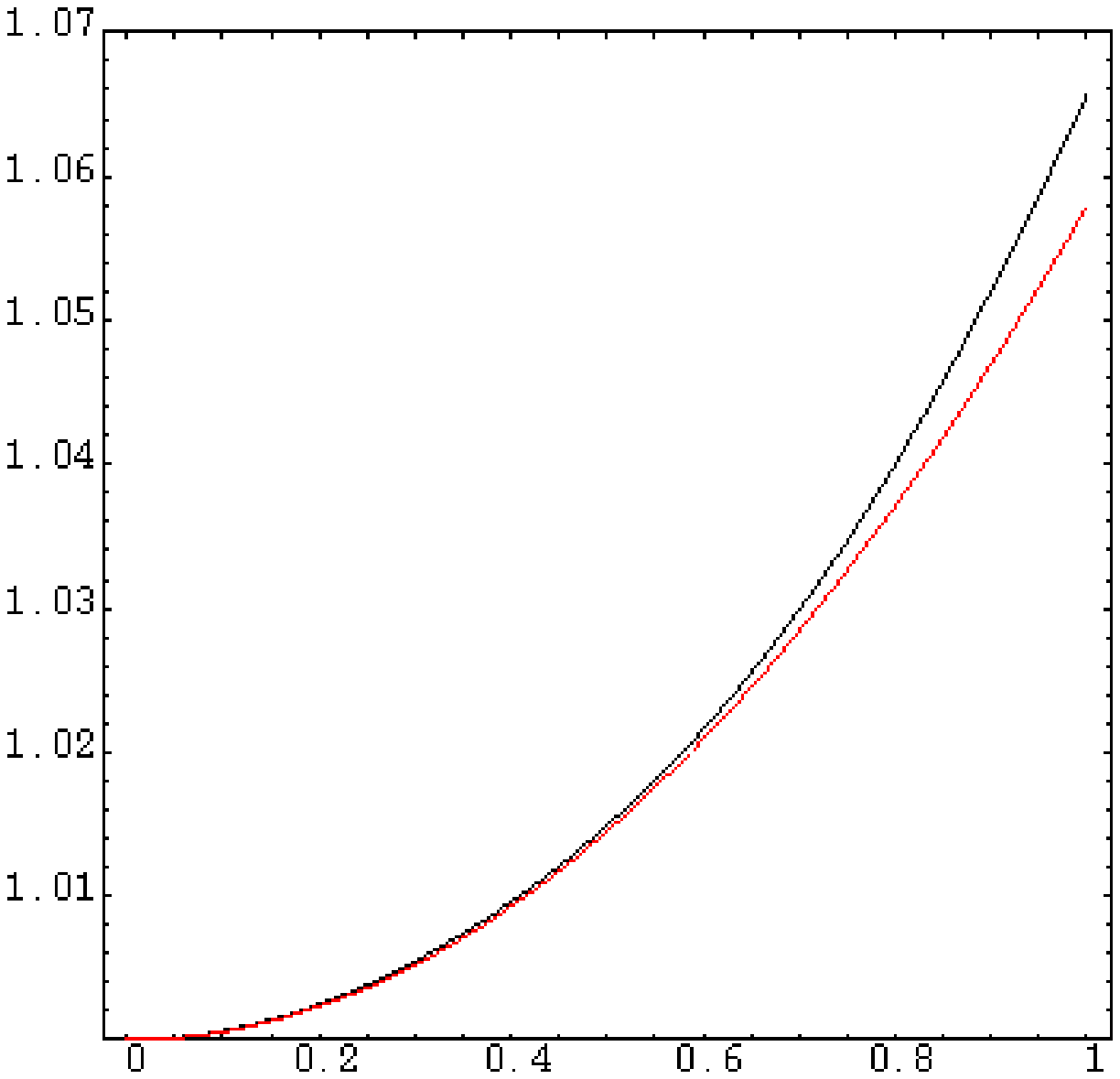}
&
\includegraphics[clip=true,height=4.5cm,
]{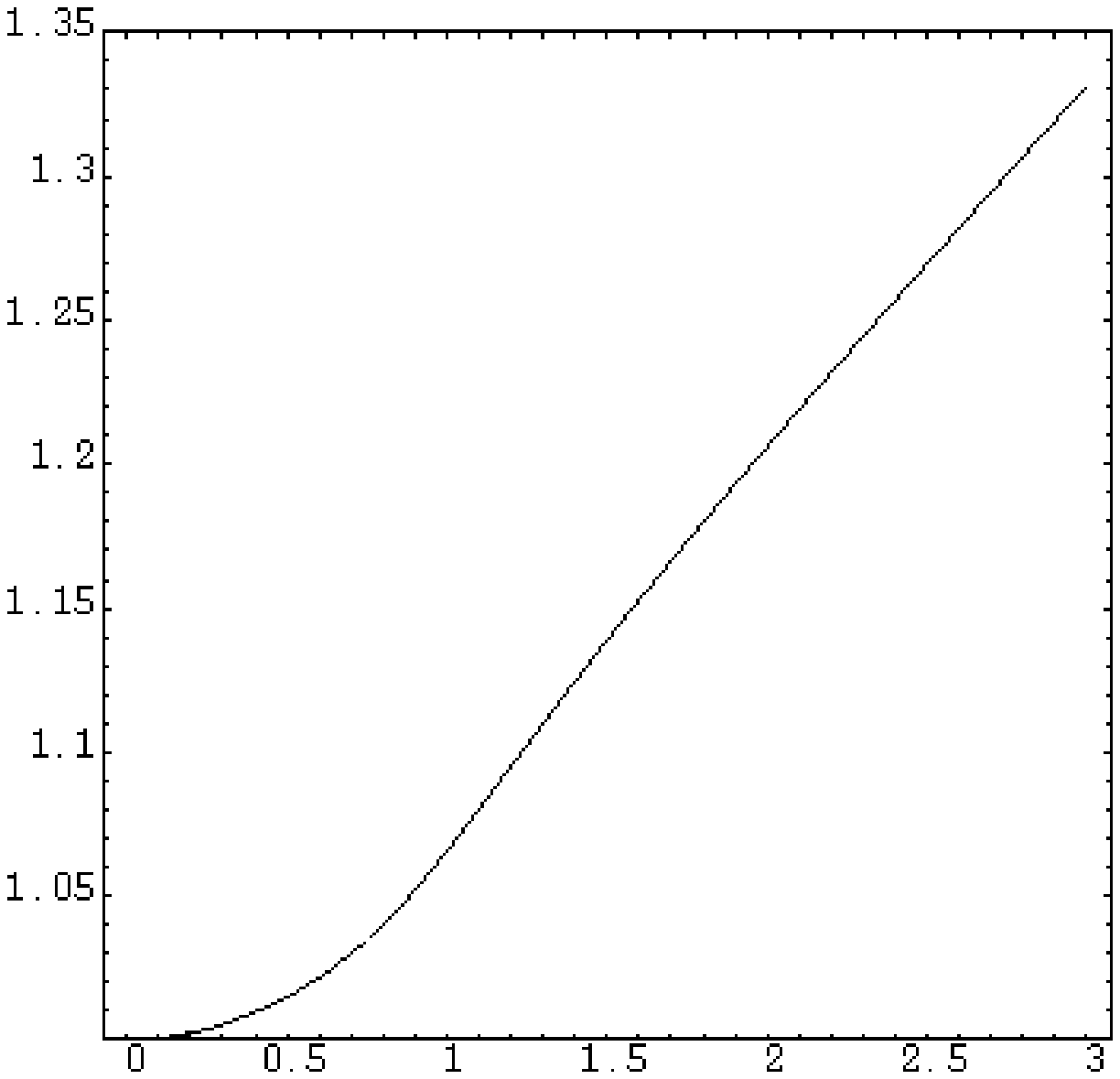}
&
\includegraphics[clip=true,height=4.5cm,
]{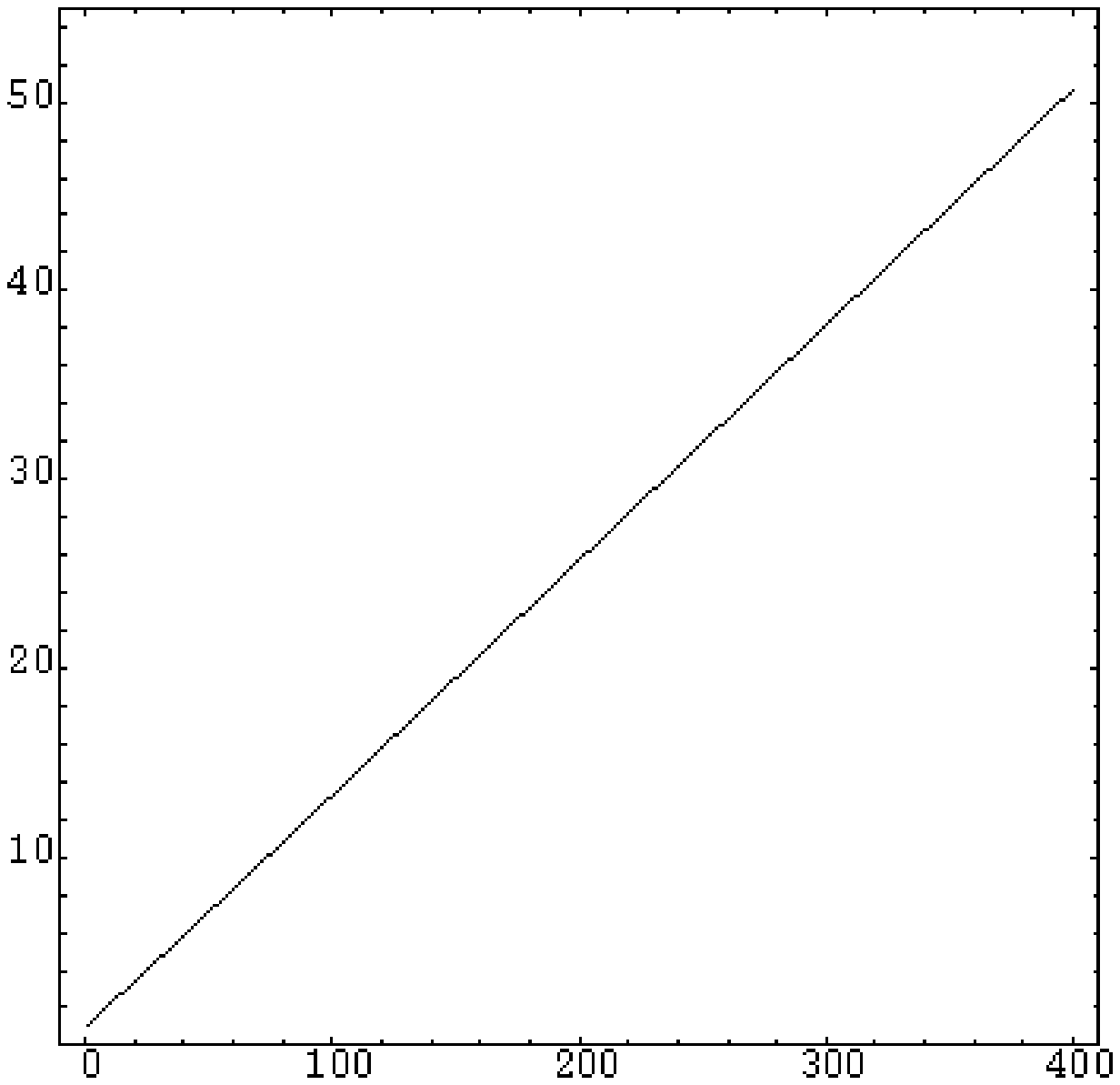}
\end{tabular}
\caption{\label{fig.tcsa3}The function $s_0(h)$ for the $v$ sector  in the ranges $h\in [0,1]$, $s_0\in [1,1.07]$; $h\in [0,3]$,
  $s_0\in [1,1.35]$; $h\in [0,400]$, $s_0\in [1,60]$ at truncation level $n_c=14$ }
\end{figure}

\begin{figure}
\begin{tabular}{lll}
\includegraphics[clip=true,trim=0 4 0 0, height=4.56cm,width=4.62cm ]{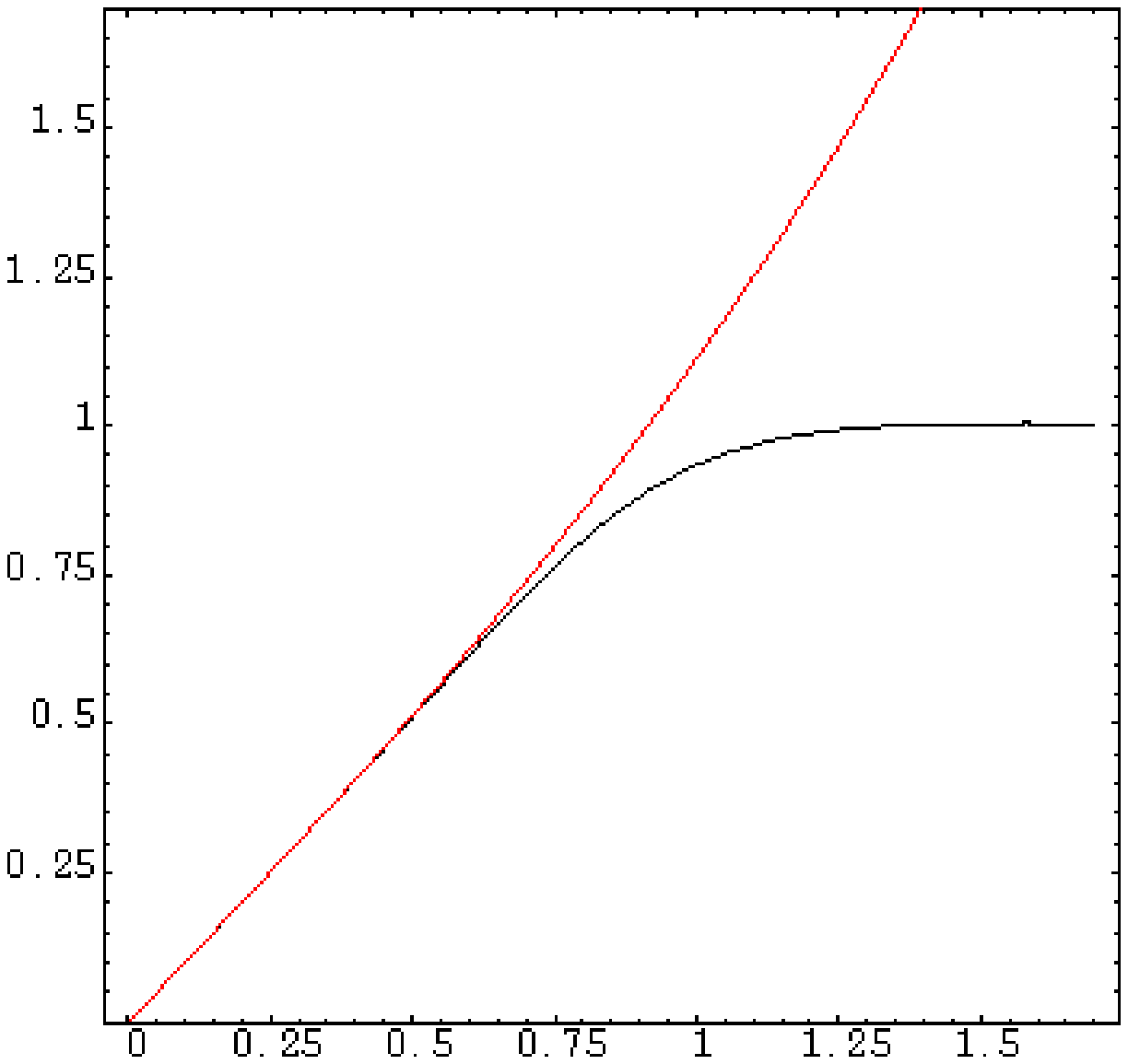}
&
\includegraphics[clip=true,height=4.5cm,
]{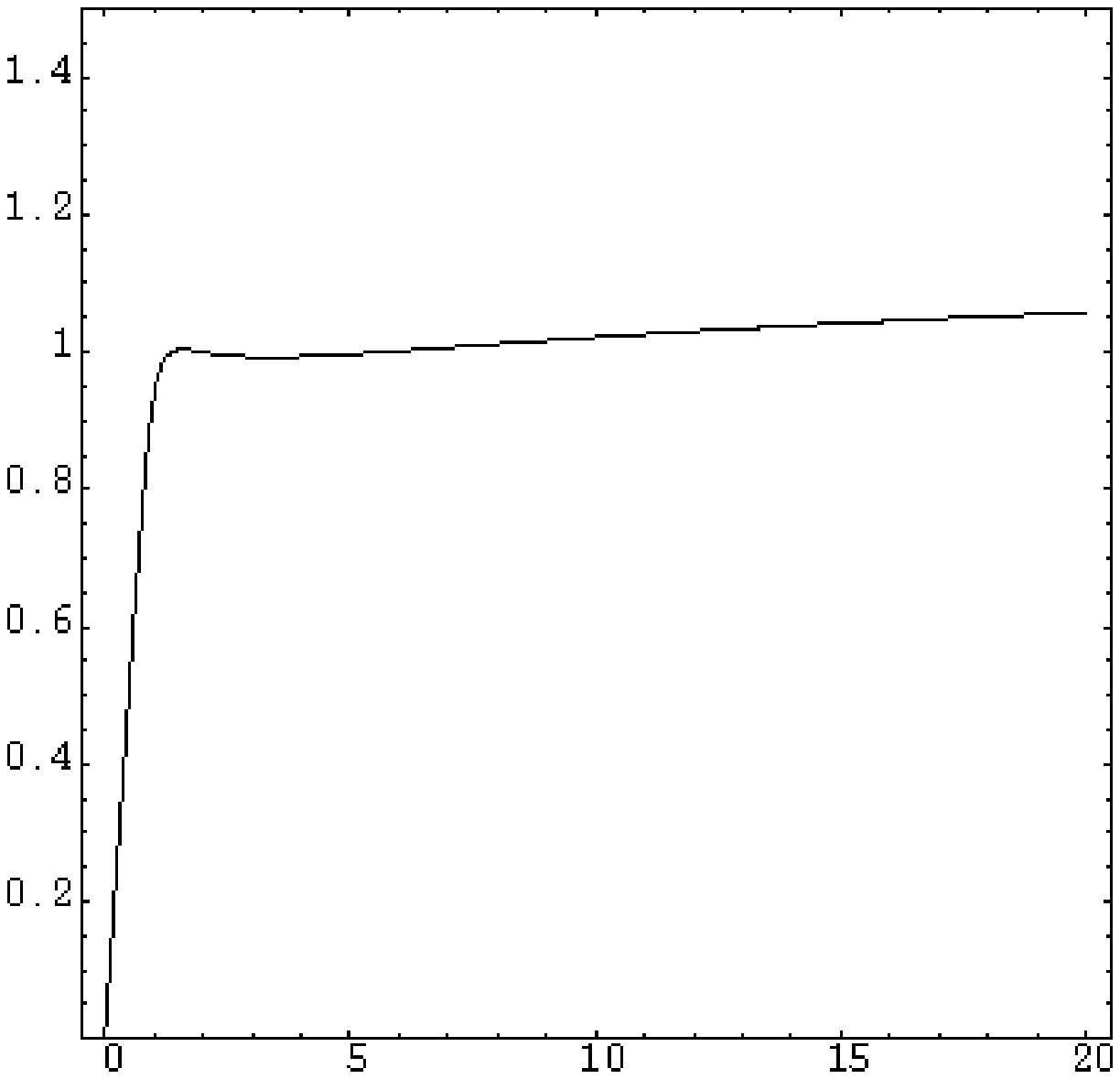}
&
\includegraphics[clip=true,height=4.5cm,
]{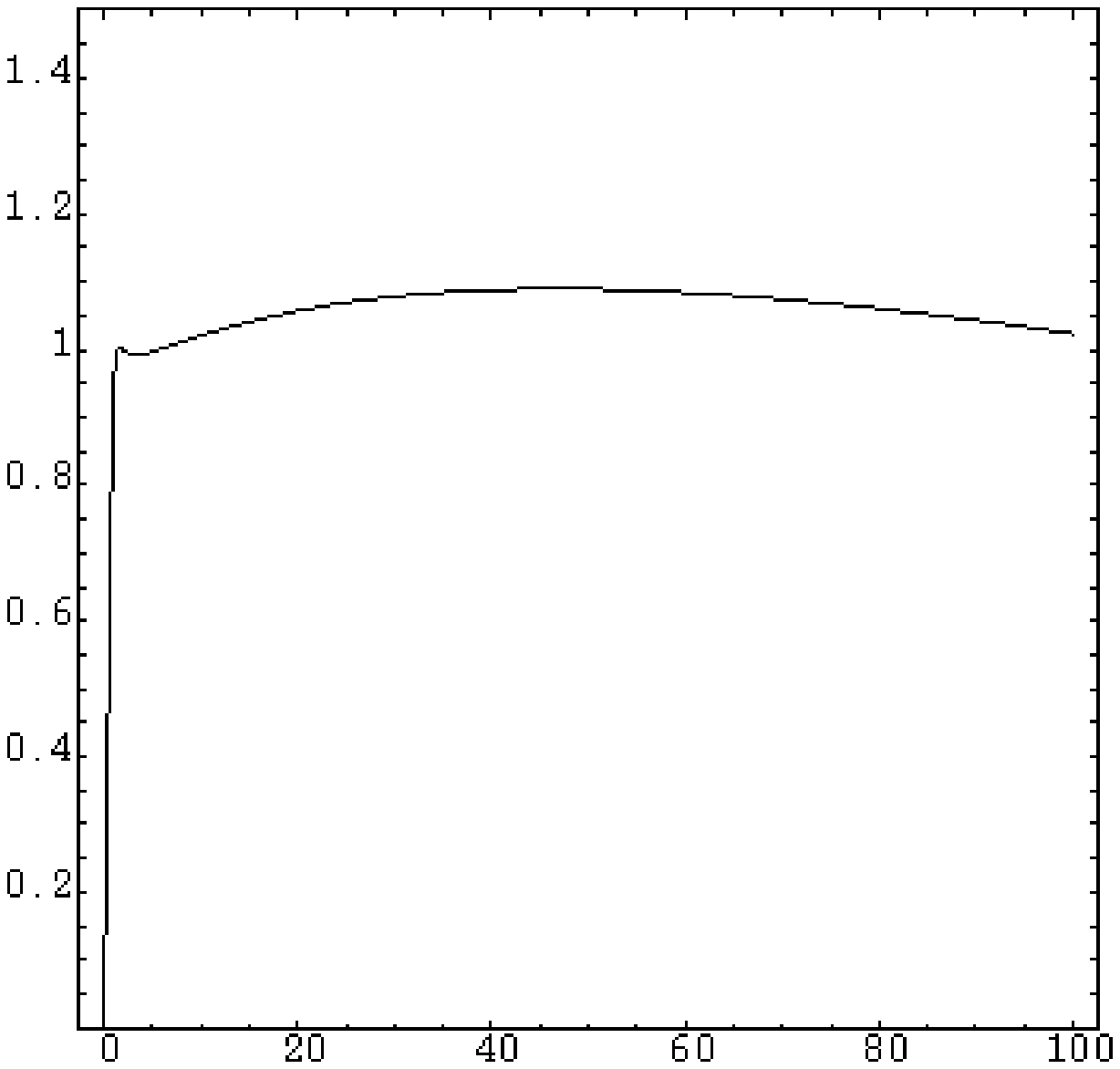}\\
\hspace{1.1mm}\includegraphics[clip=true,height=4.54cm, 
]{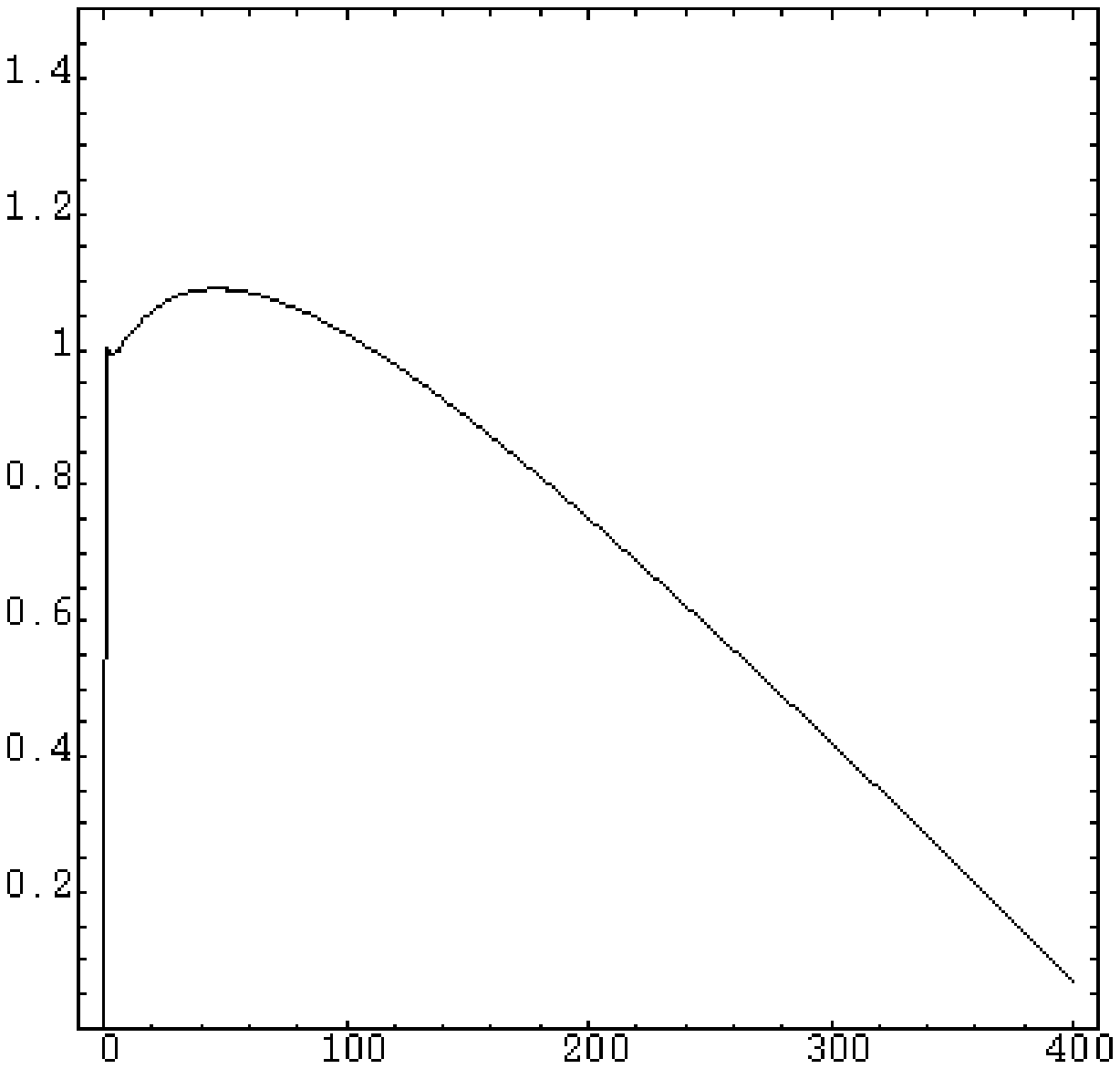} & 
\end{tabular}
\caption{\label{fig.tcsa4}The function $s_1(h)$ for the $v$ sector in the ranges $h\in [0,1.75]$, $s_1\in [0,1.7]$; $h\in [0,20]$,
  $s_1\in [0,1.5]$; $h\in [0,100]$, $s_1\in [0,1.5]$; $h\in [0,400]$, $s_1\in [0,1.5]$ at truncation level $n_c=14$  }
\end{figure}

\begin{figure}
\begin{tabular}{lll}
\includegraphics[clip=true,trim=0 7 0 0,height=4.5cm,]{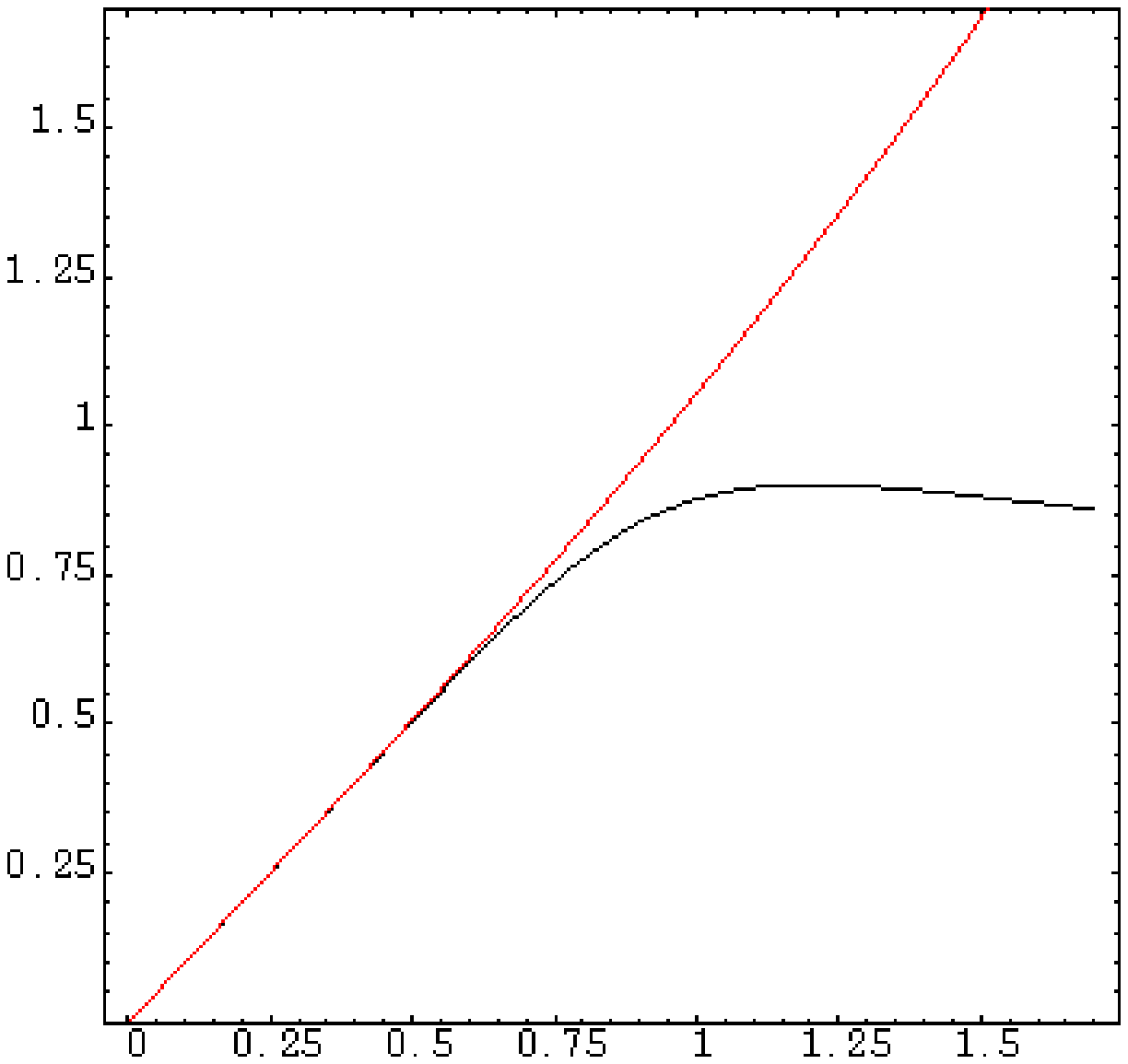}
&
\includegraphics[clip=true,height=4.5cm,
]{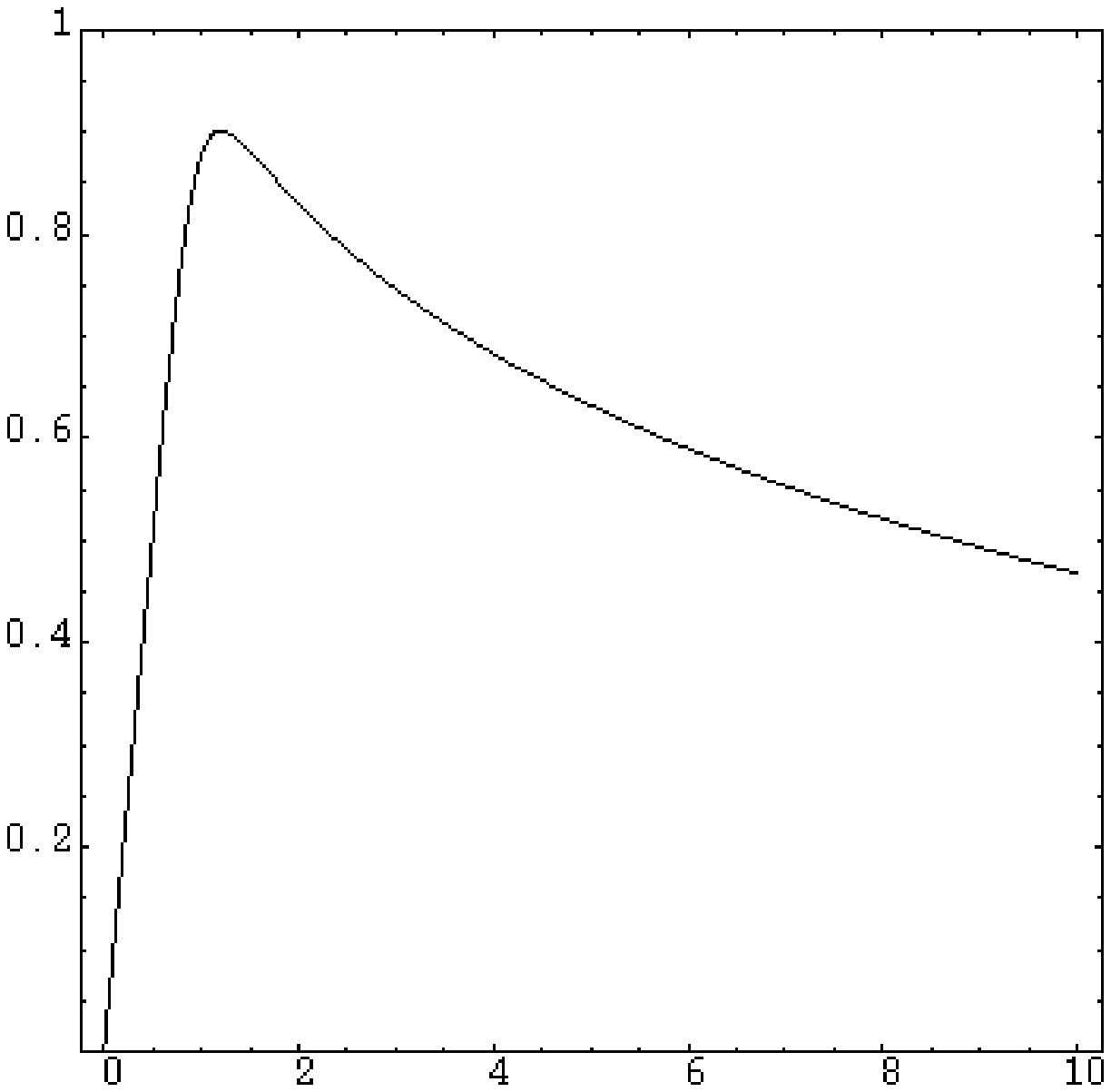}
&
\includegraphics[clip=true,height=4.5cm,
]{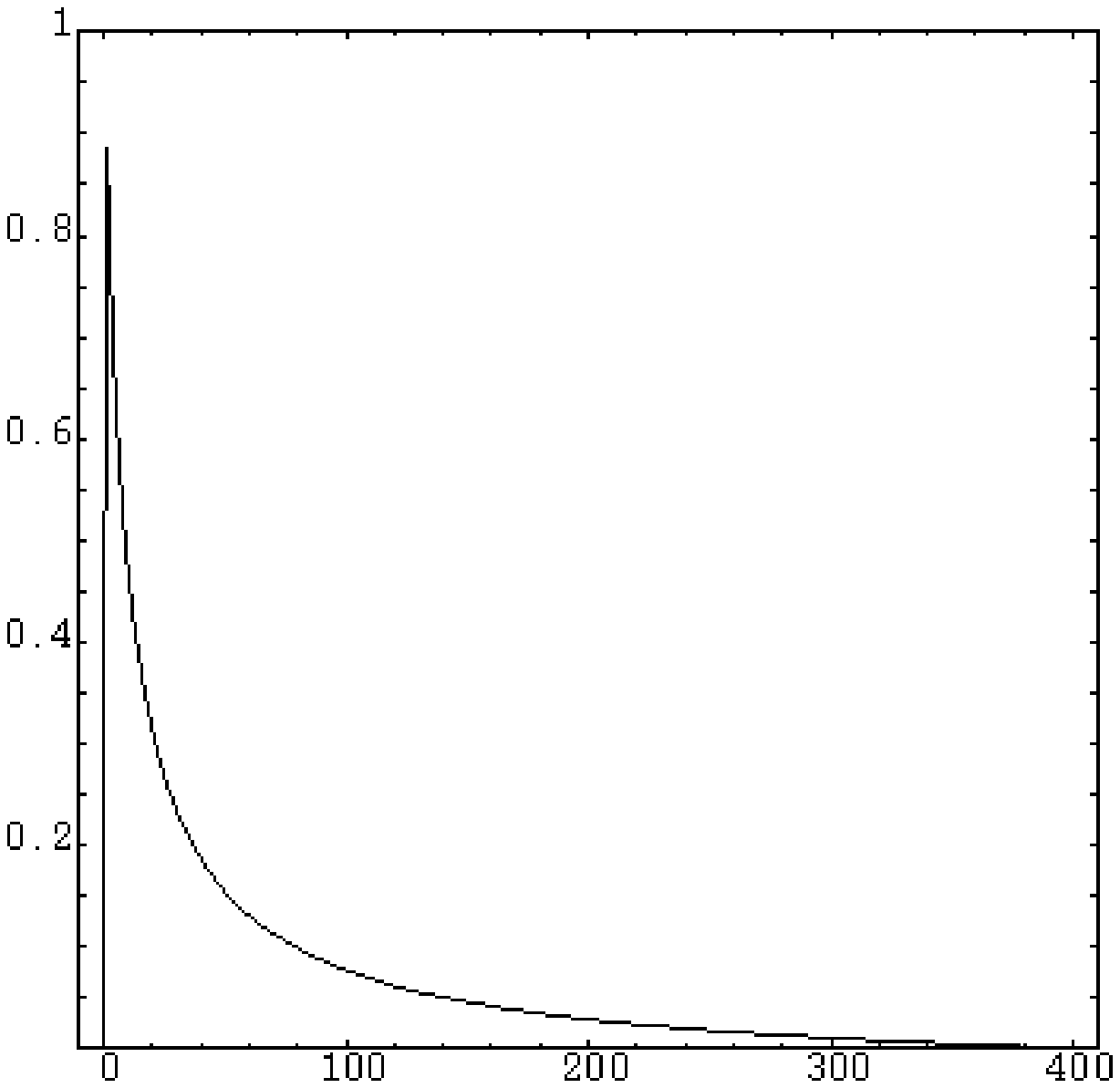}
\end{tabular}
\caption{\label{fig.tcsa5}The function $s_1(h)/s_0(h)$ for the $v$ sector  in the ranges $h\in [0,1.75]$, $s_1/s_0\in [0,1.75]$; $h\in [0,10]$,
  $s_1/s_0\in [0,1]$; $h\in [0,400]$, $s_1/s_0\in [0,1]$ at truncation level $n_c=14$ }
\end{figure}

Figures   \ref{fig.tcsa3}-\ref{fig.tcsa5} show the functions $s_0(h)$, $s_1(h)$, $s_1(h)/s_0(h)$ in
various ranges calculated in the same way as in the mode truncated case. 
The figures also show the curves given by 
(\ref{eq.crv3}), (\ref{eq.crv4}) on the left-hand side  (red/grey line).
It is
remarkable that  $s_0(h)\propto h$  for large values of $h$.
Calculations at other values of $n_c$ show that the slope of $s_0(h)$
decreases as $n_c$ is increased and it can be expected to  converge  to 0 as $n_c \to \infty$. 
$s_1(h)$ appears to tend to a constant for moderately large values of $h$.
Calculations at other values of $n_c$ show that this constant increases as
$n_c$ is increased and it can be expected to 
converge to infinity as $n_c \to \infty$. 
For large values of $h$, $s_1(h)$
decreases.  
$s_1(h)/s_0(h)$ reaches a maximum at $h\approx 1.6$ and then decreases to
 zero. Calculations at other values of $n_c$ show that the maximum value and the value of $h$ where it is reached  increase as
 $n_c$ is increased and it can be expected that both values converge to
 infinity as $n_c \to \infty$. 

\begin{figure}
\begin{tabular}{cc}
\includegraphics[clip=true,height=12cm,
]{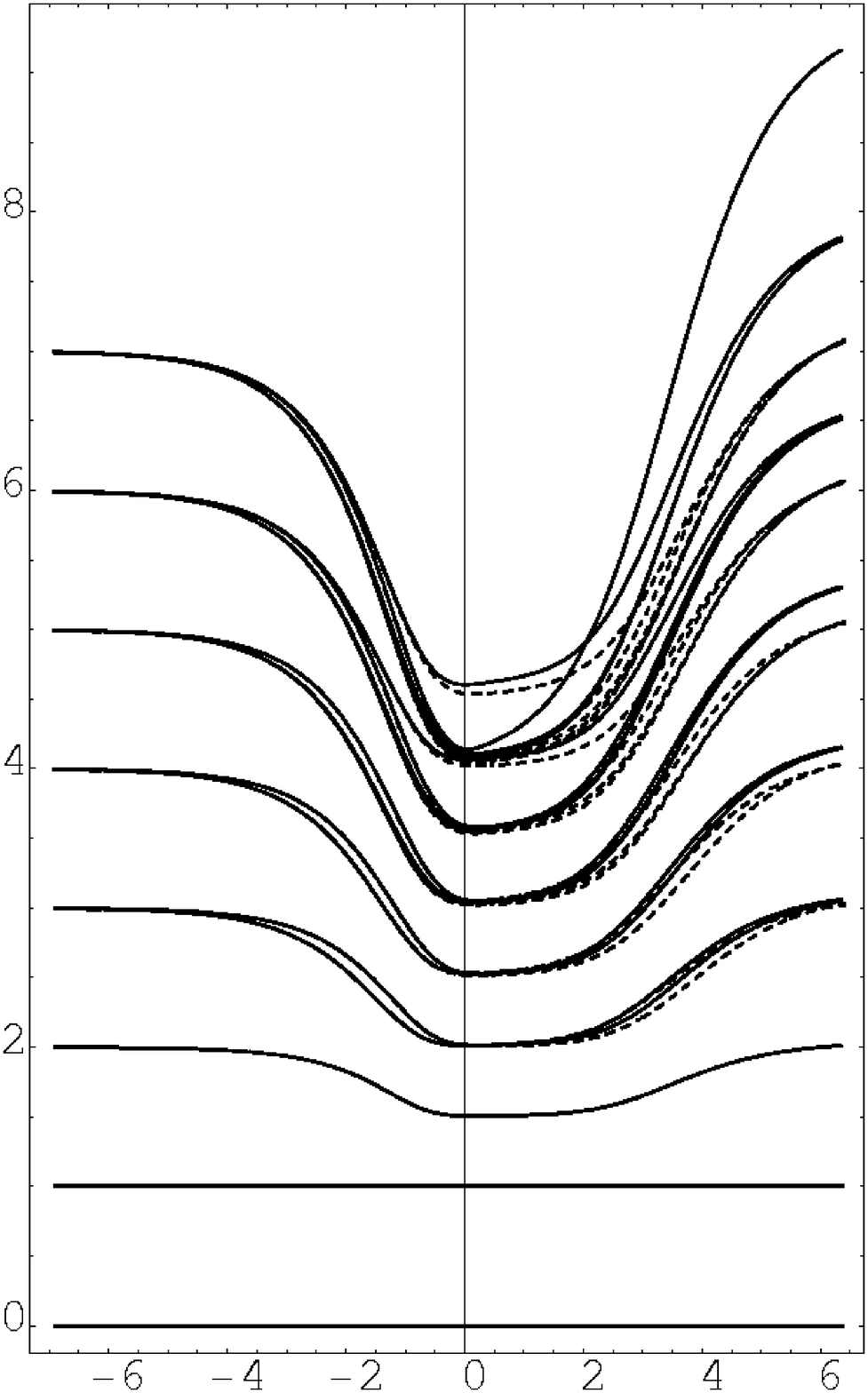}
&
\includegraphics[clip=true,height=12cm,
]{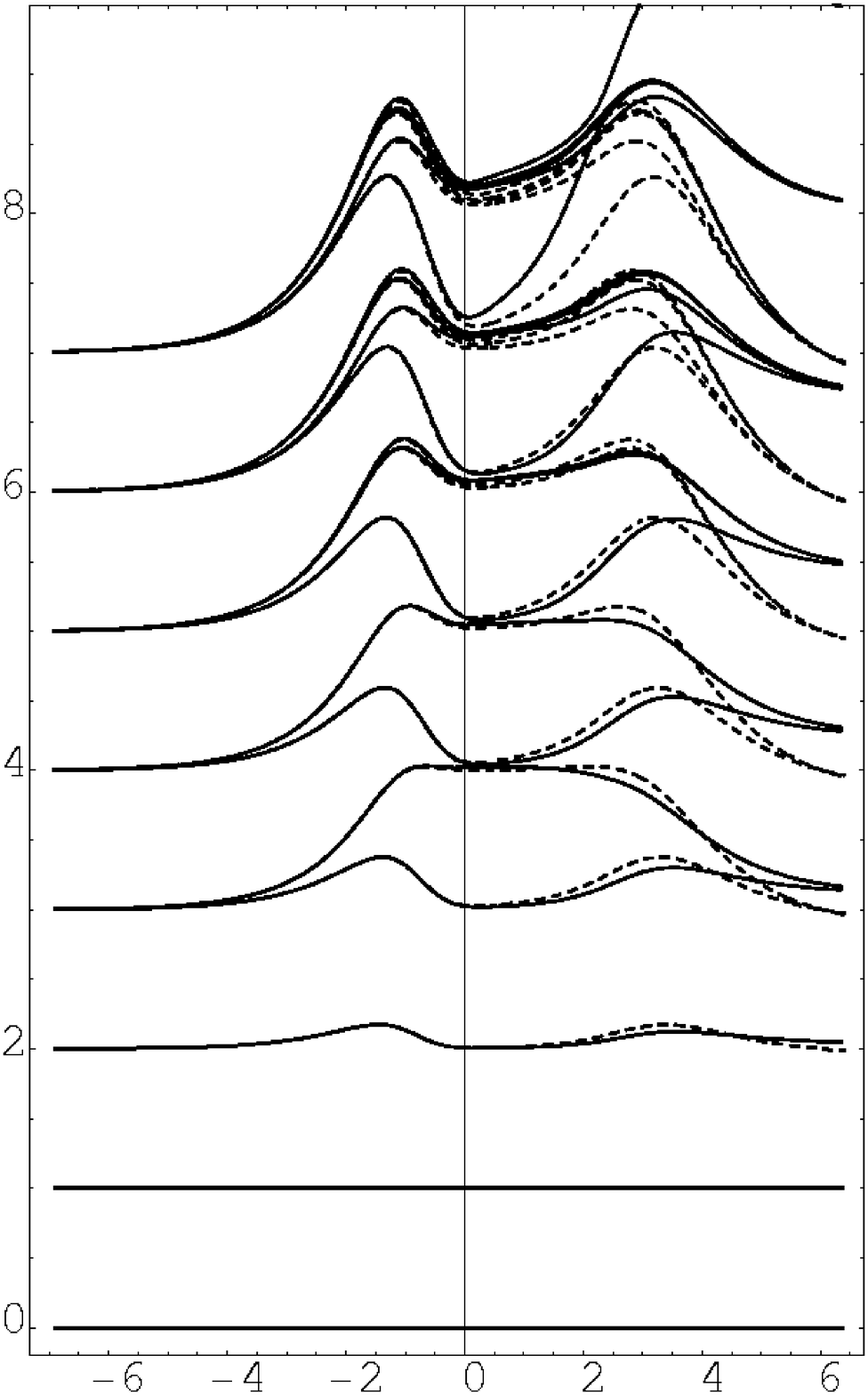}\\
(a) & (b)
\end{tabular}
\caption{\label{fig.tcsa6}The TCSA (solid lines) and rescaled exact (dashed
  lines) normalized spectra in the $v$ and $u$  sectors respectively as a
  function of $\ln (h)$ at truncation level $n_c=14$}
\end{figure}

Figure   \ref{fig.tcsa6}.a shows the normalized TCSA spectrum and the
normalized exact spectrum
rescaled by $s_0(h)$ and $s_1(h)$ (i.e.\ the normalized spectrum of $H^r$). They show  good
qualitative agreement. 

Values of the
fifth normalized energy gap $\frac{k(3,h)+k(0,h)}{k(1,h)+k(0,h)}$ of the $v$
sector are listed in Table \ref{tab.tcsa1} as in the mode truncated
case. These data show that the
rescaling significantly improves the agreement between the TCSA and exact spectra (which is especially noticeable if
$\ln(h)>-2$).

Figure  \ref{fig.tcsa6}.b  shows the normalized TCSA spectrum and the
normalized exact spectrum
rescaled by $s_0(h)$ and $s_1(h)$ in the $u$ sector. The $s_0(h)$, $s_1(h)$ functions
obtained in the $v$ sector were used for the rescaling as in the mode
truncated case. 
 Table \ref{tab.tcsa3} shows  values of the
fourth normalized energy gap $\frac{k(3,h)-k(0,h)}{k(1,h)-k(0,h)}$ of the $u$
sector. 
We see  from the table that the assumption that $s_0(h)$  and $s_1(h)$
are the same in both sectors  does not give very good result in this case, although the situation
might become better at higher truncation levels.

%% section tcsa
\clearpage

\clearpage

\clearpage

\clearpage

\begin{table}[!h]
\caption{\label{tab.tcsa1}The normalized energy gap
  $\frac{k(3,h)+k(0,h)}{k(1,h)+k(0,h)}$  in the $v$ sector:  exact, TCSA  $(n_c=14)$
   and rescaled  exact values.}\vspace{2mm}
\begin{tabular}{@{}clll}
\hline
$\ln(h)$ & Exact  & TCSA   & Rescaled  exact \\
\hline
$-6$ & 2.993681 & 2.993681 & 2.993681    \\
$-5$ & 2.982797 & 2.982797 & 2.982797   \\
$-4$ & 2.953071 & 2.953074 & 2.953073   \\
$-3$ & 2.8719584 & 2.871974 & 2.871961  \\
$-2$ & 2.66042177 & 2.660322 & 2.660236    \\
$-1$ & 2.25064420 & 2.2488 & 2.24837    \\
 0 & 2.00954942 & 2.01699 & 2.0177    \\
 1 & 2.00003474 & 2.028321 & 2.031631   \\
 2 & 2.00000008 & 2.101337 & 2.105516    \\
 3 & 2.0000000 & 2.33433 & 2.32573  \\
 4 & 2.0000000 & 2.670454 & 2.643480    \\
 5 & 2.0000000 & 2.916318 & 2.879339   \\
 6 & 2.0000000 & 3.037542 & 2.997132   \\
\hline  
\end{tabular} 
\vspace{5mm}

\caption{\label{tab.tcsa3}The normalized energy gap
 $\frac{k(3,h)-k(0,h)}{k(1,h)-k(0,h)}$  in the $u$ sector:  exact, TCSA  $(n_c=14)$ 
   and rescaled  exact values.}\vspace{2mm}
\begin{tabular}{@{}clll}
\hline
$\ln(h)$ & Exact  & TCSA  & Rescaled  exact \\
\hline
$-7$ & 3.002321 & 3.002321 & 3.002321\\
$-6$ & 3.006305 & 3.006305 & 3.006305\\
$-5$ & 3.017105 & 3.017105 & 3.017105\\
$-4$ & 3.046201 & 3.046206 & 3.046198\\
$-3$ & 3.1225067 & 3.122558 & 3.122504\\
$-2$ & 3.29172833 & 3.292196 & 3.291849\\
$-1$ & 3.32029283 & 3.31947 & 3.31865\\
 0 & 3.01716419 &  3.0268 & 3.0315\\ 
 1 & 3.00006366 &  3.03610 & 3.05522\\
 2 & 3.0000000 &  3.11033 & 3.16799\\
 3 & 3.0000000&  3.27063 & 3.36158\\
 4 & 3.0000000 &  3.28320 & 3.30254\\
 5 & 3.0000000 &  3.203134 & 3.115763\\
 6 & 3.0000000 &  3.152931 & 3.002865\\
\hline  
\end{tabular} 
\end{table}

\clearpage

\section{Discussion}
\label{sec.disc2}

We have investigated the validity of the approach (\ref{eq.i2}) for the description
of truncation effects in TCSA spectra. Comparison  with another truncation
method called mode truncation shows that the remarkably regular behaviour of the TCSA spectrum for large $h$ in the case
of the model (\ref{eq.yyy}), namely the presence of second flows, is not universal
(i.e.\ not independent of the truncation scheme). The numerical calculations show
that (\ref{eq.i2}) provides a good approximation of the truncated spectra in both the
TCSA and the mode truncation scheme. This is confirmed by perturbative
analytic 
calculations as well. The main difference between the mode truncated 
and TCSA spectra at large $h$ seems to be explicable through the different  behaviour of the function
$s_0(n_c,h)$ in the two schemes. Difference between the $s_0(n_c,h)$
functions appears also in perturbation theory. We
have shown analytically  that in the mode truncation scheme the
convergence of the truncated spectra to the exact spectra can be improved by
(at least) one order in $1/n_c$ by
the  rescaling (\ref{eq.i2}). This has also been shown in the
TCSA scheme for low orders of perturbation theory in $h$.

We have also given a quantum field theoretic
discussion of the model (\ref{eq.yyy}). In particular we have discussed (see
Appendix \ref{app.folyt}) the change of the
boundary condition satisfied by the fermion fields as the coupling constant
(or external boundary magnetic field) is
increased. Such a change, which is emphasized in the literature, seems impossible 
naively --- in our formulation at least. The paradox is resolved by the
 phenomenon that the  fermion fields (more precisely their matrix
elements between energy eigenstates) develop a discontinuity at the boundary if the coupling
constant is nonzero.

It is still an open problem  to present  an  explanation of the validity of the
approach (\ref{eq.i2}). Within the framework of  (\ref{eq.i2}) the behaviour
of the TCSA spectrum at large $h$, in particular the second flow mentioned in
the Introduction, is explained by the behaviour of $s_0$ and $s_1$ at large
$h$: $s_0 \propto h$, $s_1$ is bounded from above, therefore $s_1/s_0$ tends to
zero. It is a further problem to give an analytic derivation of this behaviour
of $s_0$ and $s_1$.  

We shall investigate the scaling properties of $s_1/s_0$,
$s_1$ and $s_0$ and the TCSA and MT spectra
in a further publication \cite{TW}.
It is also interesting to extend the investigations to other
perturbed boundary conformal minimal models, which show similar behaviour
numerically to the model that we have studied. Certain results concerning
other minimal models and scaling properties already exist
\cite{FGPW,talk}; see also \cite{CLM}.
It is a further problem to
classify the possible behaviours of truncated spectra at large $h$ for various
truncation schemes.  Finally, the quantum field
theoretic description of the model (\ref{eq.yyy}) could be developed further
and extended to case of massive particles.

\section*{Acknowledgments}

I would like to thank G\'erard Watts for proposing the problem and
for several useful discussions,  Zolt\'an Bajnok, G\'abor
Tak\'acs and L\'aszl\'o Palla for useful discussions, and the Department of
Mathematics at King's College London, where part of this work was carried out, for hospitality.
I was supported by the Hungarian fund OTKA (grants T037674 and K60040), by the EUCLID
research training network (HPRN-CT-2002-00325) at various times and by the
Marie Curie Training Site 'Strings, Branes and Boundary Conformal Field
Theory' (MCFH-2001-00296) at King's College London.

\appendix

\section{Distributions on closed line segments}
\label{sec.distr}

\noindent
The necessary formulae for the
Dirac delta $\delta(x)$ and step function $\Theta(x)$ distributions on the
closed 
interval  $[0,L]\subset \RR$ are the following:

\begin{alignat}{2}
\int_0^L\delta(x-a)f(x)\ \rmd x &=f(a) &&\quad\quad \mbox{if}\quad a\in (0,L)\\
\int_0^L\delta(x-a)f(x)\ \rmd x &=\frac{1}{2}f(0) &&\quad\quad \mbox{if}\quad a=0\\
\int_0^L\delta(x-a)f(x)\ \rmd x &=\frac{1}{2}f(L) &&\quad\quad \mbox{if}\quad a=L\\
\int_0^L\delta(x-a)f(x)\ \rmd x &=0 &&\quad\quad \mbox{if}\quad a\not\in [0,L]\ ,
\end{alignat}
where $f$ is a function defined on $[0,L]$, and $x\in [0,L]$.
\begin{equation}
\Theta(x-a)=0 \quad \mbox{if}\quad  x<a\ ,\qquad\quad
\Theta(x-a)=1 \quad \mbox{if}\quad x\ge a\ ,
\end{equation}
where $a \in \RR$, 
\begin{alignat}{2}
\partial_x \Theta(x-L) & =  2\delta(x-L) && \\
\partial_x \Theta(x-a) & =  \delta(x-a)  &&\qquad\mbox{if}\quad a\in (0,L) \\
\partial_x \Theta(x-a) & =  0  &&\qquad \mbox{if}\quad a \not\in (0,L]\ ,  
\end{alignat}
where $x \in [0,L]$. 

\begin{equation}
\label{eq.di1}
\sum_{k\in \frac{\pi\ZZ}{L}}\exp[\rmi k(x-x')]=2L\delta(x-x')
\end{equation}
\begin{equation}
\label{eq.di2}
\sum_{k\in \frac{\pi\ZZ}{L}}\exp[\rmi k(x+x')]=2L[\delta(x+x')+\delta(x+x'-2L)]\ ,
\end{equation}
where $x,x' \in [0,L]$.

\section{Extension to Section \ref{sec.eel}}
\label{app.folyt}

In this appendix we present further details for Section \ref{sec.eel}.

The anticommutators of the $b(k)$ can be obtained in the following way:
we define a scalar product on the classical complex valued solutions of the
equations of motion:
\begin{equation}
\label{scal}
\braket{\psi_1,\psi_2,a_2}{\phi_1,\phi_2,b_2}=\int_0^L
[\psi^*_1\phi_1+\psi_2^*\phi_2] \rmd x + 2La_2^*b_2\ .
\end{equation}
This product should be calculated at a fixed time. Using the equations of
motion it can be shown that the product is independent of this time. 

The essential properties of this scalar product are that it is defined by a
local expression and that the $n(k)$ are orthogonal with respect to it:
\begin{equation}
\braket{n(k_1)}{n(k_2)}=\delta_{k_1-k_2,0}(2L+\frac{\sin(2k_1L)}{k_1})\ .
\end{equation}
The creation/annihilation operators can be expressed in the following way:
\begin{equation}
\label{eq.86}
\braket{n(k)}{(\Psi_1,\Psi_2,A_2)}=b(k)\braket{n(k)}{n(k)}
\end{equation}
Using the formula (\ref{scal}) and the anticommutation relations (\ref{ac1})-(\ref{ac2}) we get
\begin{equation}
\{ b(k_1),b(k_2)\} = \delta_{k_1+k_2,0}\frac{4Lk_1}{2Lk_1+\sin(2Lk_1)}\ .
\end{equation}
We remark that 
the above scalar
product technique is also suitable for free fields on the half-line or in the
usual full Minkowski space without boundaries in arbitrary
spacetime dimensions. 

The nonzero matrix elements of the fields are 
\begin{align}
\frac{\brakettt{P}{(\Psi_1(x,t),\Psi_2(x,t),A_2(t))}{Q}}{\sqrt{|\braket{P}{P}\braket{Q}{Q}|}}
& =   n(k)
(-1)^m \sqrt{f(k)}\\
\frac{\brakettt{Q}{(\Psi_1(x,t),\Psi_2(x,t),A_2(t))}{P}}{\sqrt{|\braket{P}{P}\braket{Q}{Q}|}}
& =  n(-k)
(-1)^m \sqrt{f(k)}\ ,
\end{align}
where $\ket{Q}=b(k_1)b(k_2)\dots b(k_n)\ket{0_h}$,
$\ket{P}=b(k_1)\dots b(k_m)b(k)b(k_{m+1})\dots b(k_n)\ket{0_h}$,
\begin{equation}
f(k)=\frac{4Lk}{2Lk+\sin(2Lk)}\ ,
\end{equation}
\begin{equation}
\braket{Q}{Q}=\prod_{i=1}^n f(k_i)\ .
\end{equation}

The Hamiltonian operator can be written as 
\begin{equation}
H=\sum_{k\in S,\ k>0} \frac{k}{f(k)}b(k)b(-k)\ .
\end{equation}

The following formula can be written for $\ket{0_h}$:
\begin{equation}
N\ket{0_h}=\lim_{\alpha
  \to \infty} \rme^{-\alpha H}\ket{v}=\prod_{k\in S,\
  k>0}\left(1-\frac{1}{f(k)}b(k)b(-k)\right) \ket{v}\ ,
\end{equation}
where $N$ is a normalization factor. 
The second equation on the right-hand
side can be verified directly using the following formulae:
$[b(k_1)b(-k_1),b(k_2)b(-k_2)]=0$
if $k_1 \ne k_2$ (and $k_1,k_2 >0$), and
$(b(k)b(-k))^n=f(k)^{n-1}b(k)b(-k)$.

We remark that we have not given a mathematically completely rigorous
proof  that (\ref{eq.mode}) satisfies (\ref{ac1})-(\ref{ac2}), but we think that this would be
possible.

The expansion (\ref{eq.mode})  and (\ref{ac1})-(\ref{ac2}) imply that the following 
formulae hold:
\begin{align}
\label{eq.ortog}
\sum_{k\in S}f(k)\psi_1(k)(x,0)\psi_1(-k)(y,0)  & =   4L\delta(x-y)\\
\sum_{k\in S}f(k)\psi_2(k)(x,0)\psi_2(-k)(y,0)  & =  4L\delta(x-y)\\
\sum_{k\in S}f(k)\psi_1(k)(x,0)\psi_2(-k)(y,0)  & =  -4L[\delta(x+y)+\delta(x+y-2L)]\\
\sum_{k\in S}f(k)a_2(k)(0)\psi_1(-k)(x,0)   & =   0\\
\sum_{k\in S}f(k)a_2(k)(0)\psi_2(-k)(x,0)   & =   0\\
\sum_{k\in S}f(k)a_2(k)(0)a_2(-k)(0)   & =   2\ .  \label{eq.ortog2}
\end{align}
These formulae are generalizations of (\ref{eq.di1}), (\ref{eq.di2}).

Using the formulae
\begin{align}
a(k) & =  \frac{1}{2\sqrt{2}L}[ \int_0^L \rme^{\rmi kx} \Psi_1(x,0)\ \rmd x - \int_0^L
\rme^{-\rmi kx} \Psi_2(x,0)\ \rmd x ] \\
A_1 & =   \frac{1}{2L}[ \int_0^L  \Psi_1(x,0)\ \rmd x - \int_0^L
 \Psi_2(x,0)\ \rmd x ]
\end{align}
and (\ref{eq.mode}) we obtain the following relations:
\begin{align}
\label{eq.bg1}
a(k) & =   \frac{-1}{\sqrt{2}L}\sum_{k' \in S} b(k') \frac{\sin[(k-k')L]}{k-k'}\\
A_1 & =  \frac{-1}{L}\sum_{k' \in S} b(k') \frac{\sin[k'L]}{k'}\\
\label{eq.bg2}
A_2(0) & =   \sum_{k\in S} b(k) \frac{-\rmi\sin(kL)}{4Lh}  =  -\rmi\sum_{k\in S} b(k)
\frac{\sin(kL)}{\sqrt{kL\tan(kL)}}\ .
\end{align}
Using (\ref{eq.86}) we get the relation
\begin{eqnarray}
\label{eq.bg3}
b(k)\left( 2L+\frac{\sin(2kL)}{k} \right) = \sqrt{2} \sum_{k'\in \frac{\pi}{L}\ZZ} a(k')
\frac{-2\sin[(k'-k)L]}{k'-k} \nonumber\\
+ \frac{\rmi\sin(kL)}{2h}A_2(0)\ .
\end{eqnarray}
(\ref{eq.bg1})-(\ref{eq.bg3}) can be regarded as Bogoliubov transformation
formulae for this model.

In the  $h\to 0$ limit 
\begin{eqnarray}
b(k(n,h)) & \to & -\sqrt{2}a(k(n,0))\qquad n\ge 1\\
b(k(0,h))+b(k(0,h))^\dagger & \to & -A_1\\
\rmi(b(k(0,h))^\dagger -b(k(0,h))) & \to & A_2(0)\\
\ket{0_h} & \to & \ket{v}\ .
\end{eqnarray}

The following boundary conditions are satisfied:
\begin{equation}
\label{eq.bc0}
\brakettt{E_1}{\Psi_1(0,t)+\Psi_2(0,t)}{E_2} =  0\qquad
% \label{eq.bc00}
\brakettt{E_1}{\Psi_1(L,t)+\Psi_2(L,t)}{E_2} =  0
\end{equation}
and
\begin{alignat}{2}
\label{eq.bc1}
&\lim_{x\to 0}\brakettt{E_1}{\Psi_1(x,t)+\Psi_2(x,t)}{E_2}&&= 0    \\
& \lim_{x\to
  L}\brakettt{E_1}{\partial_x\Psi_1(x,t)-\partial_x\Psi_2(x,t)}{E_2} & &
\nonumber\\
\label{eq.bc2}
& & &\hspace{-2cm} =  16Lh^2\brakettt{E_1}{\Psi_2(L,t)-\Psi_1(L,t)}{E_2} \\
\label{eq.kkkk1}
& \lim_{h\to\infty}\lim_{x\to L}\brakettt{E_1}{\Psi_1(x,t)-\Psi_2(x,t)}{E_2}& & =
 0 \\
\label{eq.kkkk2}
& \lim_{h\to\infty}\lim_{x\to L}\brakettt{E_1}{\Psi_1(x,t)+\Psi_2(x,t)}{E_2} & &
\ne   0\ , 
\end{alignat}
where $\ket{E_1}$ and $\ket{E_2}$ are eigenstates of $H$. 
The boundary conditions (\ref{eq.bc0})  are the same as
those satisfied by the free fields. On the other hand, (\ref{eq.bc1}) and
(\ref{eq.bc2}) are similar to the boundary conditions written down in \cite{GZ,CZ,C1}.
From the
point of view of the boundary conditions one can say that the perturbation
$H_I$ induces a flow from the boundary condition
$\lim_{x \to L}\brakettt{E_1}{\Psi_1(x,t)+\Psi_2(x,t)}{E_2} =  0$ to the boundary
condition $\lim_{x\to L}\brakettt{E_1}{\Psi_1(x,t)-\Psi_2(x,t)}{E_2}  =
 0$ and the boundary condition on the left-hand side remains constant, which
 is in accordance with the literature (see e.g.\ \cite{GZ}).
The boundary condition $\brakettt{E_1}{\Psi_1(L,t)+\Psi_2(L,t)}{E_2} =  0$ is
called free spin boundary condition in the literature (e.g.\ \cite{GZ}) and 
$\brakettt{E_1}{\Psi_1(L,t)-\Psi_2(L,t)}{E_2}=0$ is called fixed spin boundary
condition. 

We remark that from
(\ref{eq.bc1}) and (\ref{eq.bc2}) and the bulk equations of motion
$(\partial_t+\partial_x)\Psi_1=0$, $(\partial_t-\partial_x)\Psi_2=0$  equation (\ref{spektr}) can be recovered.

We define the fields
\begin{equation}
\Phi_1(x,t)  =  \sum_{k\in S} -b(k)\rme^{\rmi k(t-x)}\qquad
\Phi_2(x,t)  =  \sum_{k\in S} b(k)\rme^{\rmi k(t+x)}\ .
\end{equation}
They satisfy the following equations: 
\begin{align}
(\Psi_1-\Psi_2)(L,t) & =   \frac{1}{16Lh^2}\partial_x (\Phi_2-\Phi_1)(L,t)\\
A_2(t) & =   -\frac{1}{8Lh}(\Phi_1+\Phi_2)(L,t)
\end{align}
\begin{equation}
\label{eq.h000}
H_0  =  -\frac{\rmi}{8L}\int_0^L\ \rmd x\ \Phi_1(x,0)\partial_x \Phi_1(x,0) 
 +\frac{\rmi}{8L}\int_0^L\
\rmd x\ \Phi_2(x,0)\partial_x \Phi_2(x,0)\ -\frac{1}{2}hH_I\ .
\end{equation}
Note that the energies of the modes are not in $\pi\ZZ/L$, so one
cannot conclude that the sum of the first two terms in (\ref{eq.h000}) equals to $H$.

The above equations suggest how to describe the model discussed in Section \ref{sec.eel}
and in this appendix 
as a perturbation of the $h \to \infty$ limiting model. This description
is discussed Section \ref{sec.reverse}.
The boundary conditions in this case are 
\begin{equation}
\label{eq.revbc0}
\brakettt{E_1}{\Phi_1(0,t)+\Phi_2(0,t)}{E_2} =  0\qquad
% \label{eq.revbc00}
\brakettt{E_1}{\Phi_1(L,t)-\Phi_2(L,t)}{E_2}  =  0
\end{equation}
and
\begin{alignat}{2}
\label{eq.revbc1}
& \lim_{x\to 0}\brakettt{E_1}{\Phi_1(x,t)+\Phi_2(x,t)}{E_2}&& =  0\\
&\lim_{x\to
  L}\brakettt{E_1}{\partial_x\Phi_1(x,t)-\partial_x\Phi_2(x,t)}{E_2} &&
\nonumber\\
\label{eq.revbc2}
& && \hspace{-2cm} = \frac{1}{16Lg}\brakettt{E_1}{\Phi_1(L,t)-\Phi_2(L,t)}{E_2} \\
& \lim_{g\to\infty}\lim_{x\to L}\brakettt{E_1}{\Phi_1(x,t)+\Phi_2(x,t)}{E_2} && =
 0\\
&\lim_{g\to\infty}\lim_{x\to L}\brakettt{E_1}{\Phi_1(x,t)-\Phi_2(x,t)}{E_2} 
&& \ne   0\ ,
\end{alignat}
where $\ket{E_1}$ and $\ket{E_2}$ are eigenstates of $H$.

\section{Bethe-Yang equations}
\label{app.by}

The Bethe-Yang equations can be used to give a description of the spectrum  of
models in finite volume which have factorized
scattering in their infinite volume limit. The Bethe-Yang equations for
relativistic models
defined on a cylinder are exposed, for example, in \cite{ZamTBA,KM1,key-21};  
for models defined  on the strip they
are written down in \cite{FeSa,BPTby}. It should be noted that the Bethe-Yang
equations usually
give approximate result only.  

In the case of the  model that we study the ingredients of the Bethe-Yang description are the
following: there is a single massless particle with fermionic statistics,  the
two-particle S-matrix
is a constant scalar $S(k)=-1$, where $k$ is the relative momentum. The
reflection matrix on the left-hand side can be read from (\ref{eq.bc1}), it is
$R_L(k)=-1$; the reflection matrix on the right-hand side can be read from
(\ref{eq.bc2}), it is 
\begin{equation}
R_R(k)=\frac{16Lh^2+\rmi k}{16Lh^2-\rmi k}\ .
\end{equation}

The transfer matrices for $N$-particle states are  scalars:
\begin{eqnarray}
T_i(k_1,k_2,\dots,k_N)
=R_L(k_i)R_R(k_i)\prod_{j,\ j\ne k}S(k_i+k_j)\prod_{j,\ j\ne
  k}S(k_i-k_j)=-R_R(k_i)\ ,\nonumber\\  i=1\dots N\ ,
\end{eqnarray}
where $k_1>k_2>\dots k_N \ne 0$. This very simple form is the consequence of
the simplicity of the S-matrix.
The   Bethe-Yang equations
for the momenta $k_1,k_2,\dots k_N$ of the $N$-particle   states take the form
\begin{equation}
\label{eq.BY2}
\rme^{2\rmi k_iL}T_i(k_1,k_2,\dots,k_N)=\rme^{2\rmi k_iL}\frac{\rmi
  k_i+16Lh^2}{\rmi k_i-16Lh^2}=1\
,\qquad i=1\dots N\ . 
\end{equation}    
The total energy of an $N$-particle state in the Bethe-Yang framework is 
$E=\sum_{i=1}^N k_i$.
(\ref{eq.BY2}) can be rewritten as 
$k_iL\tan(k_iL)=16L^2h^2,\ i=1\dots N$,
which has the same form as (\ref{spektr}). This means that the Bethe-Yang
description reproduces the result of Section \ref{sec.pertmod} for the
spectrum exactly.

The reverse model is similar, one can read from (\ref{eq.revbc1}) and (\ref{eq.revbc2})
that 
\begin{equation}
R_L(k)=-1\ ,\qquad R_R(k)=\frac{1-\rmi k16Lg}{1+\rmi k16Lg}\ ,
\end{equation}
and the Bethe-Yang
equations for the momenta can be written as
\begin{equation}
k_iL\tan(k_iL)=\frac{-1}{16g}\ , \qquad i=1\dots N\ ,
\end{equation}
which has the same form as (\ref{spektrrev}), i.e.\  the result of Section
\ref{sec.reverse} for the spectrum is reproduced exactly.

\section{Power series expansion of the energy levels}
\label{sec.RS}

The eigenvectors of $H_0$ suitable for Rayleigh-Schr\"odinger perturbation
theory are those introduced in (\ref{eq.freebasis}). Degenerate perturbation
theory has to be used.

The nonzero matrix elements of $H_I$ in Section \ref{sec.pertmod} are the following:
\begin{alignat}{2}
-\brakettt{Qv}{H_I}{Qv} & = \brakettt{Qu}{H_I}{Qu} && =  2\\
\brakettt{Qu}{H_I}{Pv} & = \brakettt{Pv}{H_I}{Qu} && =  2\sqrt{2}(-1)^{n+m}(-1)^{kL/\pi}\\
\brakettt{Qv}{H_I}{Pu} & =\brakettt{Pu}{H_I}{Qv} && =  -2\sqrt{2}(-1)^{n+m}(-1)^{kL/\pi}\ ,
\end{alignat}
where $P  =  a(k_1)a(k_2)\dots a(k_m)a(k)a(k_{m+1})\dots a(k_n)$, 
$Q =  a(k_1)a(k_2)\dots a(k_n)$.

We remark that certain perturbative calculations were also done in \cite{CSS}.

The eigenvalue of the state  starting from
$ a(\frac{N_1\pi}{L})
  a(\frac{N_2\pi}{L})\dots  a(\frac{N_r\pi}{L})\ket{w}$ at $h=0$ is
\begin{multline}
\label{eq.rspert}
E_{ \{ N_1,N_2,\dots ,N_r,w\} }(h) =
\frac{(N_1+N_2+\dots +N_r)\pi}{L} +2\epsilon h \\
+\left( \frac{16L}{\pi}
\left(\frac{1}{N_1}+\frac{1}{N_2}+\dots +\frac{1}{N_r}\right)  
-\sum_{n=1}^{n_m} \frac{8L}{n\pi} \right)h^2
-\epsilon  \sum_{n=1}^{n_m} \frac{32L^2}{n^2\pi^2} h^3+ \dots,
\end{multline}
where $w$ stands for $v$ or $u$, $\epsilon=-1$ if $w=v$,  
$\epsilon=1$ if $w=u$.
$n_m=n_c$ in the MT scheme ($n_c$ is the truncation parameter introduced in Section \ref{sec.modetrfree}),  
$n_m=n_c-(N_1+N_2+\dots +N_r)$ in the TCSA scheme ($n_c$ is the conformal
truncation parameter), and $n_m=\infty$ in the non-truncated case.

In the non-truncated case the  coefficient of $h^2$ is ultraviolet divergent and should be
regularized. The TCSA and the mode truncation both provide a regularization.

We remark that the above formulae show that the truncated energy gaps
converge to the non-truncated energy gaps as $1/n_c$.


\small
\begin{thebibliography}{11}


\bibitem{YZ1}Yurov V P and Zamolodchikov Al B, \emph{Truncated conformal space approach
 to scaling Lee-Yang model}, 1990 \emph{Int.\ J.\ Mod.\ Phys.} A \textbf{5} 
 3221--46 



\bibitem{FRT1}Feverati G, Ravanini F and Tak\'acs G, \emph{Nonlinear integral
  equation and finite volume spectrum of minimal models perturbed by
  Phi(1,3)}, 2000
\emph{Nucl.\ Phys.} B \textbf{570} 615--43  [hep-th/9909031];
Feverati G, Ravanini F and Tak\'acs G,
 \emph{Scaling functions in the
  odd charge sector of sine-Gordon / massive Thirring theory}, 1998 \emph{Phys.\
  Lett.} B \textbf{444} 442--50  [hep-th/9807160];
Feverati G, Ravanini F and Tak\'acs G, 
 \emph{Nonlinear integral
  equation and finite volume spectrum of sine-Gordon theory}, 1999 \emph{Nucl.\
  Phys.} B \textbf{540}  543--86 [hep-th/9805117];
Feverati G, Ravanini F and Tak\'acs G,
  \emph{Truncated conformal space
  at c=1, nonlinear integral equation and quantization rules for multi-soliton
  states}, 1998 \emph{Phys.\ Lett.} B \textbf{430} 264--73  [hep-th/9803104]




\bibitem{PT}Pozsgay B and Tak\'acs G, \emph{Characterization of resonances using
  finite size effects}, 2006 \emph{Nucl.\ Phys.} B \textbf{748}  485--523 [hep-th/0604022]



\bibitem{BPTW}Bajnok Z, Palla L, Tak\'{a}cs G and W\'{a}gner F, \emph{A nonperturbative
study of the two-frequency sine-Gordon model}, 2000 \emph{Nucl.\ Phys.}
B \textbf{601} 503--38 [hep-th/0008066] 

 
\bibitem{sajat2}T\'oth G Zs, \emph{A non-perturbative study of phase transitions
in the multi-frequency sine-Gordon model}, 2004 \emph{J.\ Phys.} A \textbf{37} 
9631--50 [hep-th/0406139]



\bibitem{Kormos}Kormos M, \emph{Boundary renormalisation group flows of unitary
  superconformal minimal models}, 2006 \emph{Nucl.\ Phys.} B \textbf{744} 
  358--79  [hep-th/0512085] 



\bibitem{RRS}Recknagel A, Roggenkamp D and Schomerus V,  
\emph{On Relevant Boundary
 Perturbations of Unitary Minimal Models}, 2000
 \emph{Nucl.\ Phys.} B \textbf{588}  552--64 [hep-th/0003110]


\bibitem{Watts3}Dorey P, Pocklington A, Tateo R and Watts G M T, \emph{TBA and TCSA
  with Boundaries and Excited States}, 1998 \emph{Nucl.\ Phys.} B \textbf{525} 
  641--63  [hep-th/9712197]



\bibitem{Watts2}Graham K, Runkel I and Watts G M T, \emph{Renormalisation Group Flows of Boundary
  Theories, 4th Annual European TMR Conference on Integrability},
  Nonperturbative Effects and Symmetry in Quantum Field Theory, Paris, France,
  7-13 Sep 2000
  [hep-th/0010082]

\bibitem{Watts4}Graham K, Runkel I and Watts G M T, \emph{Boundary Renormalisation Group Flows of
  Minimal Models}, \emph{Non-perturbative QFT methods and their applications:
  proc. of the 24th Johns Hopkins Workshop  on Current Problems in Particle
  Theory, Budapest, Hungary}, ed Z Horvath and L Palla (Singapore: World Scientific, 2001) pp 95--113 

\bibitem{Watts1}
 Dorey P, Pillin M, Pocklington A, Runkel I, Tateo R and Watts G M T, 
 \emph{Finite Size Effects in Perturbed Boundary Conformal Field Theories},
  4th Annual European TMR Conference on Integrability, Nonperturbative Effects
 and Symmetry in Quantum Field Theory, Paris, France, 7--13 Sep 2000
 [hep-th/0010278]

\bibitem{C2}Chatterjee R,  \emph{Exact Partition Function and Boundary State of 2-D
  Massive Ising Field Theory with Boundary Magnetic Field}, 1996 \emph{Nucl.\ Phys.}
  B \textbf{468}
  439--60  [hep-th/9509071]


\bibitem{GW}Graham K and Watts G M T,  \emph{Defect lines and boundary flows},
  2004 \emph{JHEP} \textbf{0404} 019  [hep-th/0306167] 

\bibitem{Fr}Fredenhagen S,   \emph{Organizing  Boundary RG Flows}, 2003 \emph{Nucl.\
  Phys.} B \textbf{660},
  436--72  [hep-th/0301229]

\bibitem{LSS}Lesage F, Saleur H and Simonetti P, \emph{Boundary flows in minimal
models}, 1998 \emph{Phys.\ Lett.} B \textbf{427}  85--92  [hep-th/9802061] 


\bibitem{CAZ}Cappelli A, D'Appollonio G and Zabzine M,  \emph{Landau-Ginzburg
description of boundary critical phenomena in two-dimensions}, 2004 \emph{JHEP}
\textbf{0404}  010 [hep-th/0312296] 


\bibitem{Kl-M}Klassen T R and Melzer E,  \emph{Spectral flow between conformal field theories
in 1+1 dimensions}, 1992 \emph{Nucl.\ Phys.} B \textbf{370}   511--50


\bibitem{CLM}Cardy J L, L\"assig M and Mussardo G, \emph{The scaling region of 
the tricritical Ising model in two-dimensions}, 1991 \emph{Nucl.\ Phys.} 
B \textbf{348} 591--618 


\bibitem{prc}Watts G M T, 2004, private communication

\bibitem{talk}Feverati G, Graham K, Pearce P A , T\'oth G Zs and Watts G, 
\emph{A Renormalisation group for TCSA}, talk presented by G. Watts at the workshop ``Integrable
Models and Applications: from Strings to Condensed Matter'', Santiago de
Compostela, Spain, 12-16 September 2005 [hep-th/0612203] 



\bibitem{Cardy1}Cardy J,  \emph{Boundary Conditions, Fusion Rules And The Verlinde
 Formula}, 1989 \emph{Nucl.\ Phys.} B \textbf{324}  581

\bibitem{CL}Cardy J and Lewellen D,  \emph{Bulk and boundary operators in conformal
field theory}, 1991 \emph{Phys.\ Lett.} B \textbf{259}  274--78


\bibitem{GZ}Ghoshal S and Zamolodchikov A,  \emph{Boundary S matrix and boundary state
in two-dimensional integrable quantum field theory}, 1994 \emph{Int.\ J.\
Mod.\ Phys.} A \textbf{9}  3841--86, Erratum-ibid.\ A \textbf{9} 4353 [hep-th/9306002] 


\bibitem{AL2}Affleck I and Ludwig A W W, \emph{Universal Noninteger ``Ground-State
  Degeneracy'' in Critical Quantum Systems}, 1991 \emph{Phys.\ Rev.\ Lett.}
  \textbf{67}  161--4


\bibitem{FGPW}Feverati G, Graham K, Pearce P A and Watts G M T, to appear


\bibitem{C1}Chatterjee R, \emph{Exact Partition Function and Boundary State of
  Critical Ising Model with Boundary Magnetic Field}, 1995 \emph{Mod.\ Phys.\
  Lett.} A \textbf{10} 
  973--84  [hep-th/9412169]


\bibitem{CZ}Chatterjee R and Zamolodchikov A,  \emph{Local Magnetization in Critical
  Ising Model with Boundary Magnetic Field}, 1993 [hep-th/9311165]


\bibitem{LMSS}LeClair A, Mussardo G, Saleur H and Skorik S,  \emph{Boundary energy
  and boundary states in integrable quantum field theories}, 1995 \emph{Nucl.\
  Phys.} B  \textbf{453}
   581--618 
 [hep-th/9503227]

\bibitem{Kon}Konechny A,  \emph{Ising model with a boundary magnetic field: an
  example of a boundary flow}, 2004 \emph{JHEP} \textbf{0412} 058
  [hep-th/0410210]


\bibitem{KLeM}Konik R, LeClair A and Mussardo G,  \emph{On Ising correlation
functions with boundary magnetic field}, 1996 \emph{Int.\ J.\ Mod.\ Phys.}  
A \textbf{11}  2765--82  [hep-th/9508099] 
 

\bibitem{FMS}Di Francesco P, Mathieu P and S\'en\'echal D, \emph{Conformal Field
  Theory}, Graduate Texts in Contemporary Physics, Springer-Verlag 

\bibitem{Ginsparg}Ginsparg P, \emph{Applied Conformal Field Theory}, 
1989 Fields, Strings and Critical Phenomena 
(Les Houches, Session XLIX, 1988) ed E Br\'ezin and J Zinn-Justen (Elsevier)
[hep-th/9108028]  

\bibitem{TW}T\'oth G Zs and Watts G M T, to appear  

\bibitem{ZamTBA}Zamolodchikov A B, \emph{Thermodynamic Bethe Ansatz In
    Relativistic Models. Scaling Three State Potts And Lee-Yang Models}, 1990 \emph{Nucl.\
Phys.} B \textbf{342} 695--720 

\bibitem{KM1}Klassen T R and Melzer E, \emph{Kinks in Finite Volume}, 1992 \emph{Nucl.\
Phys.} B \textbf{382}  441--85
[hep-th/9202034]


\bibitem{key-21}Bajnok Z, Palla L, Tak\'{a}cs G and W\'{a}gner F, \emph{The k-folded
sine-Gordon model in finite volume}, 2000 \emph{Nucl.\ Phys.} B \textbf{587}  585--618 [hep-th/0004181] 


\bibitem{FeSa}Fendley P and Saleur H, \emph{Deriving boundary S matrices}, 1994
\emph{Nucl.\ Phys.} B \textbf{428}  681--93  [hep-th/9402045]


\bibitem{BPTby}Bajnok Z, Palla L and Tak\'acs G, \emph{Boundary states and finite
size effects in sine-Gordon model with Neumann boundary condition}, 2001
\emph{Nucl.\ Phys.} B \textbf{614} 405--48 [hep-th/0106069]  

\bibitem{CSS}Caux J S, Saleur H and Siano F, \emph{The two-boundary sine-Gordon
  model}, 2003
 \emph{Nucl.\ Phys.} B \textbf{672}  411--61 [cond-mat/0306328]



\end{thebibliography}
\end{document}